\documentclass{aa} 
\usepackage{graphicx}
\usepackage{txfonts}
\usepackage{longtable}
\usepackage{subfig}
\usepackage{float}
\usepackage{booktabs}
\usepackage{natbib}
\bibpunct{(}{)}{;}{a}{}{,} % to follow the A&A style
\usepackage{amssymb}
\usepackage{pifont}% http://ctan.org/pkg/pifont
\usepackage{ gensymb }

\begin{document}
   \title{X-ray spectral variability of Seyfert 2 galaxies}

   \subtitle{}

 \author{Hern\'{a}ndez-Garc\'{i}a, L.\inst{1}; Masegosa, J.\inst{1};
   Gonz\'{a}lez-Mart\'{i}n, O.\inst{2}; M\'{a}rquez, I. \inst{1} }

   \institute{Instituto de Astrof\'{i}sica de Andaluc\'{i}a, CSIC,
     Glorieta de la Astronom\'{i}a, s/n, 18008 Granada,
     Spain\\ \email{lorena@iaa.es} \and Centro de radioastronom\'{i}a
     y Astrof\'{i}sica (CRyA-UNAM), 3-72 (Xangari), 8701, Morelia,
     Mexico \\ }

   \date{Received XXXX; accepted YYYY}

\authorrunning{Hern\'{a}ndez-Garc\'{i}a et al.}
\titlerunning{X-ray variability of Seyfert 2s}

% \abstract{}{}{}{}{} 
% 5 {} token are mandatory
 
  \abstract
  % context heading (optional)
  % {} leave it empty if necessary 
{Variability across the
  electromagnetic spectrum is a property of active galactic nuclei
  (AGN) that can help constrain the physical properties of these
  galaxies. Nonetheless, the way in which the changes happen and
  whether they occur in the same way in every AGN are still open
  questions. }
  % aims heading (mandatory)
   {This is the third in a series of papers with the aim of studying
     the X-ray variability of different families of AGN. The main
     purpose of this work is to investigate the variability pattern(s)
     in a sample of optically selected Seyfert 2 galaxies.}
  % methods heading (mandatory)
   {We use the 26 Seyfert 2s in the V\'{e}ron-Cetty and V\'{e}ron
     catalog with data available from \emph{Chandra} and/or
     \emph{XMM}--Newton public archives at different epochs, with
     timescales ranging from a few hours to years. All the spectra of
     the same source were simultaneously fitted, and we let different
     parameters vary in the model. Whenever possible, short-term
     variations from the analysis of the light curves and/or long-term
     UV flux variations were studied. We divided the sample into
     \emph{\emph{Compton}}-thick and \emph{\emph{Compton}}-thin
     candidates to account for the degree of obscuration. When
     transitions between \emph{\emph{Compton}}-thick and thin were
     obtained for different observations of the same source, we
     classified it as a changing-look candidate. }
  % results heading (mandatory)
   {Short-term variability at X-rays was studied in ten cases, but
     variations are not found. From the 25 analyzed sources, 11 show
     long-term variations. Eight (out of 11) are
     \emph{\emph{Compton}}-thin, one (out of 12) is
     \emph{\emph{Compton}}-thick, and the two changing-look candidates
     are also variable. The main driver for the X-ray changes is
     related to the nuclear power (nine cases), while variations at
     soft energies or related to absorbers at hard X-rays are less
     common, and in many cases these variations are accompanied by
     variations in the nuclear continuum. At UV frequencies, only
     NGC\,5194 (out of six sources) is variable, but the changes are
     not related to the nucleus. We report two changing-look
     candidates, MARK\,273 and NGC\,7319.}
  % conclusions heading (optional), leave it empty if necessary 
   { A constant reflection component located far away from the nucleus
     plus a variable nuclear continuum are able to explain most of our
     results. Within this scenario, the \emph{\emph{Compton}}-thick
     candidates are dominated by reflection, which suppresses their
     continuum, making them seem fainter, and they do not show
     variations (except MARK\,3), while the \emph{\emph{Compton}}-thin
     and changing-look candidates do.}


   \maketitle
%
%________________________________________________________________

\newcommand{\xmark}{\ding{55}}%
\newcommand{\cmark}{\ding{51}}%

\begin{table*}
\begin{center}
\caption{\label{properties} General properties of the sample galaxies.}
\begin{tabular}{lccccccccc} \hline
\hline
Name    & RA & DEC &  Dist.$^1$  & N$_{Gal}$ & $m_V$ & Morph.  &  HBLR & Ref.  \\
 & (J2000)  & (J2000) & (Mpc) & ($10^{20}$ cm$^{-2}$) & & type  & & \\
(1) & (2) & (3) & (4) & (5) & (6) & (7) & (8) & (9)   \\  \hline
MARK\,348   &       0 48 47.2 &   31 57 25 & 63.90  & 5.79 &  14.59   &  S0-a  &  \cmark & 1 \\ 
NGC\,424    &       1 11 27.7 &  -38  5  1 &  47.60 & 1.52 &  14.12   &  S0-a  &  \cmark & 1  \\ 
MARK\,573   &       1 43 57.8 &    2 20 59 & 71.30  & 2.52 &  14.07   & S0-a   &  \cmark & 1  \\ 
NGC\,788    &       2  1  6.5 &  - 6 48 56 & 56.10  & 2.11 &  12.76   & S0-a   &  \cmark & 1  \\ 
 ESO\,417-G06   &    2 56 21.5 &  -32 11  6 & 65.60 & 2.06 &  14.30     &  S0-a   &  - \\ 
 MARK\,1066     &    2 59 58.6 &   36 49 14 & 51.70 & 9.77 &  13.96    &  S0-a  &   \xmark & 2 \\  
 3C\,98.0      &    3 58 54.5 &   10 26  2 & 124.90 & 10.20 &  15.41    &  E  & - \\  
MARK\,3     &       6 15 36.3 &   71  2 15 & 63.20  & 9.67 &  13.34   & S0   &   \cmark & 1 \\ 
 MARK\,1210     &    8  4  5.9 &    5  6 50 & 53.60 & 3.45 &  13.70     &  -  & \cmark & 2 \\  
 NGC\,3079      &   10  1 58.5 &   55 40 50 & 19.10 & 0.89 &  12.18    & SBcd    & \xmark & 2 \\  
 IC\,2560       &   10 16 19.3 &  -33 33 59 & 34.80 & 6.40 &  13.31    &  SBb  & - \\  
 NGC\,3393      &   10 48 23.4 &  -25  9 44   & 48.70 & 6.03 &  13.95    & SBa   & - \\ 
NGC\,4507   &      12 35 36.5 &  -39 54 33 &  46.00 & 5.88 &  13.54   & Sab    &  \cmark & 1  \\  
    
 NGC\,4698      &   12 48 22.9 &    8 29 14 & 23.40 & 1.79 &  12.27    &  Sab   & - \\  
 NGC\,5194      &   13 29 52.4 &   47 11 41 & 7.85 & 1.81 &  13.47    &  Sbc   & \xmark & 2 \\   
 MARK\,268     &   13 41 11.1 &   30 22 41 & 161.50 & 1.37  &  14.66    &  S0-a   & - \\  
 MARK\,273     &   13 44 42.1 &   55 53 13 & 156.70 & 0.89  &  14.91    &  Sab   & - \\  
Circinus    &      14 13  9.8 &  -65 20 17 & 4.21  & 74.40 &  12.1    &  Sb   &   \cmark & 1 \\
 NGC\,5643      &   14 32 40.7 &  -44 10 28 & 16.90 & 7.86 &  13.60    &  Sc  & \xmark & 2 \\  
 MARK\,477     &   14 40 38.1 &   53 30 15 & 156.70 & 1.05  &  15.03    & E?    & \cmark & 2 \\  
 IC\,4518A      &   14 57 41.2 &  -43  7 56 & 65.20 & 8.21  &  15.      &  Sc  & - \\   
 ESO\,138-G01   &   16 51 20.5 &  -59 14 11 & 36.00 & 13.10 &  13.63    &  E-S0  & - \\  
 NGC\,6300      &   17 16 59.2 &  -62 49  5 & 14.43 & 7.76 &  13.08    &  SBb   & - \\  
 NGC\,7172      &   22  2  1.9 &  -31 52  8 & 33.90 & 1.48   &  13.61    &  Sa  & \xmark & 2 \\  
NGC\,7212   &      22  7  2.0 &   10 14  0 & 111.80  & 5.12 &  14.8    &  Sb  &  \cmark & 1  \\
 NGC\,7319      &   22 36  3.5 &   33 58 33 & 77.25 & 6.15 &  13.53    & Sbc    & - \\ 
\hline
\end{tabular}
\caption*{  (Col. 1) Name, (Col. 2) right ascension, (Col. 3) declination, (Col. 4) distance, (Col. 5) galactic absorption, (Col. 6) aparent magnitude in the Johnson filter V from \cite{veron2010}, (Col. 7) galaxy morphological type from Hyperleda, (Col. 8) hidden broad polarized lines detected, and (Col. 9) its refs.: (1) \cite{veron2010}; and (2) \cite{gu2002}. }
\end{center}
\vspace*{-0.5cm}
\footnote*{All distances are taken from the NED and correspond to the average redshift-independent distance estimates.}
\end{table*}

\section{\label{intro}Introduction}

It is widely accepted that active galactic nuclei (AGN) are powered by
accretion onto a supermassive black hole
\citep[SMBH,][]{rees1984}. Among them, the different classes of
Seyfert galaxies (type 1/type 2) have led to postulating a unified
model (UM) for all AGN \citep{antonucci1993, urrypadovani1995}.  Under
this scheme, the SMBH is fed by the accretion disk that is surrounded
by a dusty torus. This structure is responsible for obscuring the
region where the broad lines are produced (known as broad line region,
BLR) in type 2 objects, while the region where the narrow lines are
produced (narrow line region, NLR) is still observed at optical
frequencies. The difference between type 1 and 2 objects is therefore
due to orientation effects.

In agreement with the UM, the type 1/type 2 classifications at X-ray
frequencies are based on the absorption column density, $N_{H}$,
because it is related with the obscuring material along our line of
sight \citep{maiolino1998}; therefore, we observe a Seyfert 1 if
$N_{H} < 10^{22} cm^{-2}$, i.e., unobscured view of the inner parts of
the AGN, and a type 2 if the column density is higher, i.e., obscured
view through the torus \citep[e.g.,][]{risaliti2002b}. When $N_{H} >
1.5 \times 10^{24} cm^{-2}$, the absorbing column density is higher
than the inverse of the \emph{\emph{Compton}}-scattering
cross-section, and the sources are known as
\emph{\emph{Compton}}-thick \citep{maiolino1998}.

In fact, X-rays are a suitable tool for studying AGN because they are
produced very close to the SMBH and because of the much smaller effect
of obscuration at these frequencies than at UV, optical, or
near-IR. Numerous studies have been made at X-ray frequencies to
characterize the spectra of Seyfert galaxies
\citep[e.g.,][]{turner1997,risaliti2002,guainazzi2005a,guainazzi2005b,panessa2006,cappi2006,noguchi2009,lamassa2011,brightman2011}. The
present work is focused on Seyfert 2 galaxies, which represent
$\sim$ 80\% of all AGN \citep{maiolino1995}. The works mentioned above
have shown that the spectra of these objects are characterized by a
primary power-law continuum with a photoelectric cut-off, a thermal
component, a reflected component, and an iron emission line at 6.4
keV. It is important to appropiately account for the physical
parameters of their spectra in order to constrain physical properties
of the nuclei.

Given that variability across the electromagnetic spectrum is a
property of all AGN, understanding these variations offers an
exceptional opportunity to constrain the physical characteristics of
AGN, which are known to show variations on timescales ranging from a
few days to years \citep{bradley1997}.  The first systematic
variability study of Seyfert 2 galaxies was performed by
\cite{turner1997} using \emph{ASCA} data. Their results show that
short-term variability (from hours to days) is not common in 
Seyfert 2s, in contrast to what is observed in Seyfert 1
\citep[e.g.,][]{nandra1997}. Because these galaxies are obscured by
the torus, the lack of variations could come from these sources being
reflection-dominated, as shown by some authors that studied
\emph{\emph{Compton}}-thick sources
\citep{awaki1991,lamassa2011,matt2013,arevalo2014b}. However, a number
of Seyfert 2s actually do show variations. The study of the
variability has been approached in different ways from the analysis of
the light curves to study of short-term variations \citep{awaki2006},
through count-rate or flux variations \citep{isobe2005,trippe2011}, or
comparisons of spectra of the same source at different epochs
\citep{lamassa2011, marinucci2013,marchese2014}.  The observed
variations may be related with absorbing material that crosses our
line of sight \citep{risaliti2002b,risaliti2010} and/or can be
intrinsic to the sources
\citep{evans2005,sobolewska2009,braito2013}. A few Seyfert 2s
also showed changes from being reflection-dominated to
transmission-dominated objects, so were called changing-look objects
\citep{guainazzi2002,guainazzi2002a,matt2003,risaliti2010}.

Although it is well established that a number of Seyfert 2s are
variable, it is unknown whether the same kind of variation is common
for all the nuclei or, more important, what drives those
variations. It is the purpose of this paper to systematically study
the variability pattern at X-rays in Seyfert 2 nuclei. This is the
third in a series of papers aimed at studying the X-ray variability in
different families of AGN. In \cite{lore2013,lore2014}, this study was
made for LINERs, while the study of Seyfert 1 and the comparison
between different families of AGN will be presented in forthcoming
papers.

This paper is organized as follows. In Sec. \ref{sample} the sample
and the data are presented, and data reduction is explained in
Sect. \ref{reduction}. The methodology used for the analysis is
described in Sect. \ref{method}, including individual and simultaneous
spectral fittings, comparisons using data with different instruments,
long-term X-ray and UV variations, short-term X-ray variations, and
\emph{\emph{Compton}}-thickness analysis.  The results derived from
this work are explained in Sect. \ref{results} and are discussed in
Sect. \ref{discusion}. Finally, the main conclusions are summarized in
Sect. \ref{conclusion}.

%__________________________________________________________________

\section{\label{sample}Sample and data}

We used the 13th edition of the V\'{e}ron-Cetty and V\'{e}ron
catalogue \citep{veron2010}, which contains quasars and active
galactic nuclei. We selected galaxies located at redshift below 0.05
and classified as Seyfert 2 (S2) or objects with broad polarized
Balmer lines detected (S1h). Indeed, S1h objects are those optically
classified as Seyfert 2 that show broad lines in polarized light, which
is the reason for their selection. This subsample includes 730 S2 and
27 S1h.

We searched for all the publicly available data for sources with
observations in more than one epoch with \emph{Chandra} and/or
\emph{XMM}--Newton using the
HEASARC\footnote{http://heasarc.gsfc.nasa.gov/} browser up to May
2014. This first selection includes 73 nuclei. To be able to properly
fit and compare spectra at different epochs, we selected sources with
a minimum of 400 number counts in the 0.5-10.0 keV energy band, as
required to use the $\chi^2$-statistics. Thirty-four galaxies and nine
observations did not met this criterium and were excluded from the
sample. Objects affected by a pileup fraction higher than 10\% were also
removed, which made us exclude three objects and 14
observations. 
  
For the
remaining 36 nuclei we searched for their optical classifications in
the literature with the aim of including only pure Seyfert 2 objects in
the sample. Nine galaxies were excluded following this condition:
NGC\,4258, and NGC\,4374 (S1.9 and L2 in \citealt{ho1997}), 3C\,317.0
and 3C\,353.0 (LINERs in NED\footnote{http://ned.ipac.caltech.edu/}),
NGC\,7314 (S1.9 in \citealt{liu2005}), MCG-03.34.064 (S1.8 in
\citealt{aguero1994}), NGC\,5252 (S1.9 in \citealt{osterbrock1993}),
and NGC\,835 and NGC\,6251 (LINERs in
\citealt{omaira2009a}). NGC\,4472 was also excluded because its
classification is based on the upper limits of line intensity ratios
\citep{ho1997}, and other classifications have been found in the
literature \citep[e.g.,][]{boisson2004}.

The final sample of Seyfert 2 galaxies contains 26 objects, 18
classified as S2 and 8 classified as S1h in \cite{veron2010}. However,
we revisited the literature to search for hidden broad-line-region
(HBLR, an usual name for S1h) and non-hidden broad-line-region (NHBLR)
objects
\citep[e.g.,][]{tran1992,tran1995,moran2000,lumsden2001,gu2002}. We
found two additional HBLR (MARK\,1210 and MARK\,477) and five NHBLR
(MARK\,1066, NGC\,3079, NGC\,5194, NGC\,5643, and NGC\,7172)
sources. We did not find information about the remaining 11 nuclei, so
we assumed they are most probably not observed in polarized light.

The final sample of Seyfert 2s in our work thus contains 26
objects (including 10 HBLR and five NHBLR). The target galaxies and
their properties are presented in Table \ref{properties}.  Tables are
in Appendix \ref{tables}, and notes on the individual nuclei in
Appendix \ref{indivnotes} and images at different wavelenths in
Appendix \ref{multiimages}.

\section{\label{reduction}Data reduction}

\subsection{Chandra data}

\emph{Chandra} observations were obtained from the ACIS instrument
\citep{garmire2003}. Data reduction and analysis were carried out in a
systematic, uniform way using CXC Chandra Interactive Analysis of
Observations (CIAO\footnote{http://cxc.harvard.edu/ciao4.4/}), version
4.3. Level 2 event data were extracted by using the task { \sc
  acis-process-events}.  Background flares were cleaned using the task
{ \sc
  lc\_clean.sl}\footnote{http://cxc.harvard.edu/ciao/ahelp/lc\_clean. html},
which calculates a mean rate from which it deduces a minimum and
maximum valid count rate and creates a file with the periods that are
considered by the algorithm to be good.

Nuclear spectra were extracted from a circular region centered on the
positions given by NED\footnote{http://ned.ipac.caltech.edu/}. We
chose circular radii, aiming to include all possible photons, while
excluding other sources or background effects. The radii are in the
range between 2-5$\arcsec$ (or 4-10 pixels, see
Table~\ref{obsSey}). The background was extracted from circular
regions in the same chip that are free of sources and close to the
object.

For the source and background spectral extractions, the {\sc
  dmextract} task was used. The response matrix file (RMF) and
ancillary reference file (ARF) were generated for each source region
using the {\sc mkacisrmf} and {\sc mkwarf} tasks,
respectively. Finally, the spectra were binned to have a minimum of 20
counts per spectral bin using the {\sc grppha} task (included in {\sc
  ftools}), to be able to use the $\chi^2$ statistics.

\subsection{XMM-Newton data}

\emph{XMM}-Newton observations were obtained with the EPIC pn camera
\citep{struder2001}. The data were reduced in a systematic, uniform
way using the Science Analysis Software
(SAS \footnote{http://xmm.esa.int/sas/}), version 11.0.0. First,
good-timing periods were selected using a method that maximizes the
signal-to-noise ratio of the net source spectrum by applying a
different constant count rate threshold on the single events, E $>$ 10
keV field-of-view background light curve.  We extracted the spectra of
the nuclei from circles of 15--30$\arcsec$ (or 300-600 px) radius
centered on the positions given by NED, while the backgrounds were
extracted from circular regions using an algorithm that automatically
selects the best area - and closest to the source - that is free of
sources. This selection was manually checked to ensure the best
selection for the backgrounds.

The source and background regions were extracted with the {\sc
  evselect} task. The response matrix files (RMF) and the ancillary
response files (ARF) were generated using the {\sc rmfgen} and {\sc
  arfgen} tasks, respectively. To be able to use the $\chi^2$
statistics, the spectra were binned to obtain at least 20 counts per
spectral bin using the {\sc grppha} task.

\subsection{ \label{sigma} Light curves}

Light curves in three energy bands (0.5--2.0 keV, 2.0--10.0 keV, and
0.5--10 keV) for the source and background regions as defined above
were extracted using the {\sc dmextract} task (for \emph{XMM}-Newton)
and {\sc evselect} task (for \emph{Chandra}) with a 1000~s bin. To be
able to compare the variability amplitudes in different light curves
of the same object, only those observations with a net exposure time
longer than 30 ksec were taken into account. For longer observations,
the light curves were divided into segments of 40 ksec, so in some
cases more than one segment of the same light curve can be extracted.
Intervals with “flare”-like events and/or prominent
decreasing/increasing trends were manually rejected from the source
light curves.  We notice that after excluding these events, the
exposure time of the light curve could be shorter, thus we recall that
only observations with a net exposure time longer than 30 ksec were
used for the analysis. The light curves are shown in Appendix
\ref{lightcurves}. We recall
that these values are used only for visual inspection of the data and
not as estimators of the variability (as in \citealt{lore2014}).

\section{\label{method}Methodology}

The methodology is explained in \cite{lore2013} and
\cite{lore2014}. In contrast to the study of LINER nuclei, we added a
new model (namely 2ME2PL), and a cold reflection component for the
individual spectral fittings and an analysis of the
\emph{\emph{Compton}}-thickness for the Seyfert galaxies.
Additionally, we changed the way we estimate the nuclear contribution
in \emph{XMM}--Newton spectra to perform the simultaneous fit using
different instruments (see Sect. \ref{simult}).  A comparison with a
sample of LINERs will be performed in a forthcoming paper.  For
clarity, we recall the procedure below.

\subsection{\label{indiv}Individual spectral analysis}

An individual spectral analysis allowed us to select the best-fit
model for each data set.  We added a new model with respect to
previous works (2ME2PL), including an additional thermal component to
the more complex model, ME2PL, to explain the two ionized zones
observed in some Seyfert galaxies
\citep[e.g.,][]{netzer1997,bianchi2010}.  Then, we also added a cold
reflection component (PEXRAV in XSPEC, \citealt{pexrav1995}) to the
best-fit model to check whether this component improves the fit.  We
used XSPEC\footnote{http://heasarc.nasa.gov/xanadu/xspec/} version
12.7.0 to fit the data with six different models:

\begin{itemize}
\item[$\bullet$] 
PL: A single power law representing the continuum of
  a non-stellar source. The empirical model is

$e^{N_{Gal} \sigma (E)} \cdot e^{N_{H} \sigma (E(1+z))}[N_{H}] \cdot Norm e^{-\Gamma}[\Gamma, Norm].$
\vspace*{0.2cm}
\item[$\bullet$] ME: The emission is dominated by hot diffuse gas, i.e., a thermal plasma. A MEKAL (in XSPEC) model is used to fit the spectrum. The model is

$e^{N_{Gal} \sigma (E)} \cdot e^{N_{H} \sigma (E(1+z))}[N_{H}] \cdot MEKAL[kT, Norm].$
\vspace*{0.2cm}

\item[$\bullet$] 2PL: In this model the primary continuum is an
  absorbed power law representing the non stellar source, while the
  soft energies are due to a scattering component that is represented
  by another power law. Mathematically the model is explained as

$e^{N_{Gal} \sigma (E)} \big( e^{N_{H1} \sigma (E(1+z))}[N_{H1}] \cdot Norm_1 e^{-\Gamma}[\Gamma, Norm_1] + e^{N_{H2} \sigma (E(1+z))}[N_{H2}] \cdot Norm_2 e^{-\Gamma}[\Gamma, Norm_2]\big)$.
\vspace*{0.2cm}

\item[$\bullet$] MEPL: The primary continuum is represented by an
  absorbed power law, but at soft energies a thermal plasma dominates
  the spectrum. Empirically it can be described as

$e^{N_{Gal} \sigma (E)} \big(e^{N_{H1} \sigma (E(1+z))}[N_{H1}] \cdot MEKAL[kT, Norm_1] + e^{N_{H2} \sigma (E(1+z))}[N_{H2}] \cdot Norm_2 e^{-\Gamma}[\Gamma, Norm_2]\big)$.
\vspace*{0.2cm}

\item[$\bullet$] ME2PL: This is same model as MEPL, but an additional
  power law is required to explain the scattered component at soft
  energies, so mathematically it is

$e^{N_{Gal} \sigma (E)} \big( e^{N_{H1} \sigma (E(1+z))}[N_{H1}] \cdot Norm_1 e^{-\Gamma}[\Gamma, Norm_1] + MEKAL[kT] + e^{N_{H2} \sigma (E(1+z))}[N_{H2}] \cdot Norm_2 e^{-\Gamma}[\Gamma, Norm_2]\big)$.
\vspace*{0.2cm}

\item[$\bullet$] 2ME2PL: The hard X-ray energies are represented by an
  absorbed power law, while the spectrum shows a complex structure at
  soft energies, where a composite of two thermal plasmas plus a power
  law are required. In Seyfert galaxies, at least two ionized phases
  (a warm and a hot) are required to properly fit their spectra
  \citep{netzer1997}, which is confirmed by high resolution data
  \citep[e.g.,][]{bianchi2010, marinucci2011}.  Ideally, the spectral
  fit should be made by using photoionization models to fit high
  quality data (e.g., RGS) and then use the obtained spectral
  parameters to fit lower quality data, as in \cite{bianchi2010} or
  \cite{omaira2010}. We tried to use photoionized models using Cloudy
  to fit the soft emission. We found that, due to the low resolution
  of our data, these models fit the data similarly to MEKAL
  models. Therefore, for simplicity, in this work we represent the
  photoionized gas by two thermal plasmas plus Gaussian lines when
  required (see below).  The power law at soft energies represents the
  scattering component.  Although this is probably a simple model for
  fitting the complexity of the spectra, the data analyzed in this
  work do not have enough spectral resolution to properly fit the data
  with more realistic models, and therefore this model is enough for
  our purposes. It is represented as

$e^{N_{Gal} \sigma (E)} \big( e^{N_{H1} \sigma (E(1+z))}[N_{H1}] \cdot Norm_1 e^{-\Gamma}[\Gamma, Norm_1] + MEKAL[kT_1] + MEKAL[kT_2] + e^{N_{H2} \sigma (E(1+z))}[N_{H2}] \cdot Norm_2 e^{-\Gamma}[\Gamma, Norm_2]\big)$.

\item[$\bullet$] (Best-fit model) + PEXRAV: From the six models
  described above, we selected the one that provided the best fit to
  the data and added a reflection component (we have chosen PEXRAV
  within XSPEC) to account for a plausible contribution of this
  component in highly obscured Seyfert 2s. The parameters of the
  MEKAL component(s) were frozen to the best-fit values. In this model
  the absorbed power law at hard energies represents the transmitted
  component, while the PEXRAV is indicative of the reflected fraction
  from the primary continuum alone, by setting the reflection scaling
  factor to 1.  The spectral index was set to be that of the power
  law(s), the exponential cutoff was fixed to 200 keV, and the
  inclination angle to 45$\degree$. These parameters are based on
  typical values obtained from X-ray analyses at harder energies
  \citep[e.g.,][]{guainazzi2005a, matt2004, akylas2009,
    noguchi2009}. The free parameters in this model are therefore
  $N_{H1},N_{H2},\Gamma, Norm_1, Norm_2$, and $Norm_{pex}$.  It is
  worth noting that we tried similar models to fit the data, such as
  exchanging the hard PL by PEXRAV or by an absorbed PEXRAV, and
  obtained very similar results, but the model explained above allowed
  the use of the F test to check for eventual improvements in the
  fits.

\end{itemize}

\noindent In the equations above, $\sigma (E)$ is the photo-electric
cross-section, $z$ is the redshift, and $Norm_i$ are the
normalizations of the power law, the thermal component or the
reflected component (i.e., $Norm_1$, $Norm_2$, and $Norm_{pex}$). For
each model, the parameters that vary are written in brackets.  The
Galactic absoption, $N_{Gal}$, is included in each model and fixed to
the predicted value (Col. 5 in Table \ref{properties}) using the tool
{\sc nh} within {\sc ftools} \citep{dickeylockman1990, kalberla2005}.
Even if not included in the mathematical form above, all the models
include three narrow Gaussian lines to take the iron lines at 6.4 keV
(FeK$\alpha$), 6.7 keV (FeXXV), and 6.95 keV (FeXXVI) into account. In
a few cases, additional Gaussian lines were required at soft energies
from a visual inspection, including Ne X at 1.2 keV, Mg XI at 1.36
keV, Si XIII at 1.85 keV, and S XIV at 2.4 keV.

The $\chi^2/d.o.f$ and F test were used to select the simplest model
that represents the data best.

\subsection{\label{simult} Simultaneous spectral analysis}

Once the individual best-fit model is selected for each observation,
and if the models are different for the individual observations, then
the most complex model that fits each object was chosen.  This model
was used to simultaneously fit spectra obtained at different dates of
the same nuclei. Initially, the values of the spectral parameters were
set to those obtained for the spectrum with the largest number counts
for each galaxy. To determine whether spectral variations are observed
in the data, this simultaneous fit was made in three steps:

\begin{itemize}

\item[0.] SMF0 (Simultaneous fit 0): The same model was used with all
  parameters linked to the same value to fit every spectra of the same
  object, i.e., the non-variable case.
\item[1.] SMF1: Using SMF0 as the baseline for this step, we let the
  parameters $N_{H1}$, $N_{H2}$, $\Gamma$, $Norm_1$, $Norm_2$,
  $Norm_{pex}$, $kT_1$, and $kT_2$ vary individually.  The best fit
  was selected for the $\chi^2_r$ closest to unity that improved SMF0
  (using the F test).
\item[2.] SMF2: Using SMF1 as the baseline for this step (when SMF1
  did not fit the data well), we let two parameters vary, the one that
  varied in SMF1 along with any of the other parameters of the
  fit. The $\chi^2_r$ and F test were again used to confirm an
  improvement in the fit.
\end{itemize}

When data from the same instrument were available at different epochs,
this method was applied separately for \emph{Chandra} and/or
\emph{XMM}--Newton. However, in some cases only one observation was
available per instrument. Instead of directly comparing the spectra
from different instruments, we tried to decontaminate the extranuclear
emission in \emph{XMM}--Newton data, to make sure that the emission
included in the larger aperture did not produce the observed
variability.  This additional analysis was performed by extracting an
annular region from \emph{Chandra} data, fitting the models explained
above to its spectrum, and selecting the one that best fits the
annular region. This model was later incorporated into the
\emph{XMM}--Newton spectrum (with its parameters frozen), so the
parameters of the nuclear emission can be estimated.  We determined
the contribution by the annular region to the \emph{Chandra} data from
the number counts (i.e., model-independent) in the 0.5-10.0 keV energy
band, and this percentage was used to estimate the number counts in
the nuclear region of \emph{XMM}--Newton data. Following the same
criteria as we used to select the data (see Sect. \ref{sample}), data
from different instruments were compared when the number counts in the
nuclear \emph{XMM}--Newton spectrum was more than 400 counts. We note
that this procedure differs from the one used in
\cite{lore2013,lore2014}.  When multiple observations of the same
object and instrument were available, we compared the data with the
closest dates (marked with $c$ in Table \ref{obsSey}).

\subsection{Flux variability}

The luminosities in the soft and hard X-ray energy bands were computed
using XSPEC for both the individual and the simultaneous fits. For
their calculation, we took the distances from NED, corresponding to
the average redshift-independent distance estimate for each object,
when available, or to the redshift-estimated distance otherwise;
distances are listed in Table \ref{properties}.

When data from the optical monitor (OM) onboard \emph{XMM}--Newton
were available, UV luminosities (simultaneously to X-ray data) were
estimated in the available filters.  We recall that UVW2 is centered
at 1894$\AA$ (1805-2454) $\AA$, UVM2 at 2205$\AA$ (1970-2675) $\AA$,
and UVW1 at 2675$\AA$ (2410-3565) $\AA$.  We used the OM observation
FITS source lists
(OBSMLI)\footnote{ftp://xmm2.esac.esa.int/pub/odf/data/docs/XMM-SOC-GEN-ICD-0024.pdf}
to obtain the photometry. When OM data were not available, we searched
for UV information in the literature. We note that in this case, the
X-ray and UV data might not be simultaneous (see Appendix
\ref{indivnotes}).

We assumed an object to be variable when the square root of the
squared errors was at least three times smaller than the difference
between the luminosities \citep[see][for details]{lore2014}.

\subsection{\label{short}Short-term variability}

Firstly, we assumed a constant count rate for segments of 30-40 ksec
of the observation in each energy band and calculated
$\rm{\chi^2/d.o.f}$ as a proxy to the variations. We considered the
source as a candidate for variability if the count rate differed from
the average by more than 3$\rm{\sigma}$ (or 99.7\% probability).

Secondly, and to be able to compare the variability amplitude of the
light curves between observations, we calculated the normalized excess
variance, $\rm{\sigma_{NXS}^2}$, for each light curve segment with
30-40 ksec following prescriptions in \cite{vaughan2003} \citep[see
  also][]{omaira2011a,lore2014}. We recall that $\rm{\sigma_{NXS}^2}$
is related to the area below the power spectral density (PSD) shape.

When $\rm{\sigma_{NXS}^2}$ was negative or compatible with zero within
the errors, we estimated the 90\% upper limits using Table 1 in
\cite{vaughan2003}. We assumed a PSD slope of -1, the upper limit from
\cite{vaughan2003}, and we added the value of
1.282err($\rm{\sigma_{NXS}^2}$) to the limit to account for Poisson
noise.  For a number of segments, N, obtained from an individual light
curve, an upper limit for the normalized excess variance was
calculated. When N segments were obtained for the same light curve and
at least one was consistent with being variable, we calculated the
normalized weighted mean and its error as the weighted variance.

We considered short-term variations for $\rm{\sigma_{NXS}^2}$
detections above 3$\sigma$ of the confidence level.

\subsection{\label{thick}\emph{Compton} thickness}

Highly obscured AGN are observed through the dusty torus, in some
cases with column densities higher than $1.5 \times 10^{24} cm^{-2}$
(the so-called \emph{\emph{Compton}}-thick). In these cases the
primary emission can be reflected at energies $\sim$ 10 keV. Since the
primary continuum cannot be directly observed, some indicators using
X-rays and [O III] data have been used to select candidates
\citep{ghisellini1994, bassani1999, panessabassani2002, cappi2006}.

To properly account for the slope of the power law, $\Gamma$, and the
equivalent width of the iron line, EW(FeK$\alpha$), an additional
analysis was performed.  We fit the 3-10 keV energy band of each
spectrum individually with a PL model (see Sect. \ref{indiv}) to
obtain the values of $\Gamma$ and EW(FeK$\alpha$).
\emph{\emph{Compton}}-thick candidates can be selected by using three
different criteria:

\begin{itemize}
\item[$\bullet$] \underline{$\Gamma < 1$} : since the transmitted
  component is suppressed below 10 keV, a flattening of the observed
  spectrum is expected \citep{cappi2006,omaira2009b}.
\item[$\bullet$] \underline{EW(FeK$\alpha)>$ 500 eV} : if the nuclear
  emission is obscured by a \emph{\emph{Compton}}-thick column
  density, the primary continuum underneath the FeK$\alpha$ line is
  strongly suppressed, and the equivalent width of the line enhanced
  to $\sim$keV \citep{krolik1994,ghisellini1994}.
\item[$\bullet$] \underline{$F(2-10 keV)/F_{[O III]}$ $<$ 1} : since
  the primary continuum is suppressed, the X-ray luminosity is
  underestimated, so when comparing with an isotropic indicator of the
  AGN power (as is the case for the [O III] emission line), the ratio
  between the two values decreases
  \citep{bassani1999,guainazzi2005a,cappi2006,omaira2009b}. Thus, we
  have used this ratio to select \emph{\emph{Compton-thick}}
  candidates, where the extinction-corrected [O III] fluxes were
  obtained from the literature (and corrected when needed following
  \citealt{bassani1999}), and the hard X-ray luminosities, $L(2-10
  keV)$, from the individual fits were used (see Table
  \ref{lumincorrSey}) for the calculation.
\end{itemize}

We considered that a source is a \emph{\emph{Compton}}-thick candidate
when at least two of the three criteria above were met. Otherwise, the
source is considered to be a \emph{\emph{Compton}}-thin
candidate. When different observations of the same source result in
different classifications, the object was considered to be a
changing-look candidate.

The spectral fits reported in Sects. \ref{indiv} and \ref{simult} are
performed with the spectral indices of the soft, $\Gamma_{soft}$, and
the hard, $\Gamma_{hard}$, power laws tied to the same value. When a
source is \emph{\emph{Compton}}-thick, its spectrum is characterized
by a flat power law at hard energies (see above), whereas the slope of
the power law is dominated by the scattered component if we tied
$\Gamma_{soft}=\Gamma_{hard}$, giving an unrealistic steep power-law
index. Thus, the simultaneous analysis was repeated by leaving
$\Gamma_{soft}$ and $\Gamma_{hard}$ free for the objects classified as
\emph{\emph{Compton}}-thick candidates. We first made the SMF1 with
$\Gamma_{hard}$ vary and found that this component does not vary in
any case. The values of $\Gamma_{hard}$ obtained for the
\emph{\emph{Compton}}-thick candidates following this procedure are
reported in Table \ref{ew} (Col. 9). We checked that the rest of the
parameters in the model are consistent with those reported in Table
\ref{bestfitSey} within the uncertainties. The same procedure was
applied to \emph{\emph{Compton}}-thin candidates, and compatible
values of $\Gamma_{soft}$ and $\Gamma_{hard}$ were found. It is worth
pointing out that it is not within the scope of this work to obtain
the best spectral parameters for each source, but to obtain their
variability patterns.  Thus, we have kept the same general analysis
for all the objects (i.e., with $\Gamma_{soft}=\Gamma_{hard}$,
although we notice that this is not the case for
\emph{\emph{Compton}}-thick candidates), but this procedure does not
affect the main results presented in this paper.

\begin{figure*}
\centering
\subfloat{\includegraphics[width=0.30\textwidth]{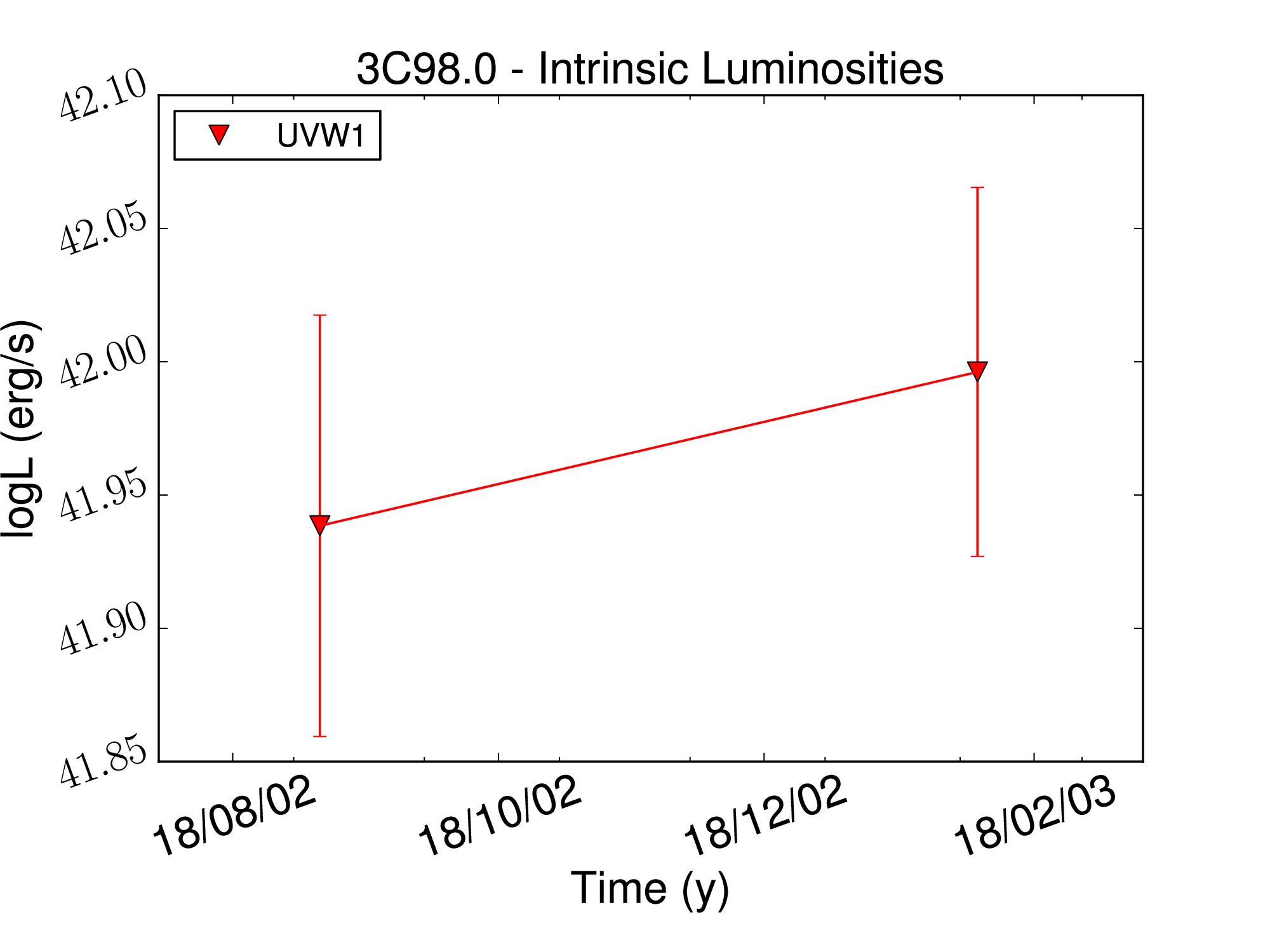}}
\subfloat{\includegraphics[width=0.30\textwidth]{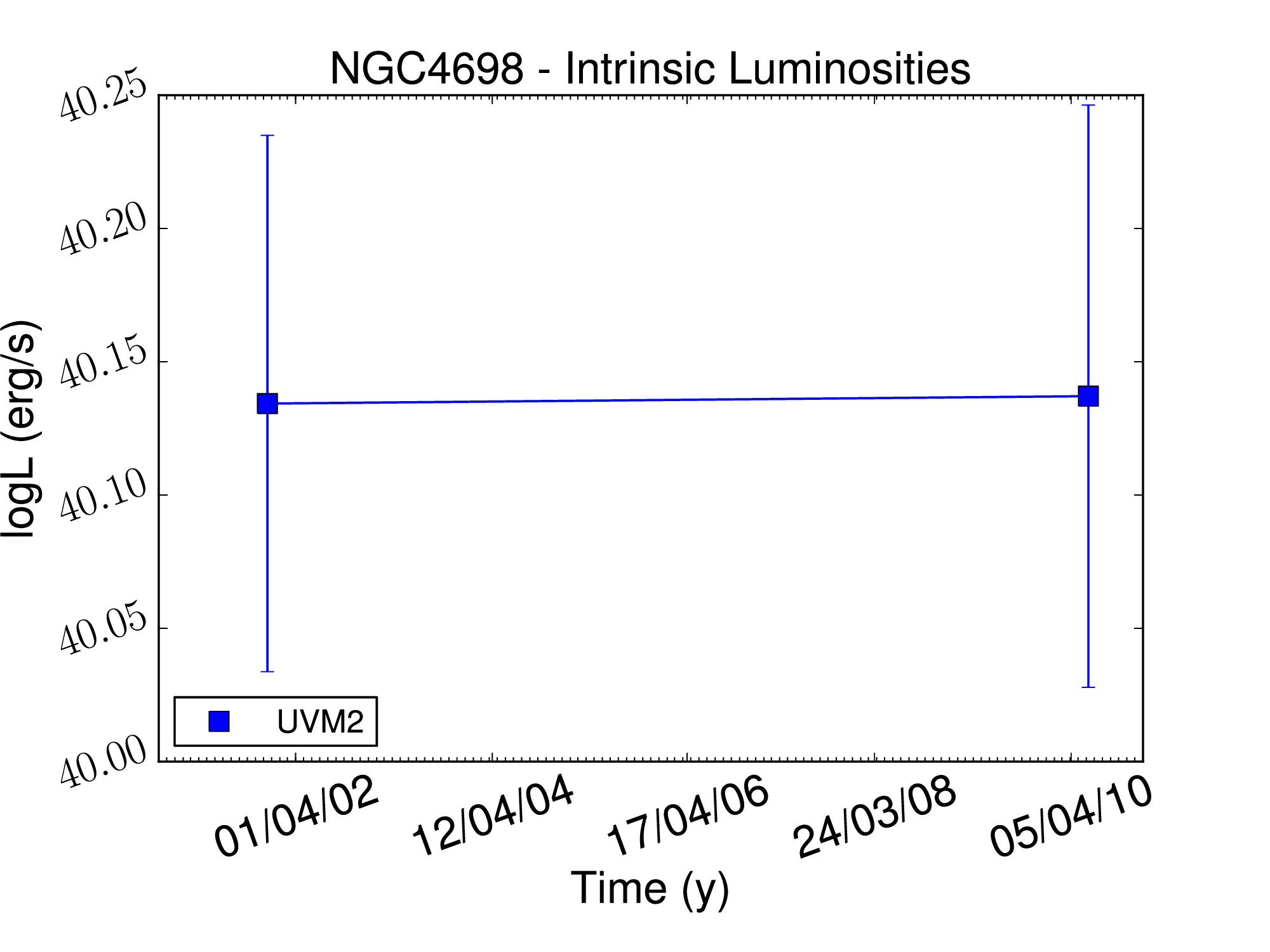}}
\subfloat{\includegraphics[width=0.30\textwidth]{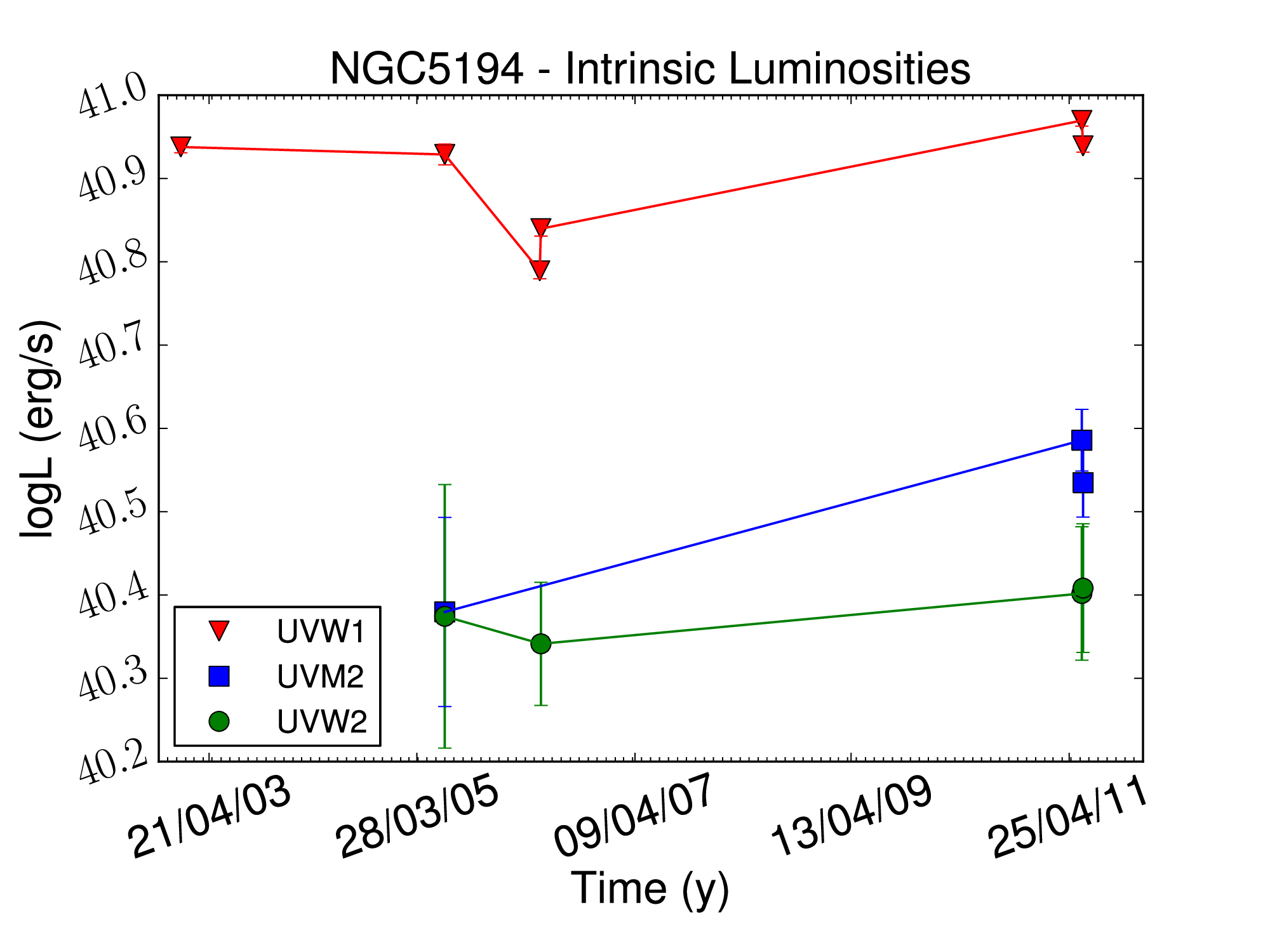}}

\subfloat{\includegraphics[width=0.30\textwidth]{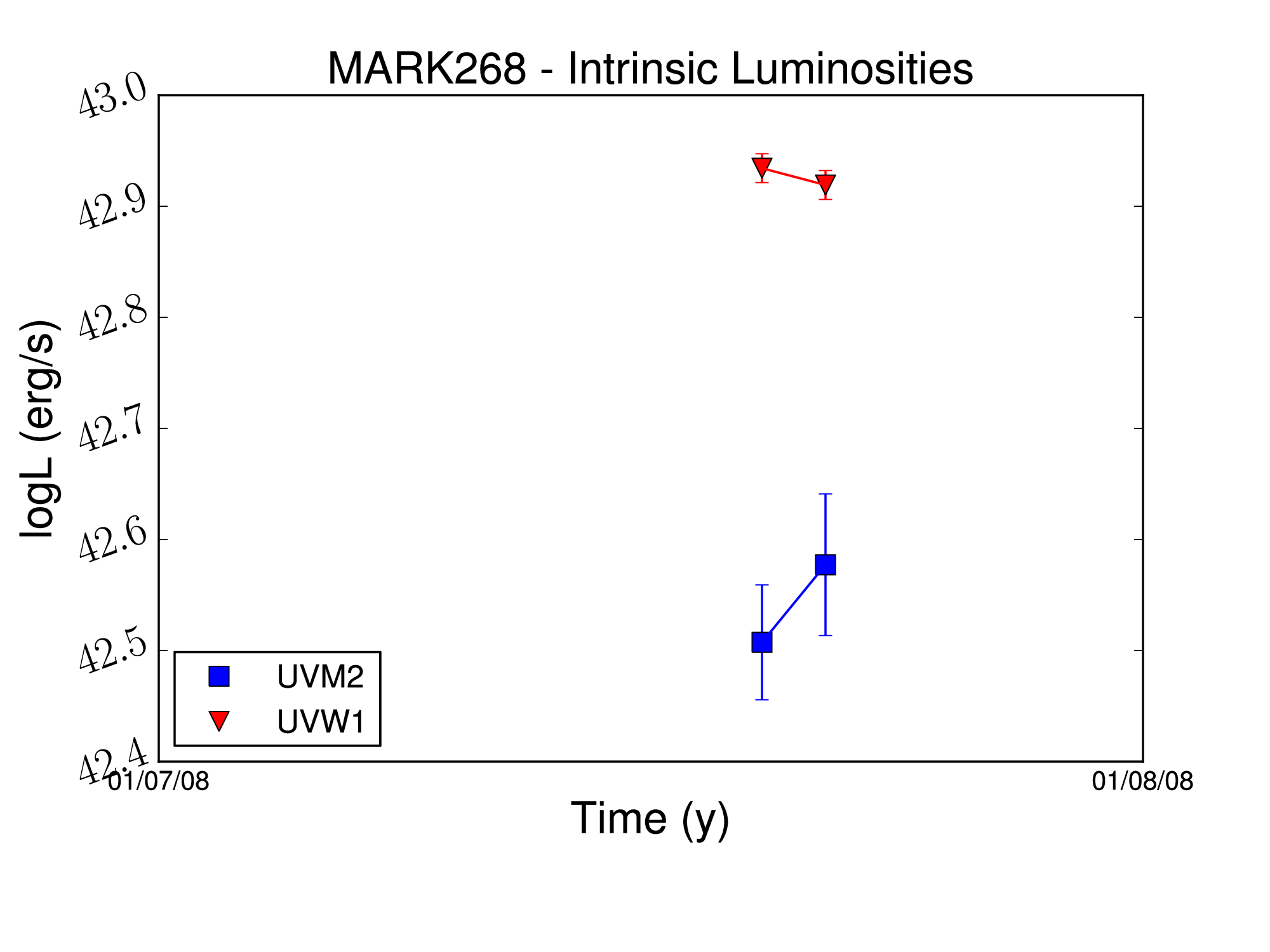}}
\subfloat{\includegraphics[width=0.30\textwidth]{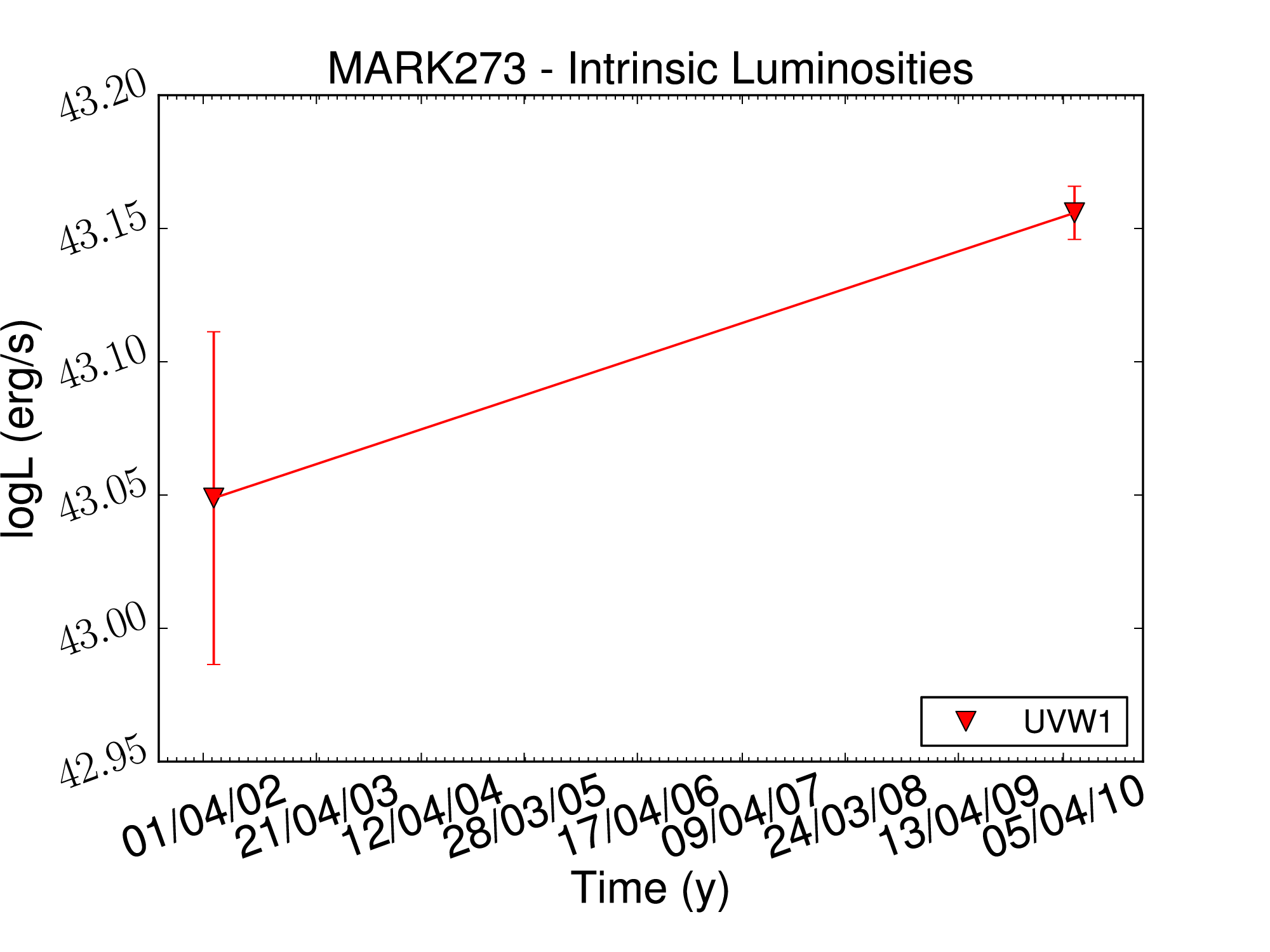}}
\subfloat{\includegraphics[width=0.30\textwidth]{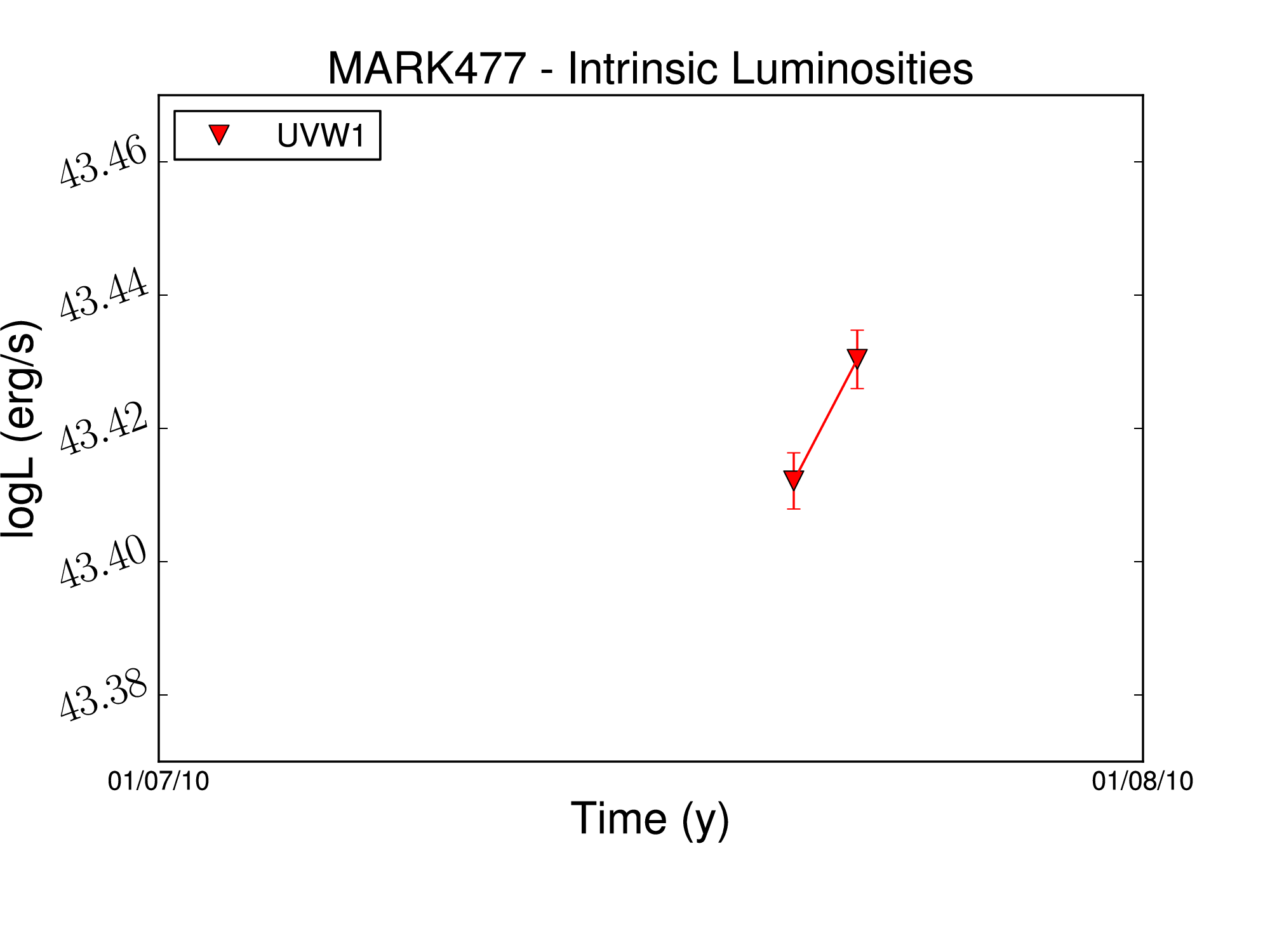}}
\caption{UV luminosities obtained from the data with the OM camera onboard \emph{XMM}--Newton, when available. Different filters have been used; UVW1 (red triangles), UVW2 (green circles), and UVM2 (blue squares).}
\label{luminUVfigSey}
\end{figure*}

\section{\label{results}Results}

In this section we present the results for the
variability analysis of the Seyfert 2 galaxies individually (see
Sect. \ref{ind}), as well as the general results, including the
characterization of the spectra of Seyfert 2s
(Sect. \ref{shape}), the long-term variability (Sect. \ref{spectral}),
first for the whole sample in general and later divided into
subsamples, X-ray short-term variations (Sect. \ref{lightcurve}), and
flux variations at UV frequencies (Sect. \ref{flux}). The main results
of the analysis are summarized in Table \ref{variab}. Individual notes
on each galaxy and comparisons with previous works can be found in
Appendix \ref{indivnotes}.

\begin{figure*}
\centering
\subfloat{\includegraphics[width=0.30\textwidth]{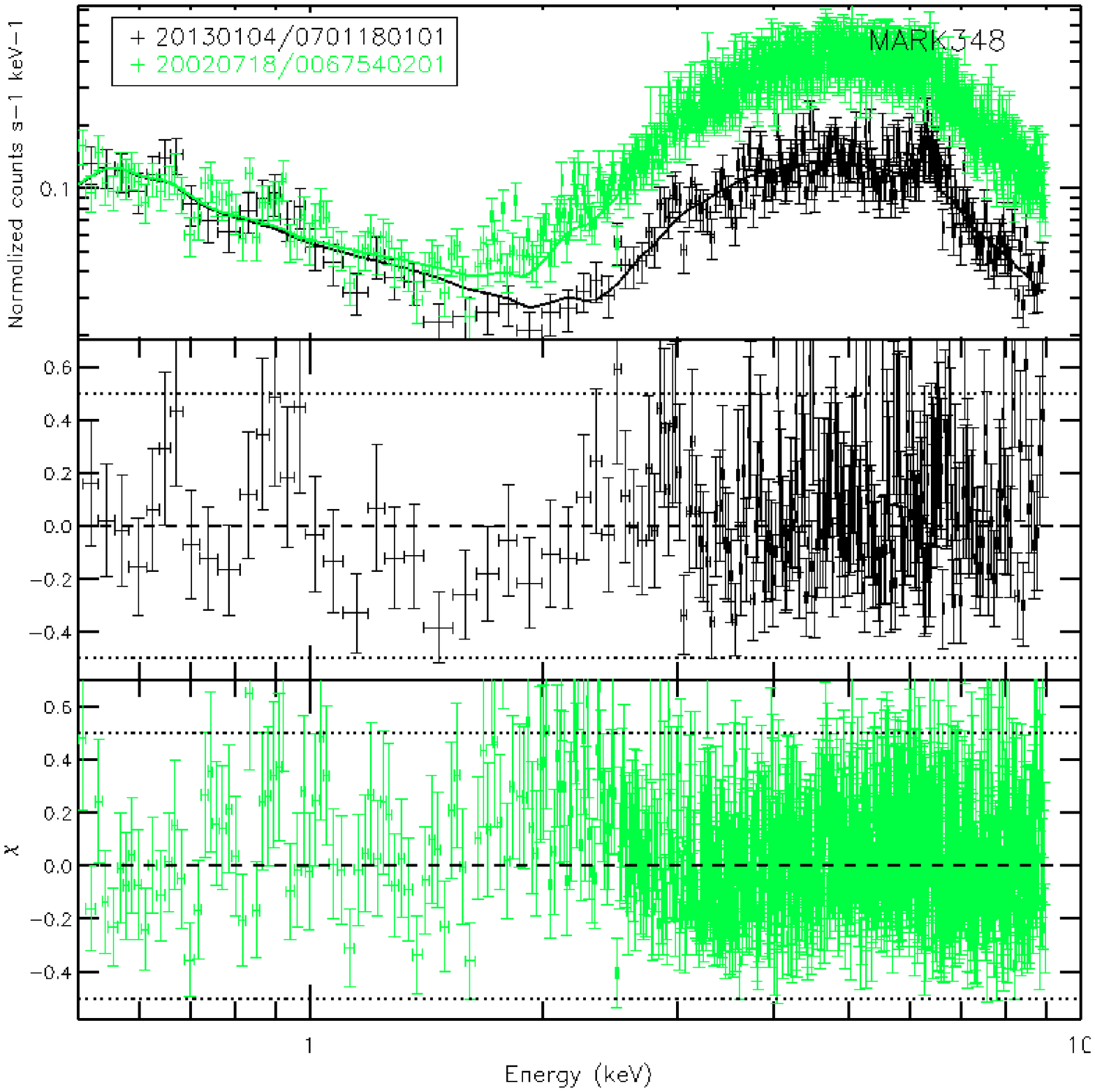}}
\subfloat{\includegraphics[width=0.30\textwidth]{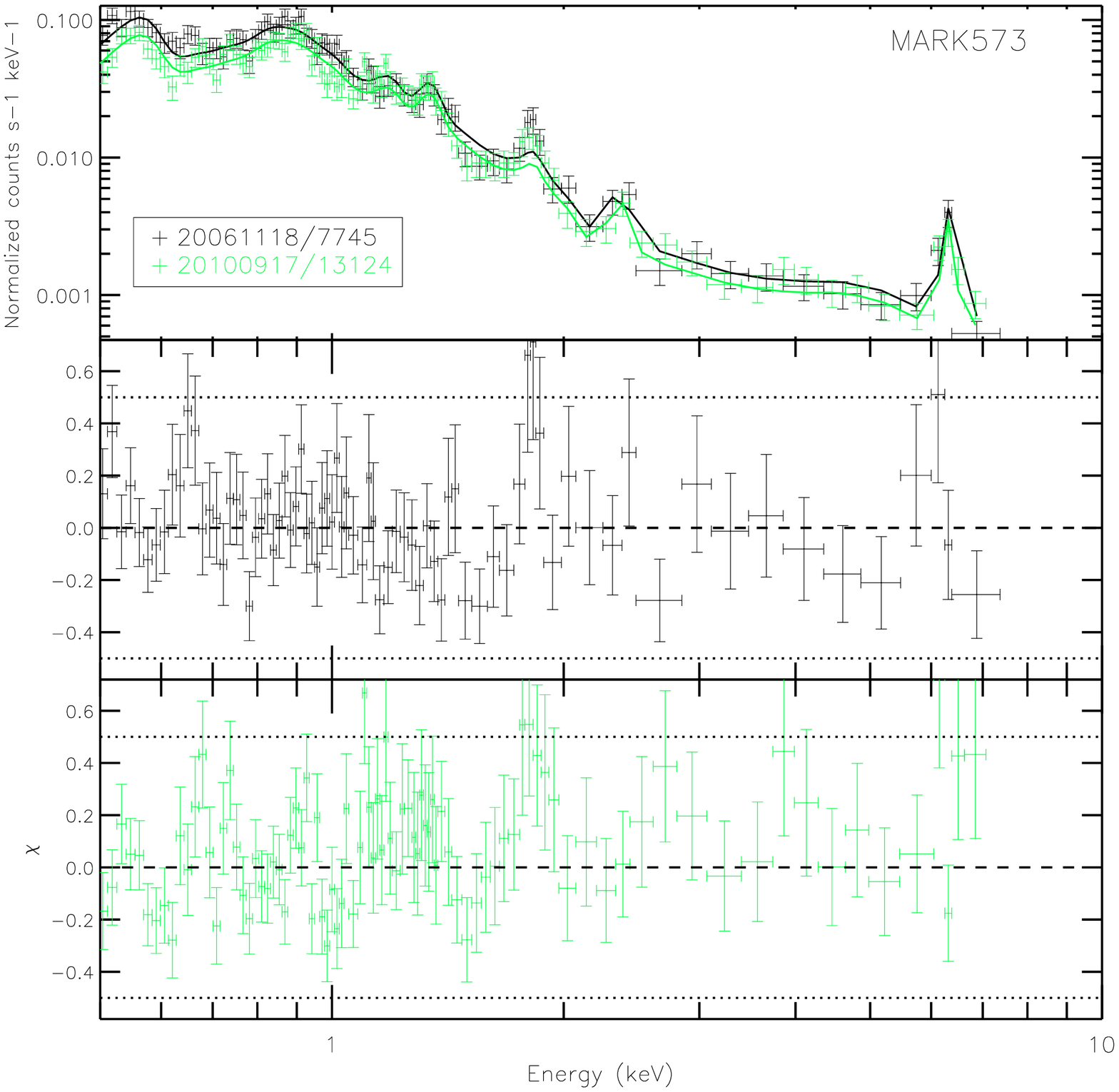}}
\subfloat{\includegraphics[width=0.30\textwidth]{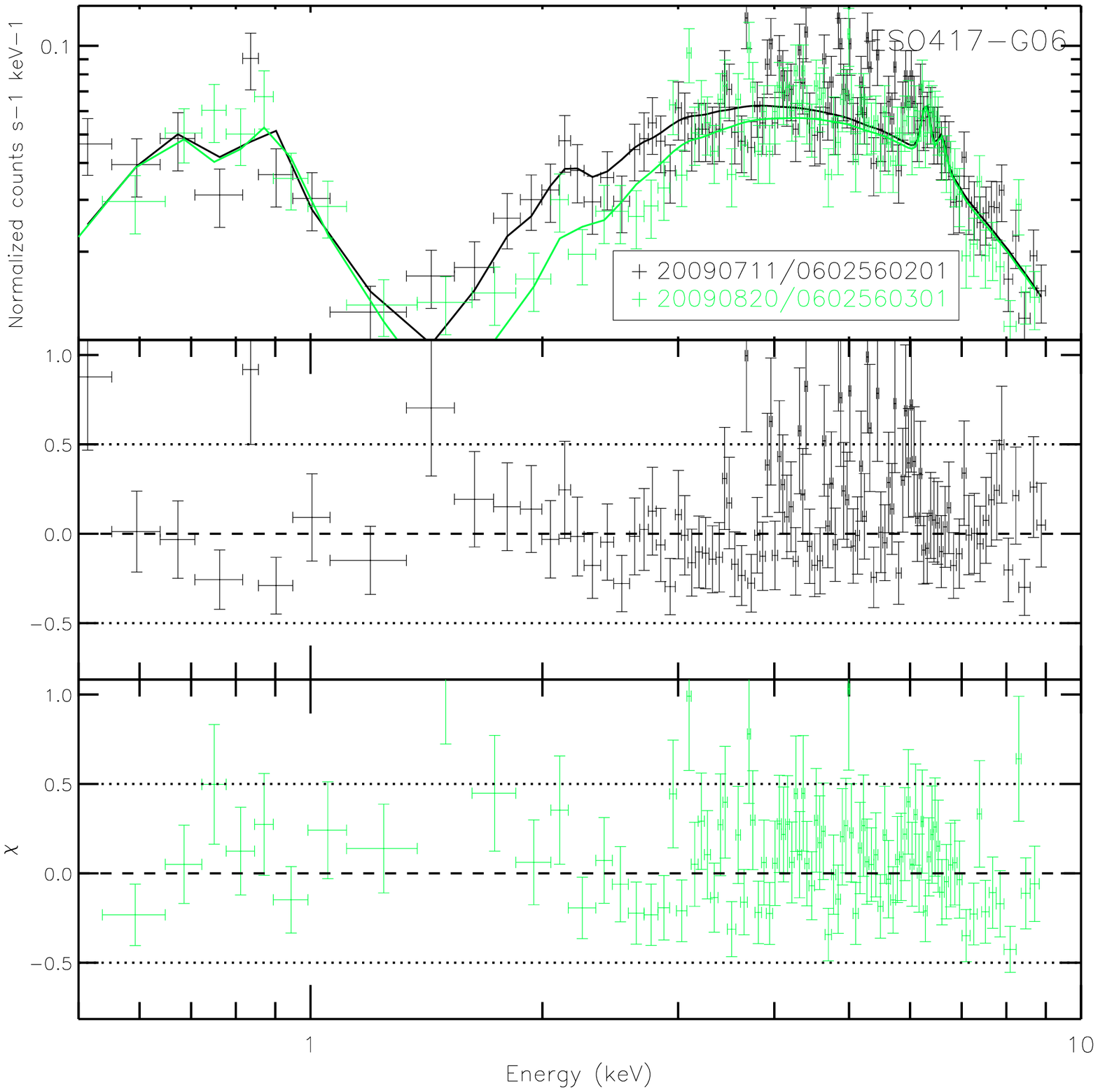}}

\subfloat{\includegraphics[width=0.30\textwidth]{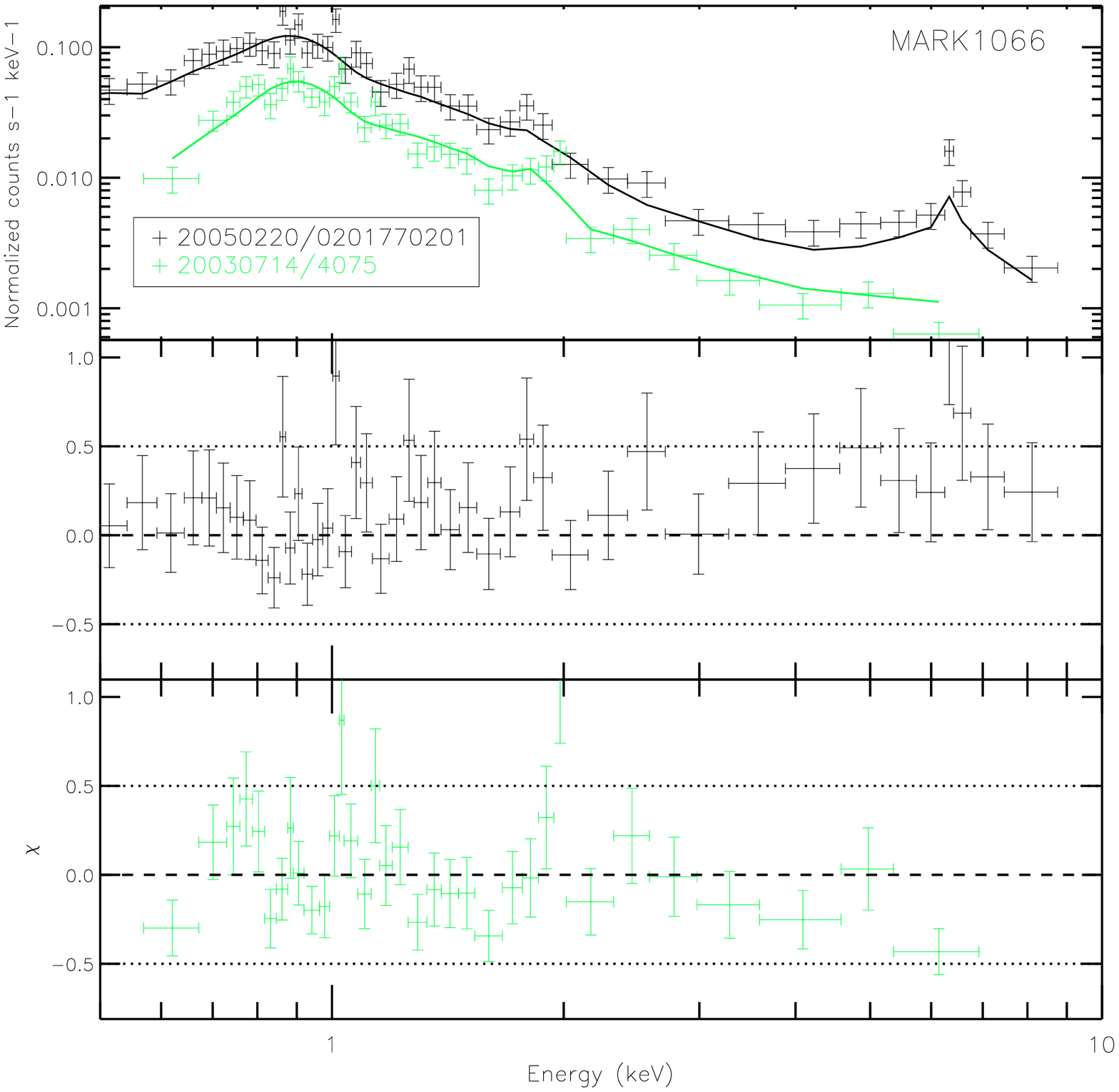}}
\subfloat{\includegraphics[width=0.30\textwidth]{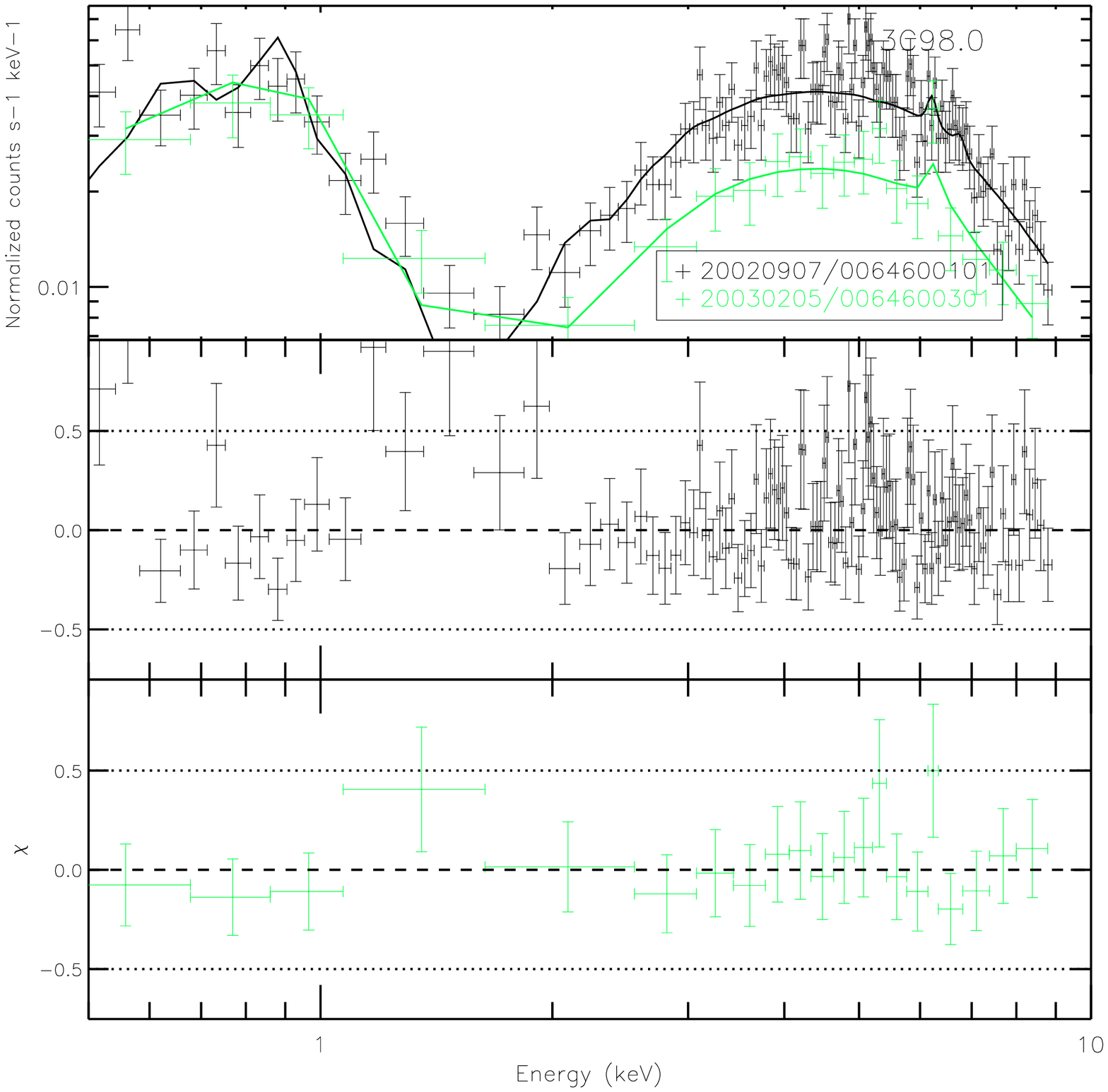}}
\subfloat{\includegraphics[width=0.30\textwidth]{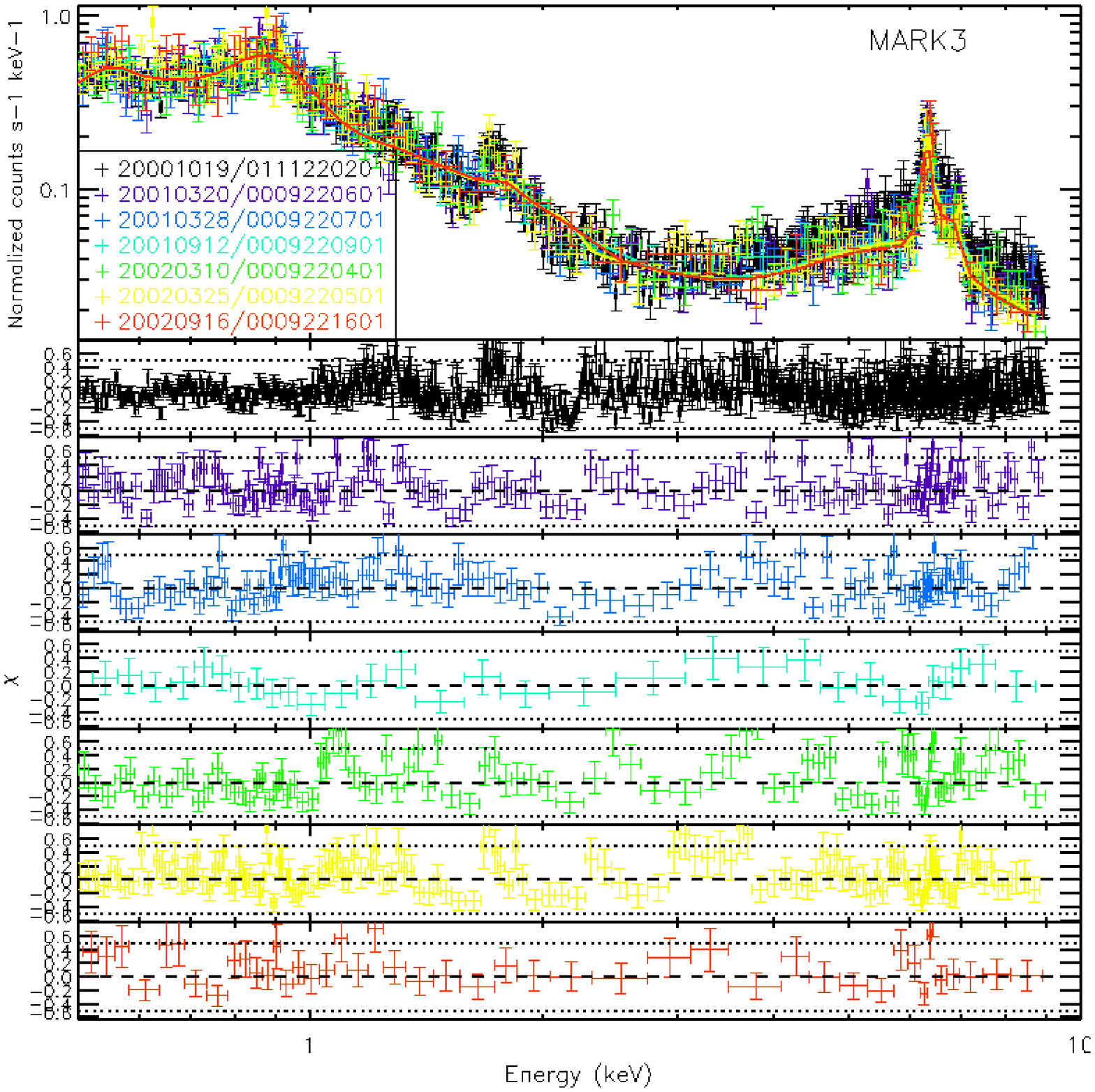}}

\subfloat{\includegraphics[width=0.30\textwidth]{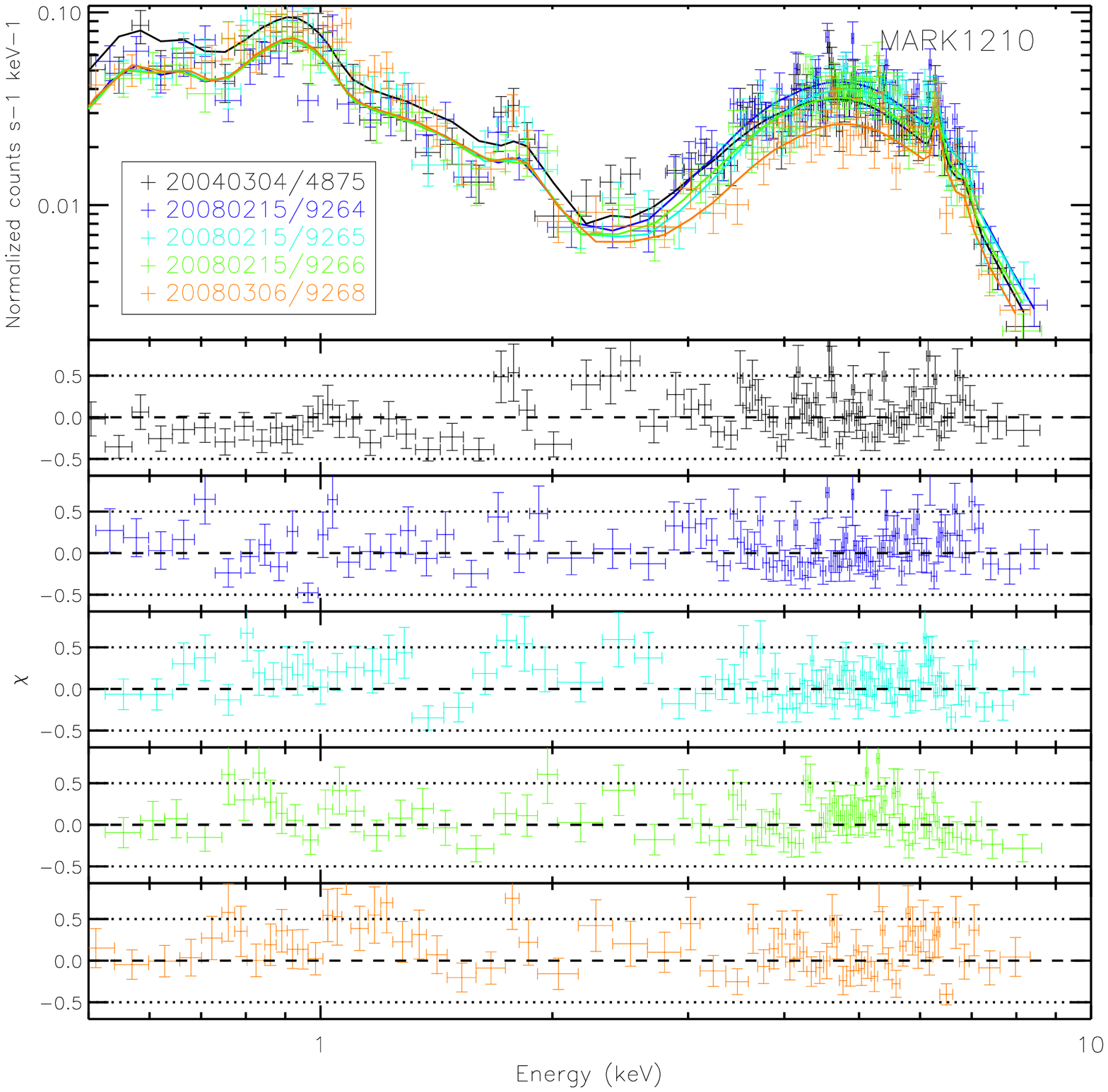}}
\subfloat{\includegraphics[width=0.30\textwidth]{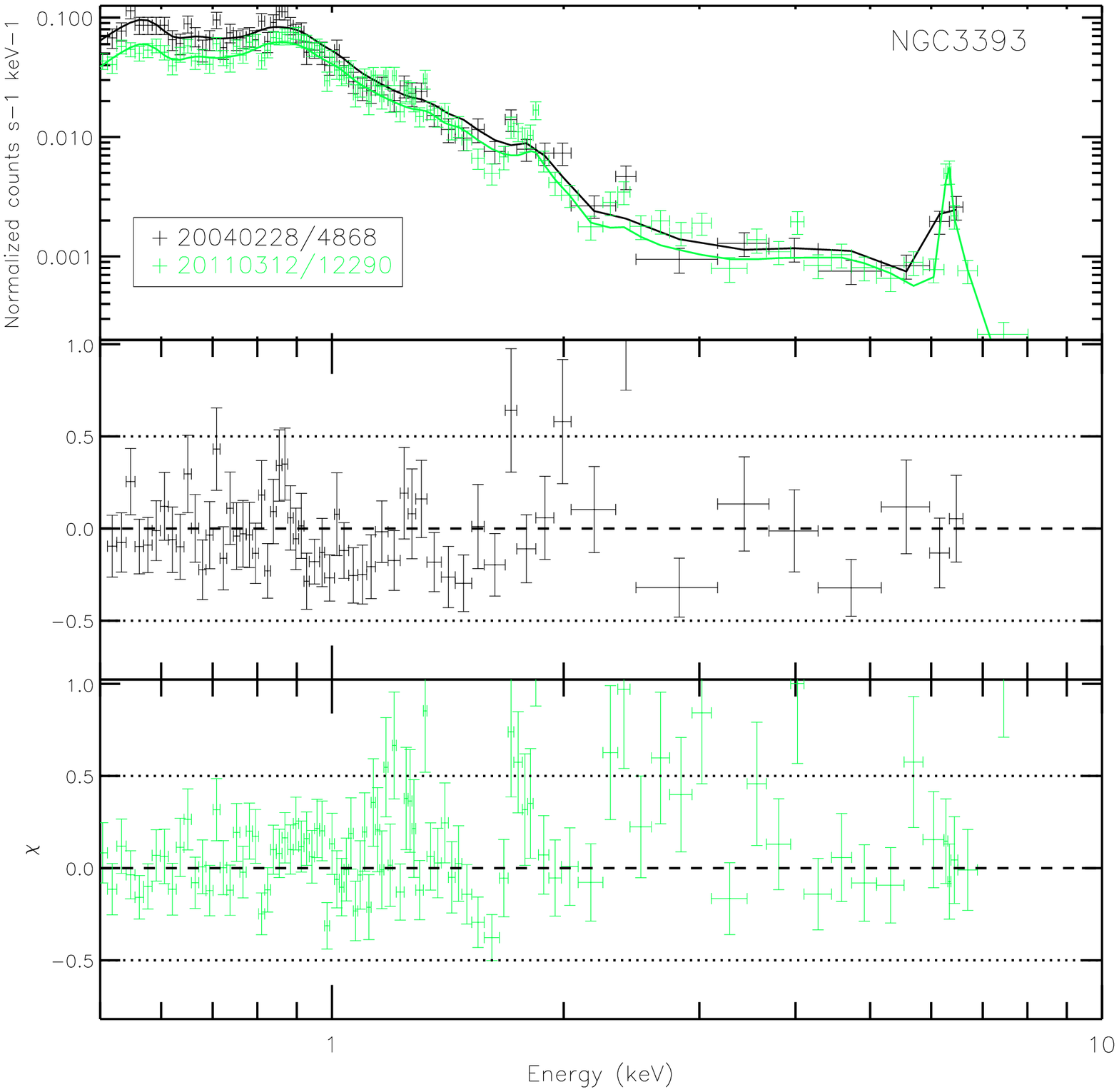}}
\subfloat{\includegraphics[width=0.30\textwidth]{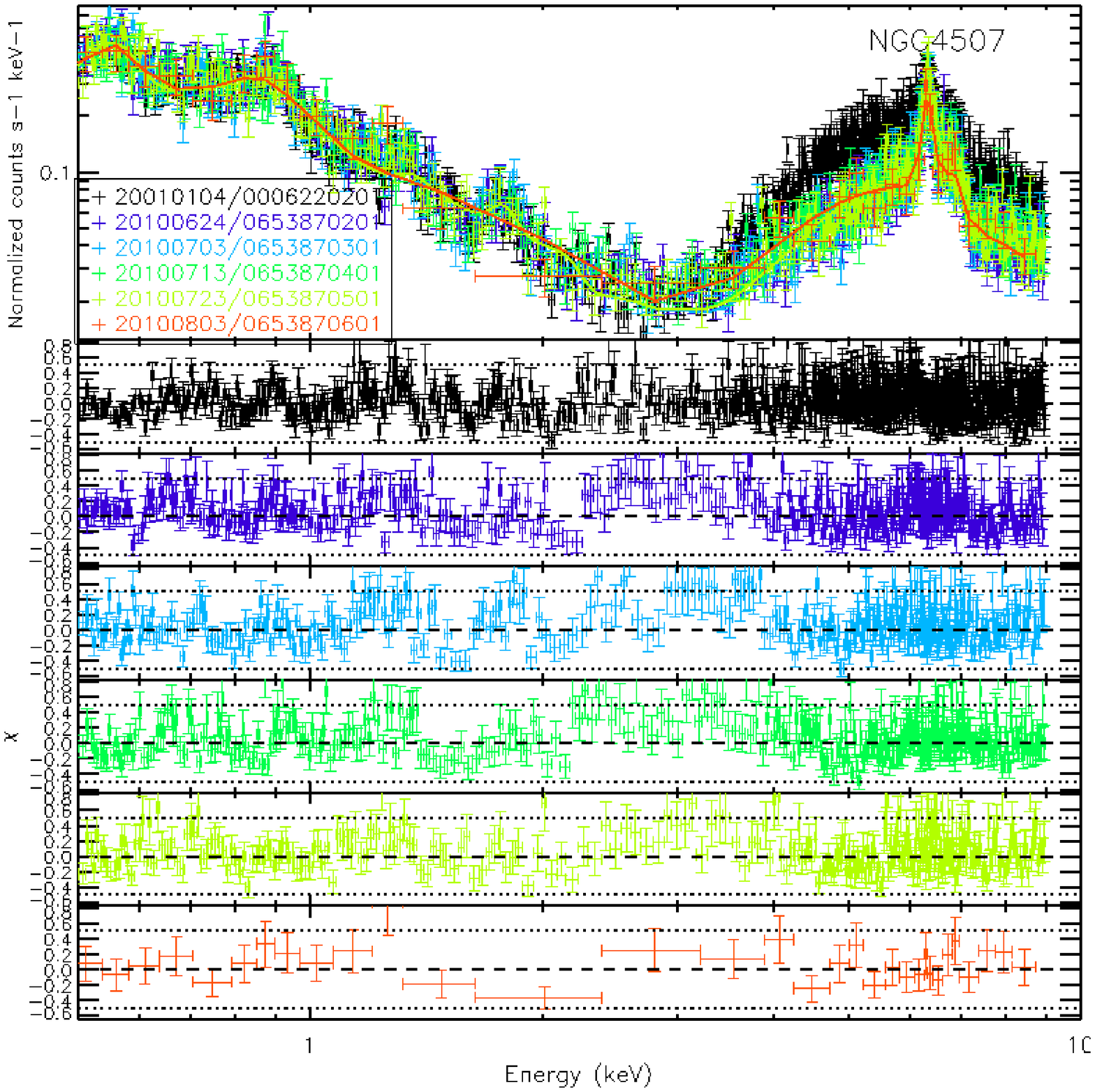}}

\subfloat{\includegraphics[width=0.30\textwidth]{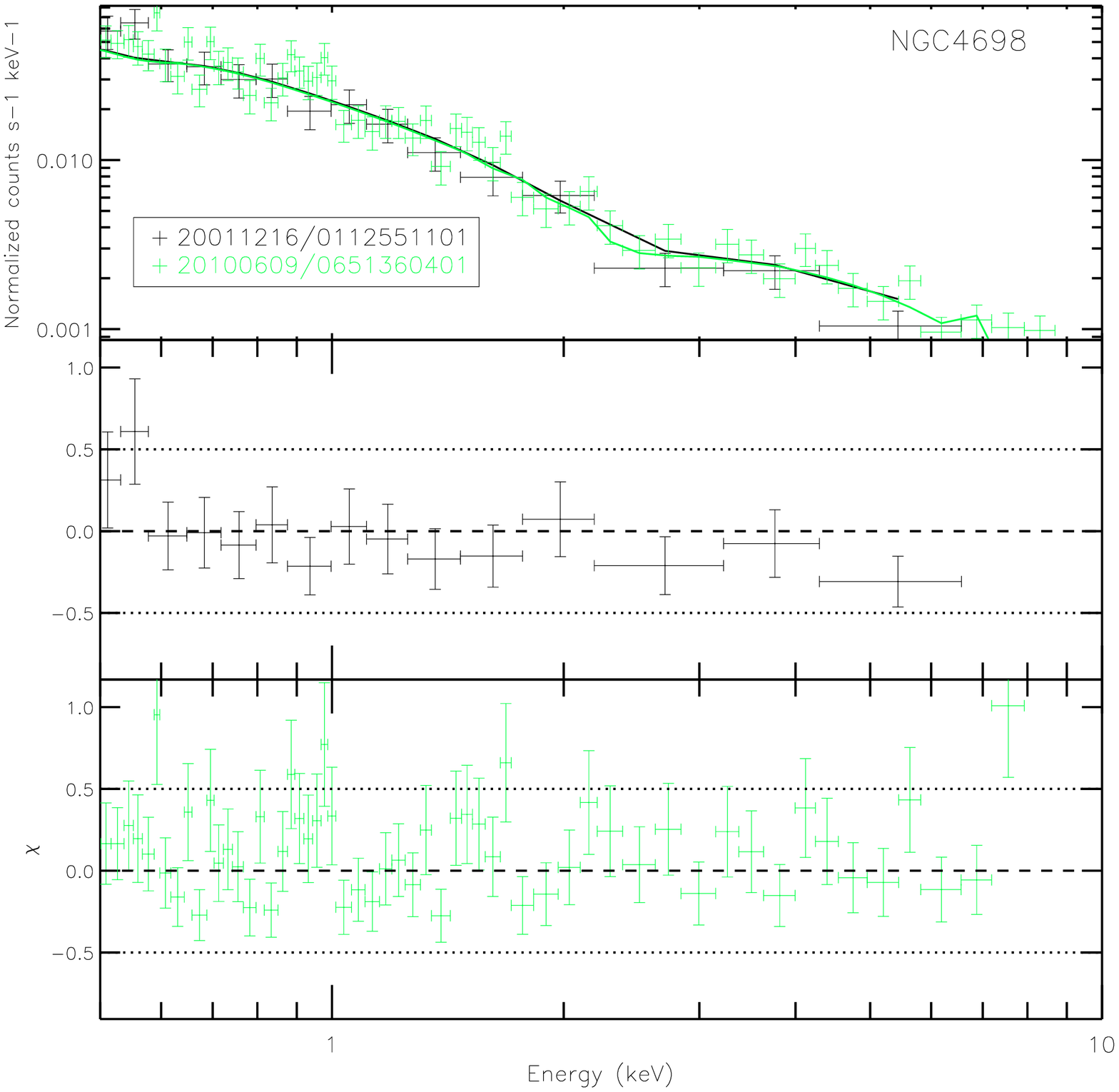}}
\subfloat{\includegraphics[width=0.30\textwidth]{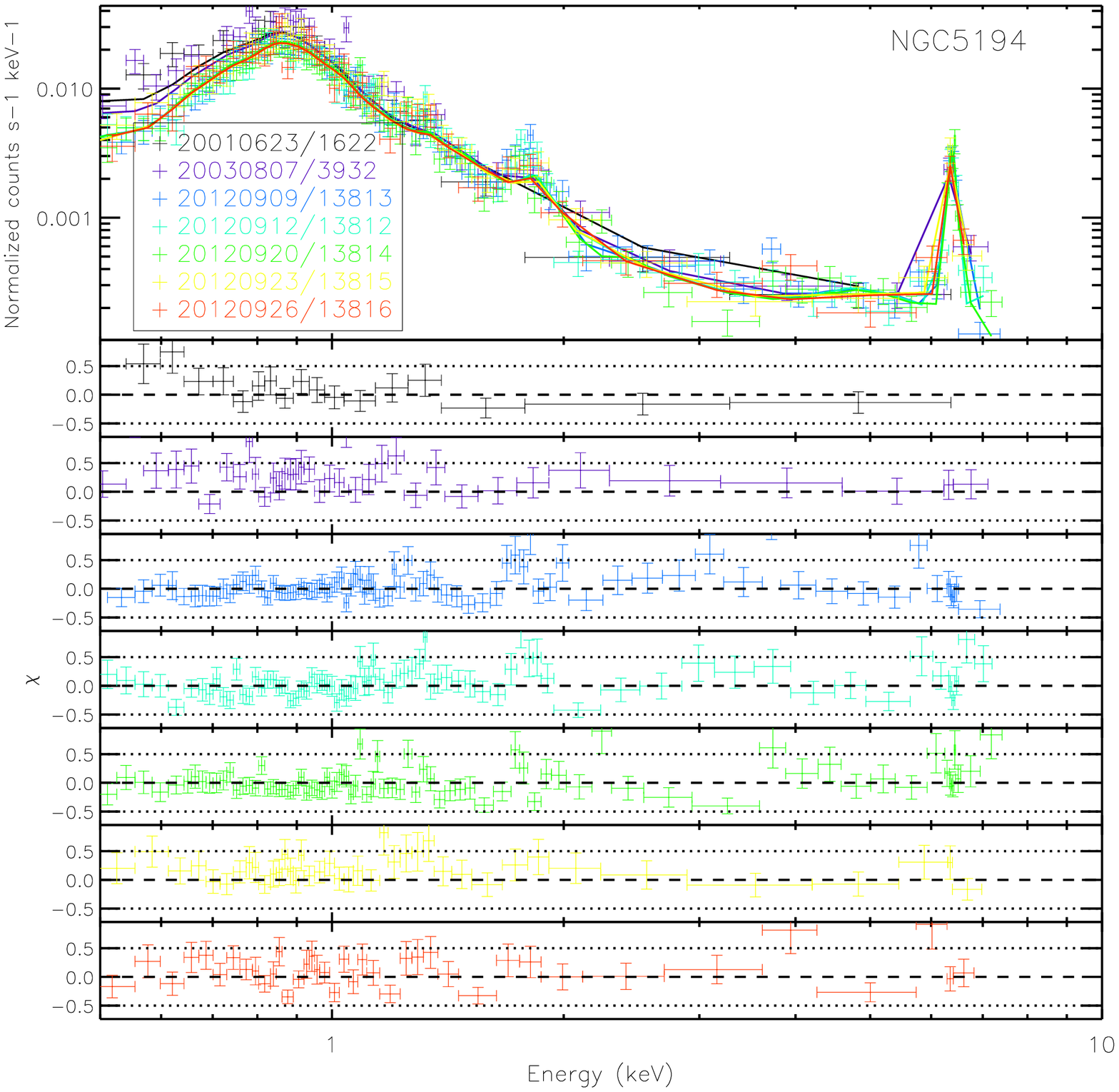}}
\subfloat{\includegraphics[width=0.30\textwidth]{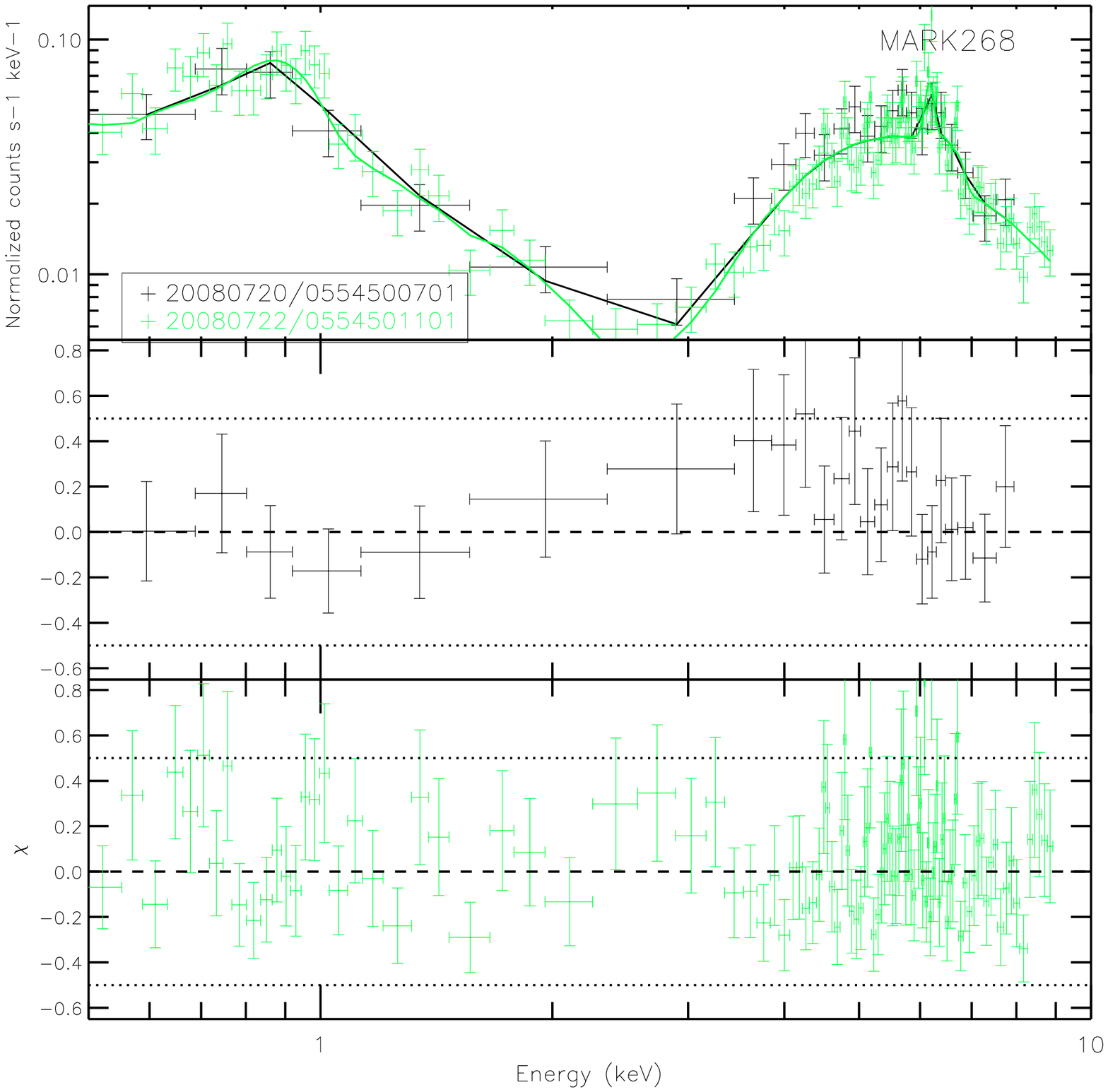}}
\caption{For each object, (top): simultaneous fit of X-ray spectra;
  (from second row on): residuals in units of $\sigma$. The legends
  contain the date (in the format yyyymmdd) and the obsID. Details are
  given in Table \ref{properties}.}
\label{bestfitSeyim}
\end{figure*}

\begin{figure*}
\setcounter{figure}{1}
\centering
\subfloat{\includegraphics[width=0.30\textwidth]{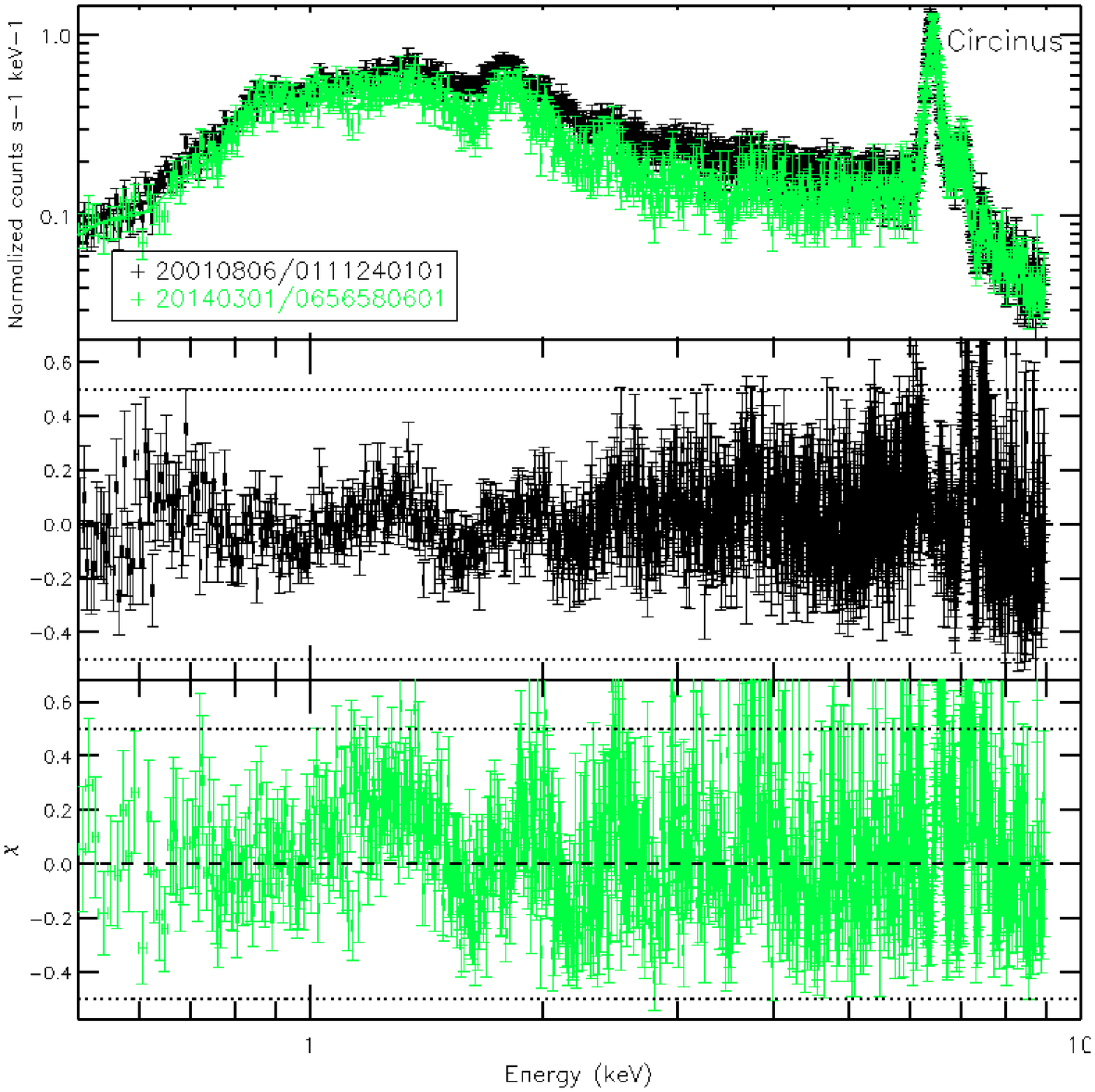}}
\subfloat{\includegraphics[width=0.30\textwidth]{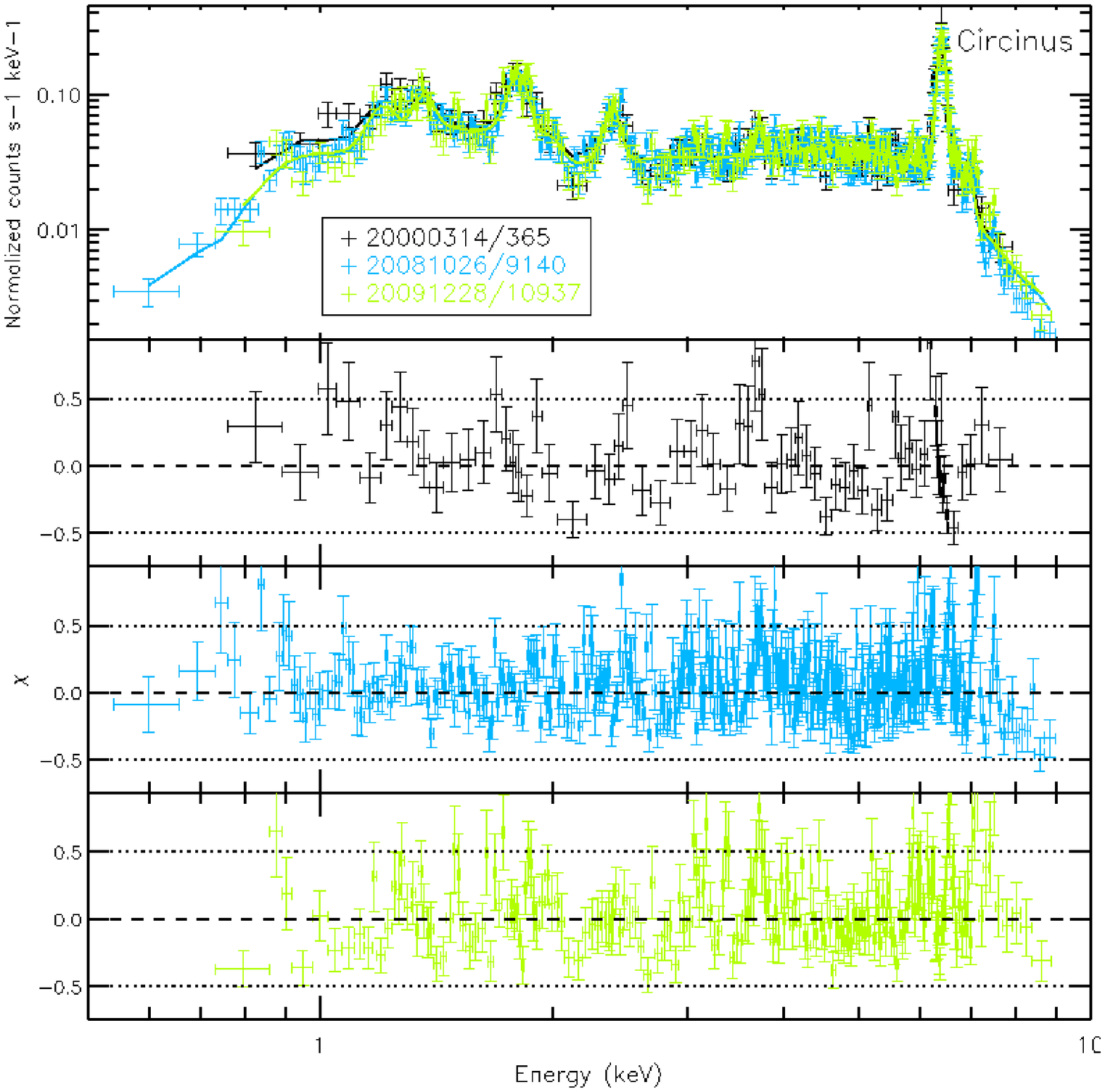}}
\subfloat{\includegraphics[width=0.30\textwidth]{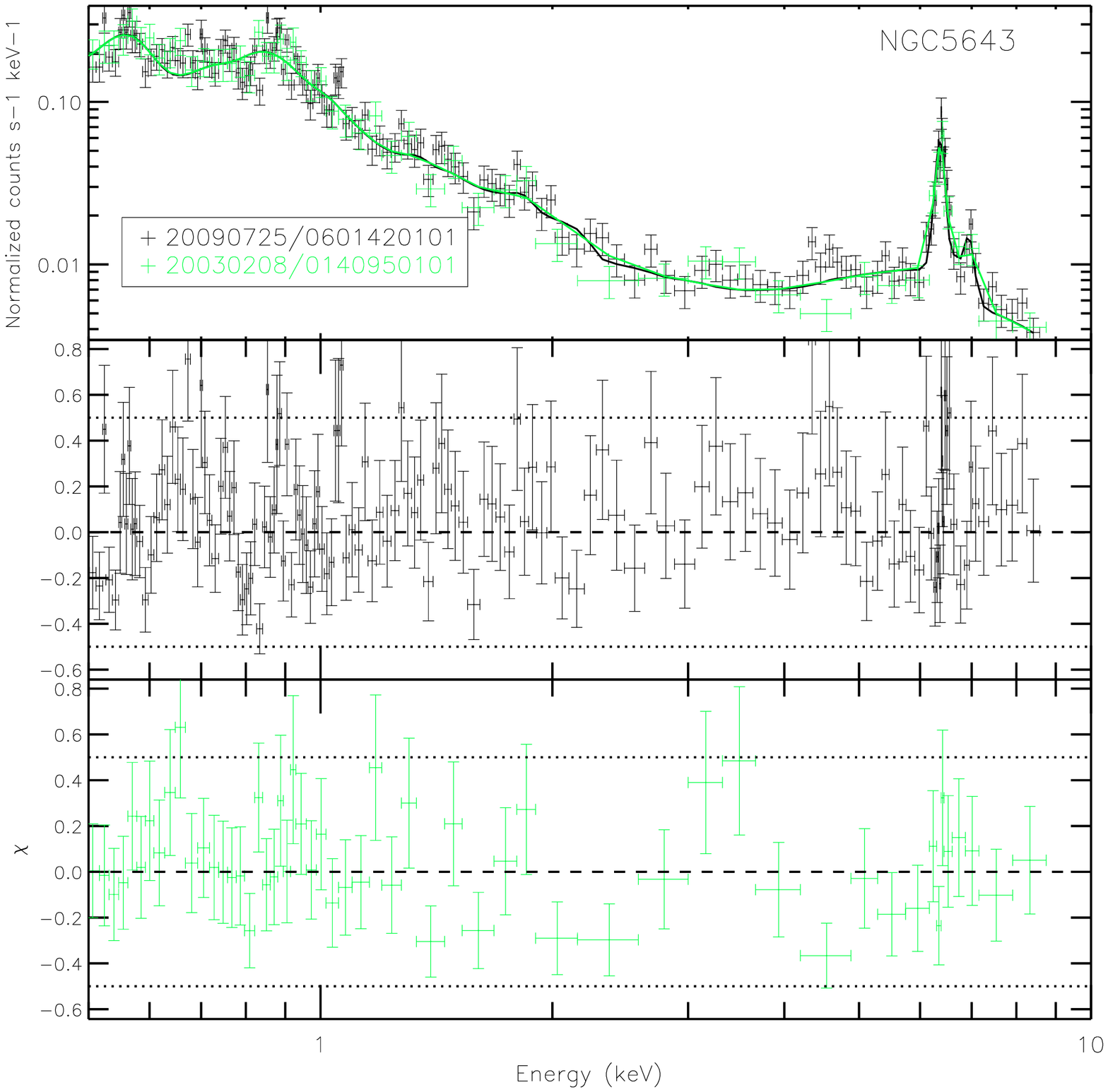}}

\subfloat{\includegraphics[width=0.30\textwidth]{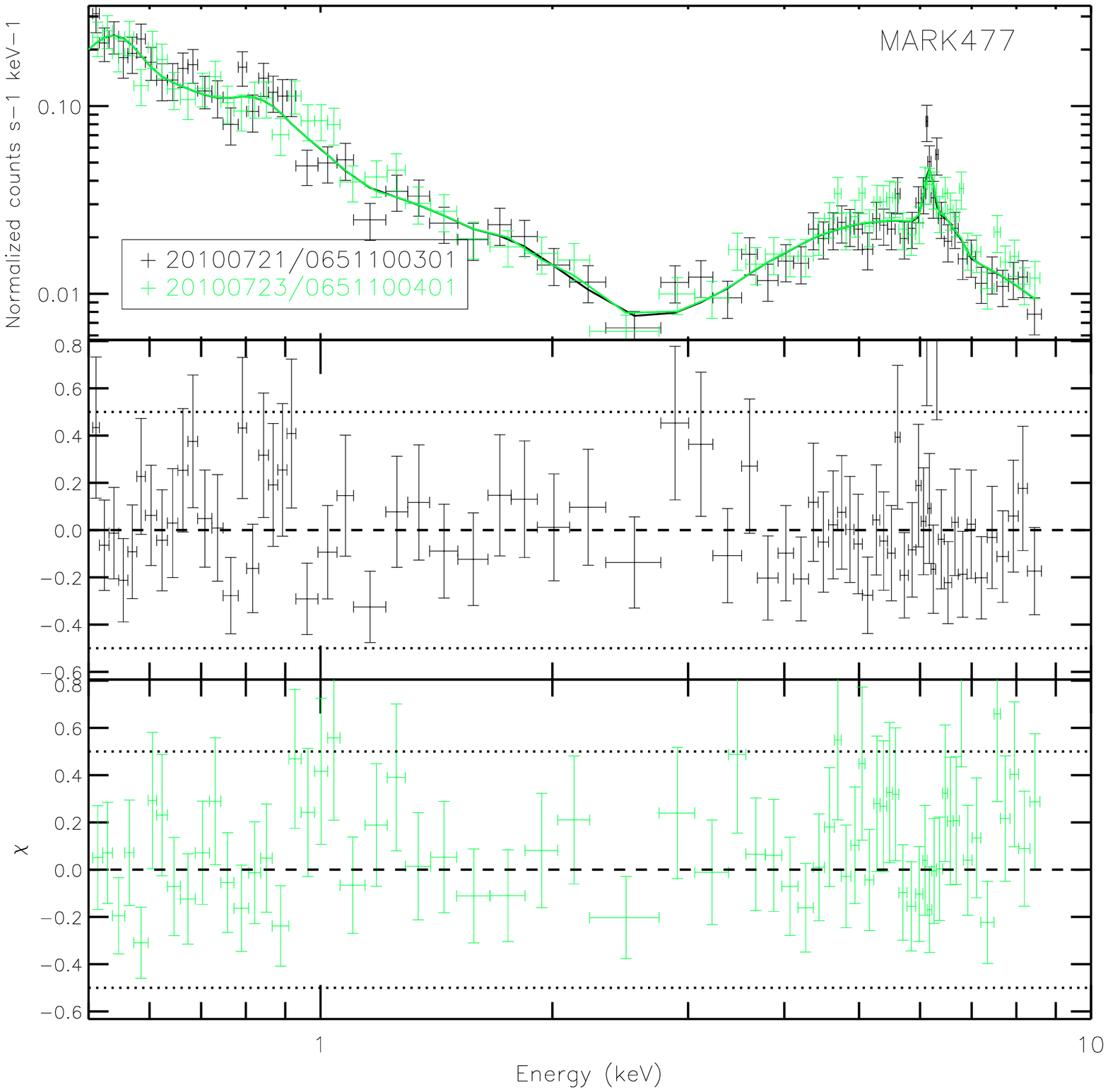}}
\subfloat{\includegraphics[width=0.30\textwidth]{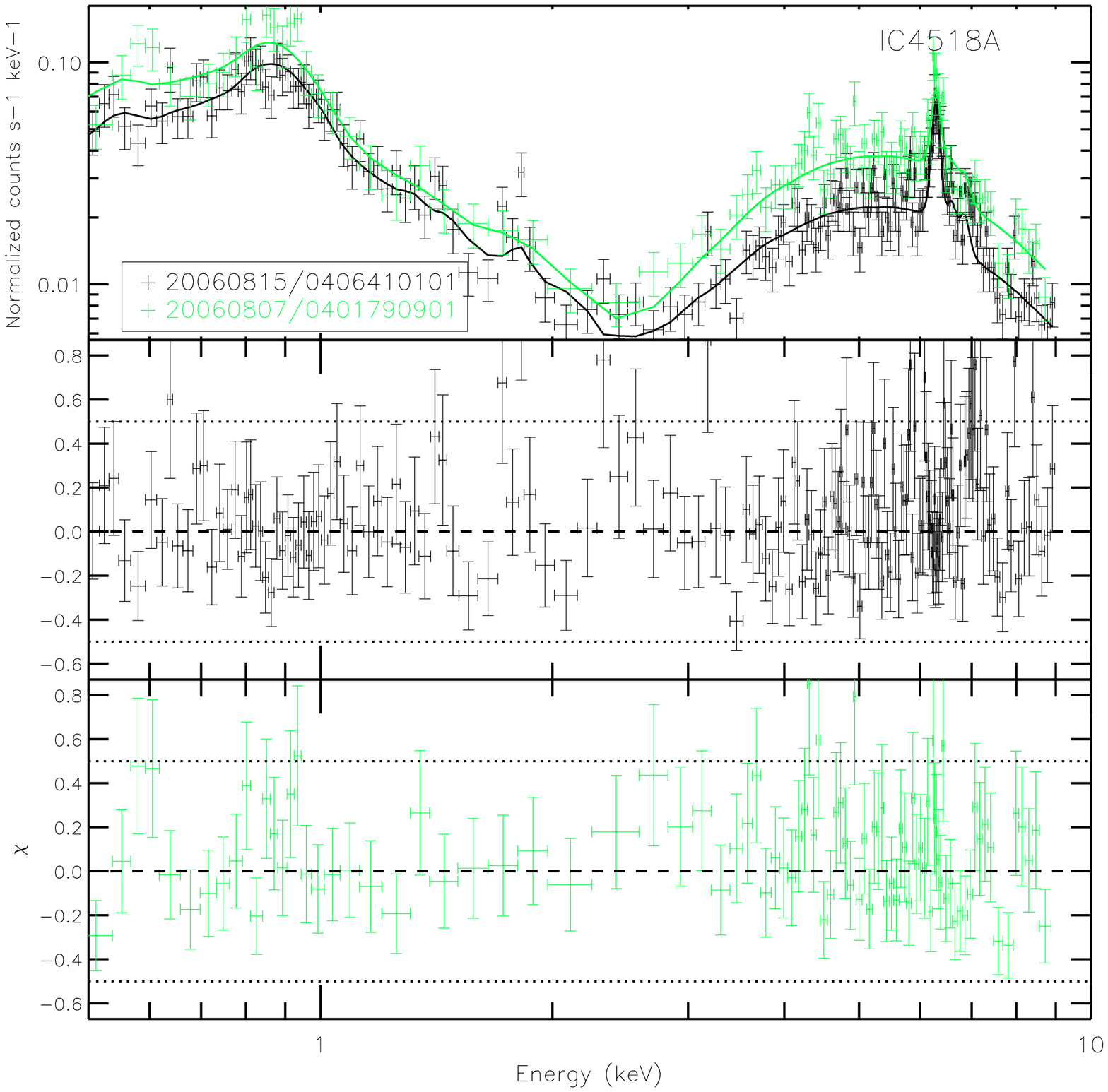}}
\subfloat{\includegraphics[width=0.30\textwidth]{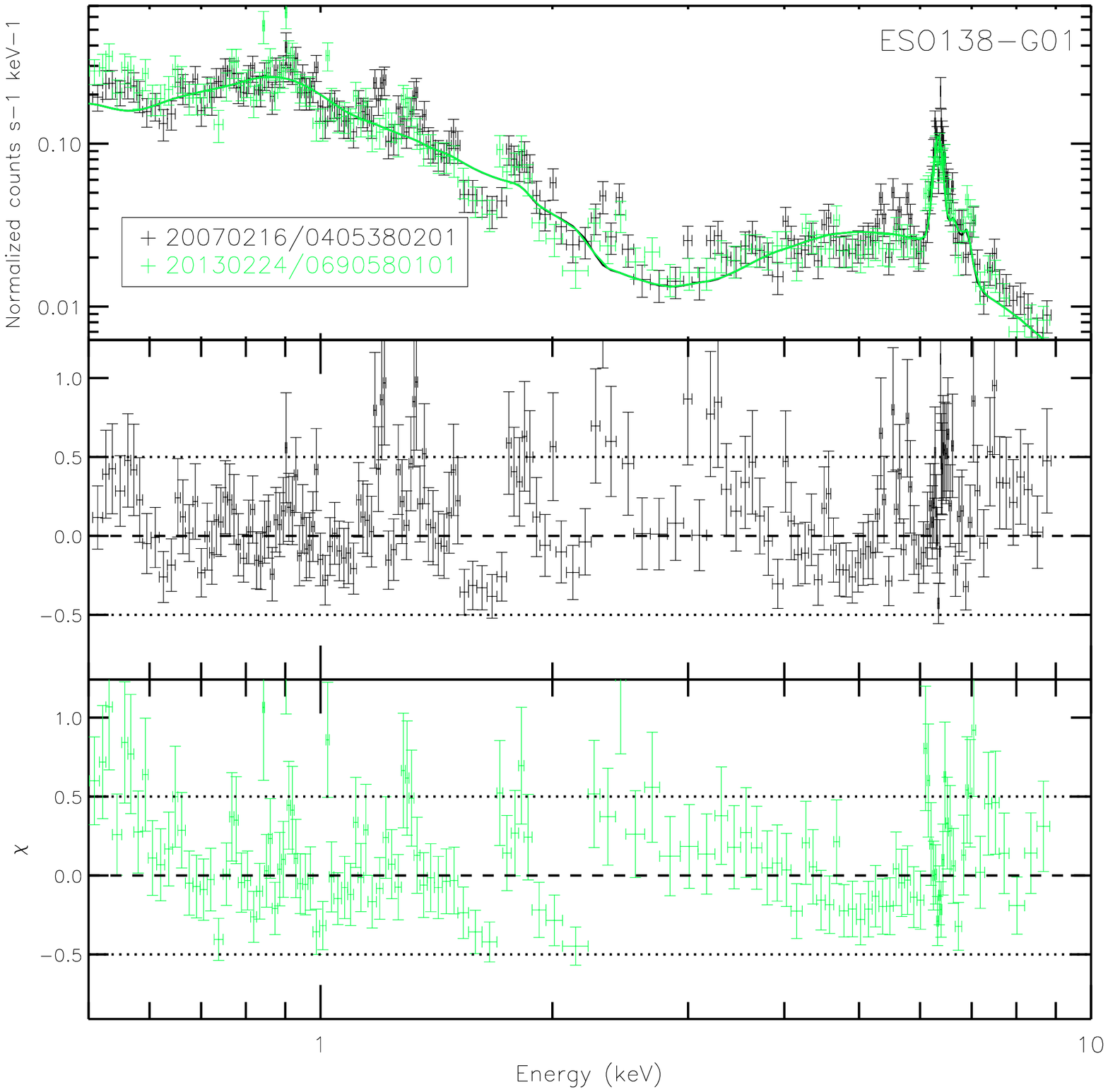}}

\subfloat{\includegraphics[width=0.30\textwidth]{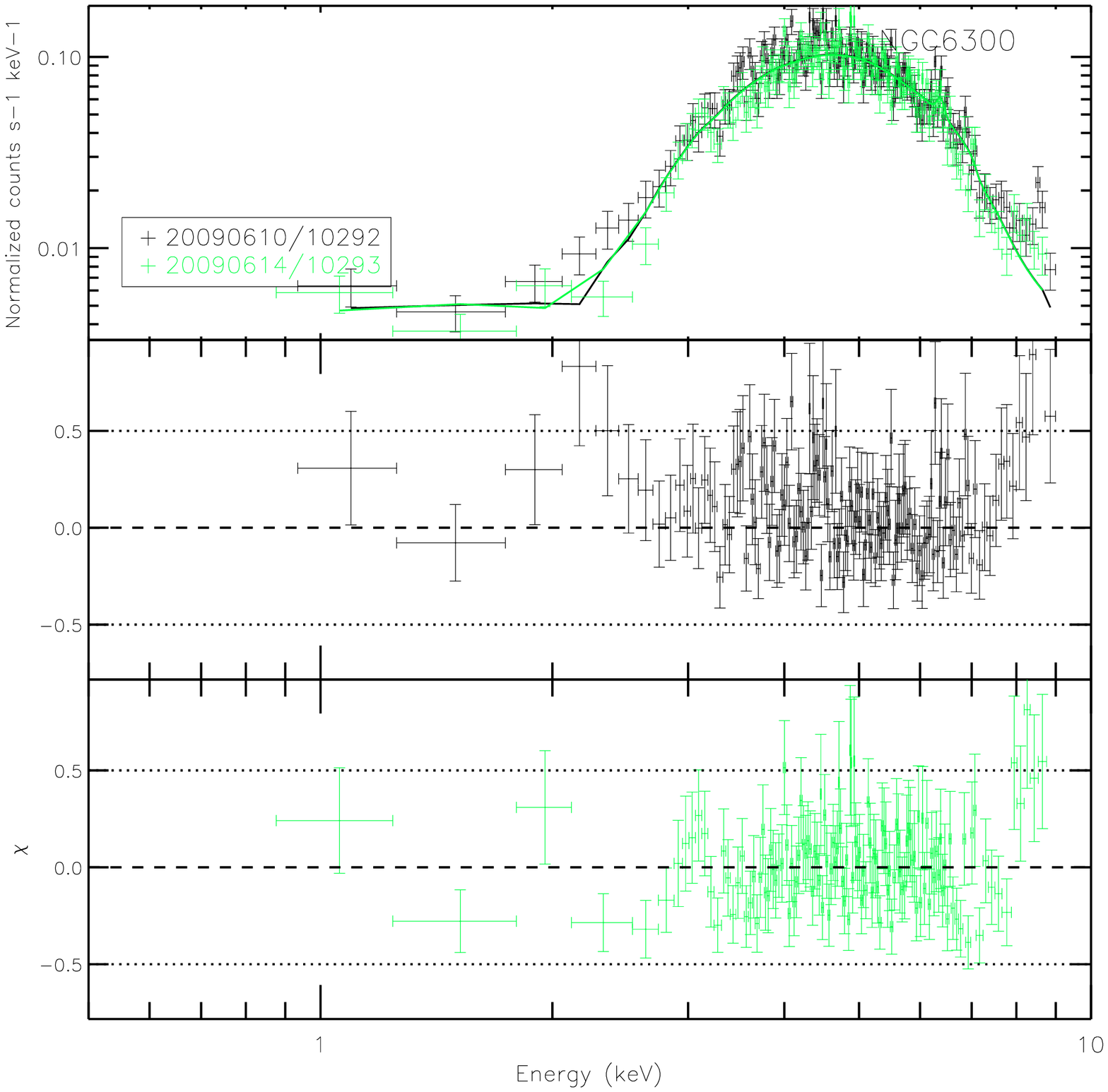}}
\subfloat{\includegraphics[width=0.30\textwidth]{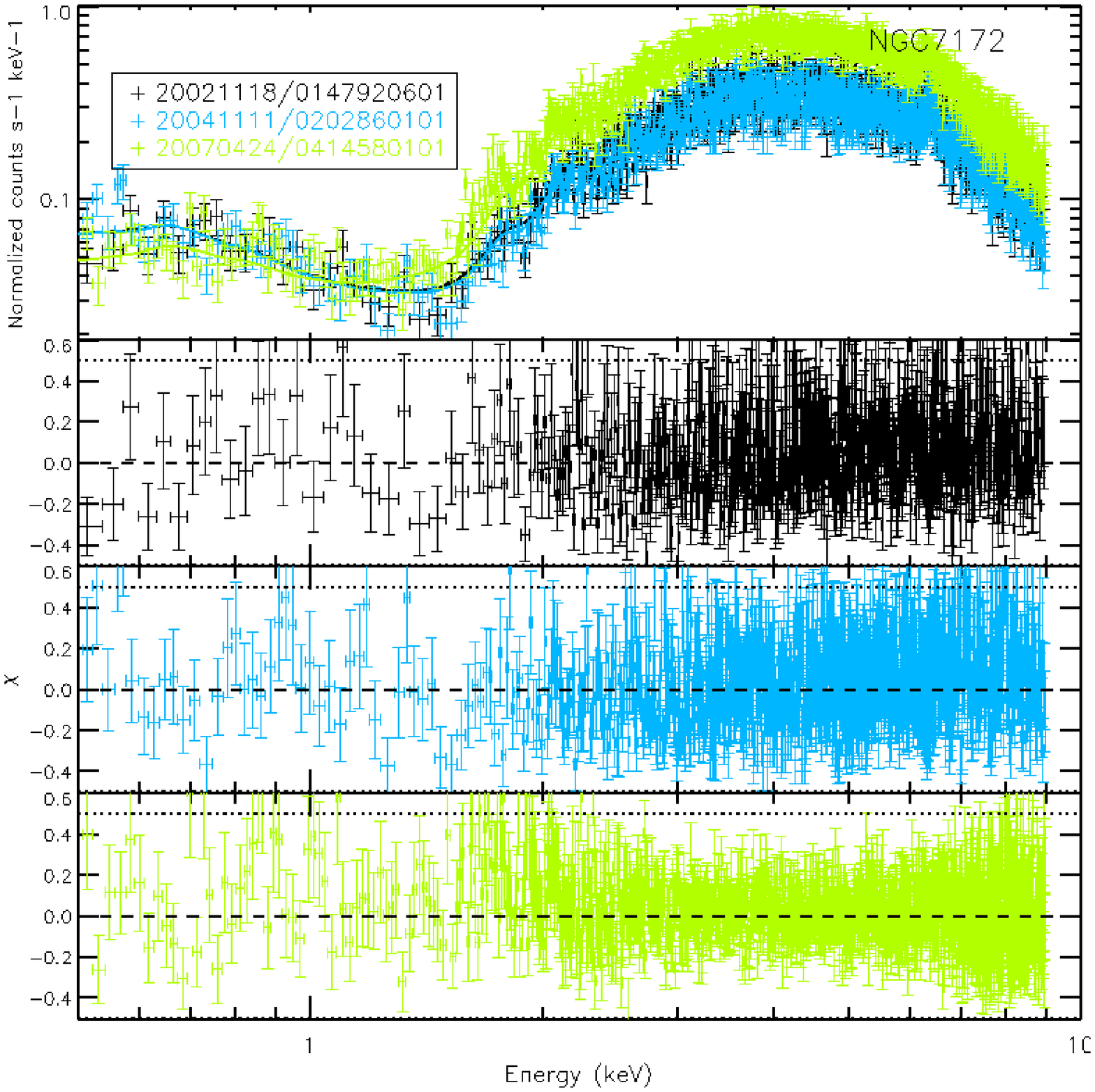}}
\subfloat{\includegraphics[width=0.30\textwidth]{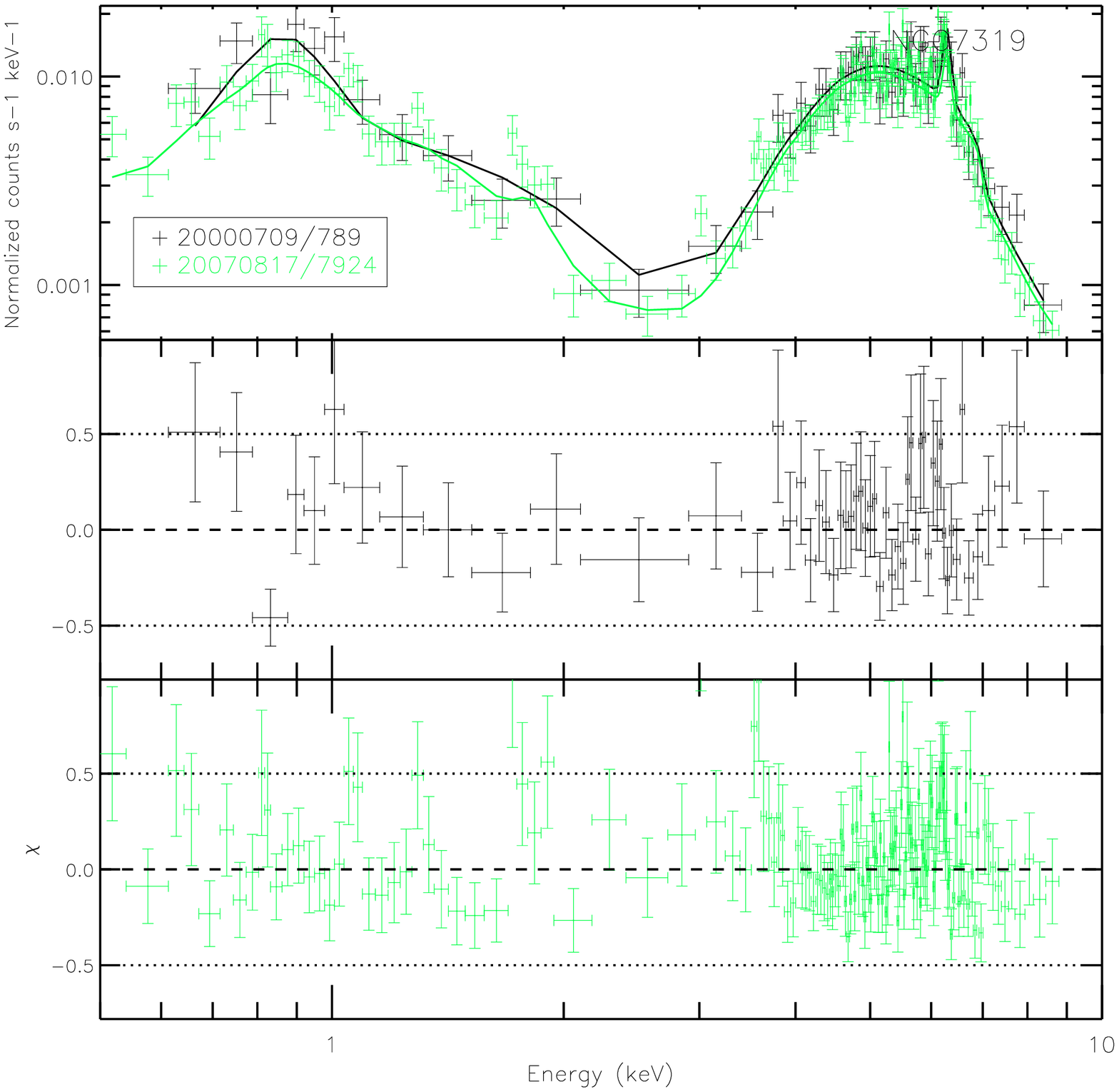}}
\caption{Cont.}
\end{figure*}

\subsection{\label{ind}Individual objects}

\begin{figure*}
\centering
\subfloat{\includegraphics[width=0.30\textwidth]{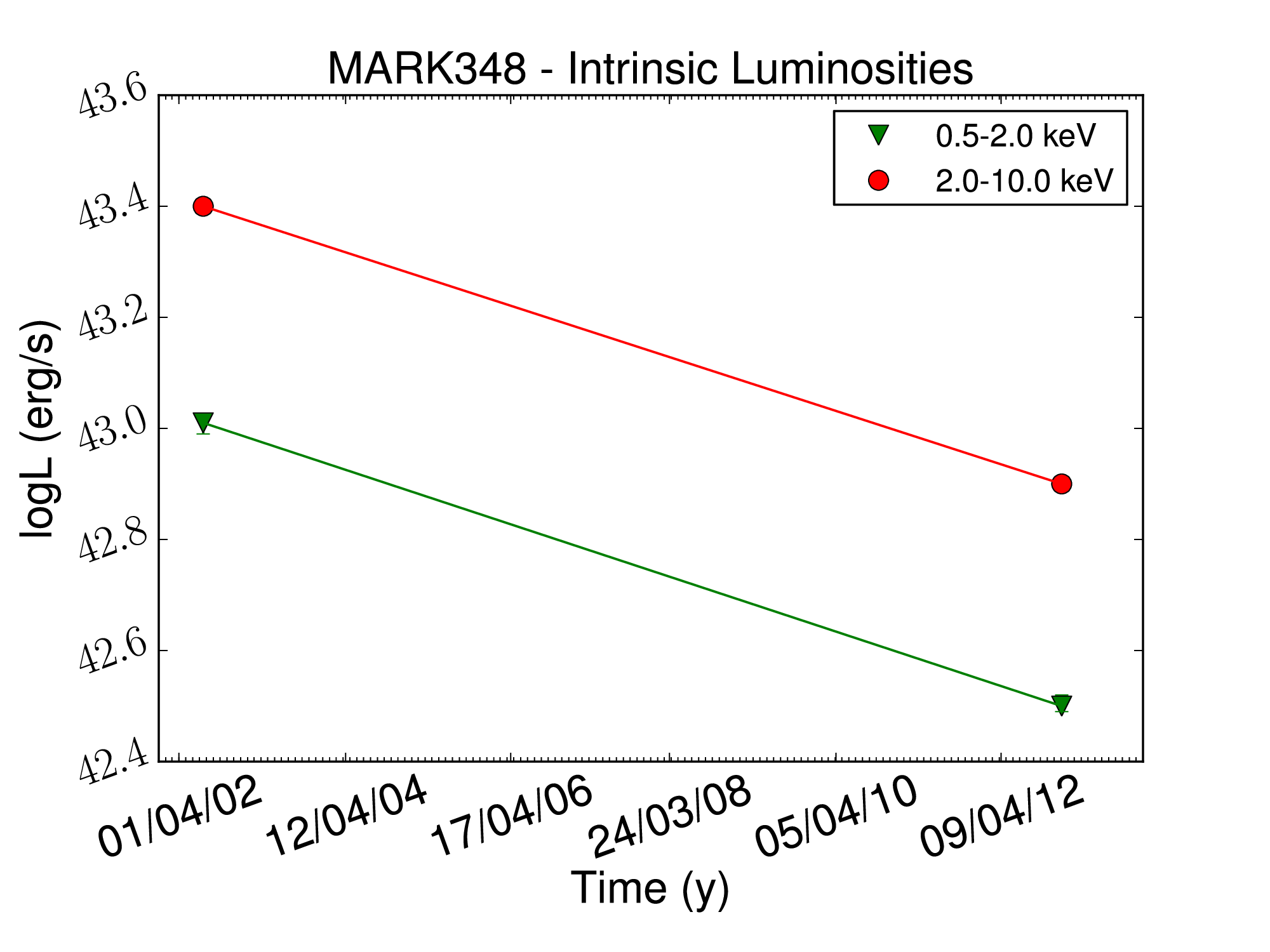}}
\subfloat{\includegraphics[width=0.30\textwidth]{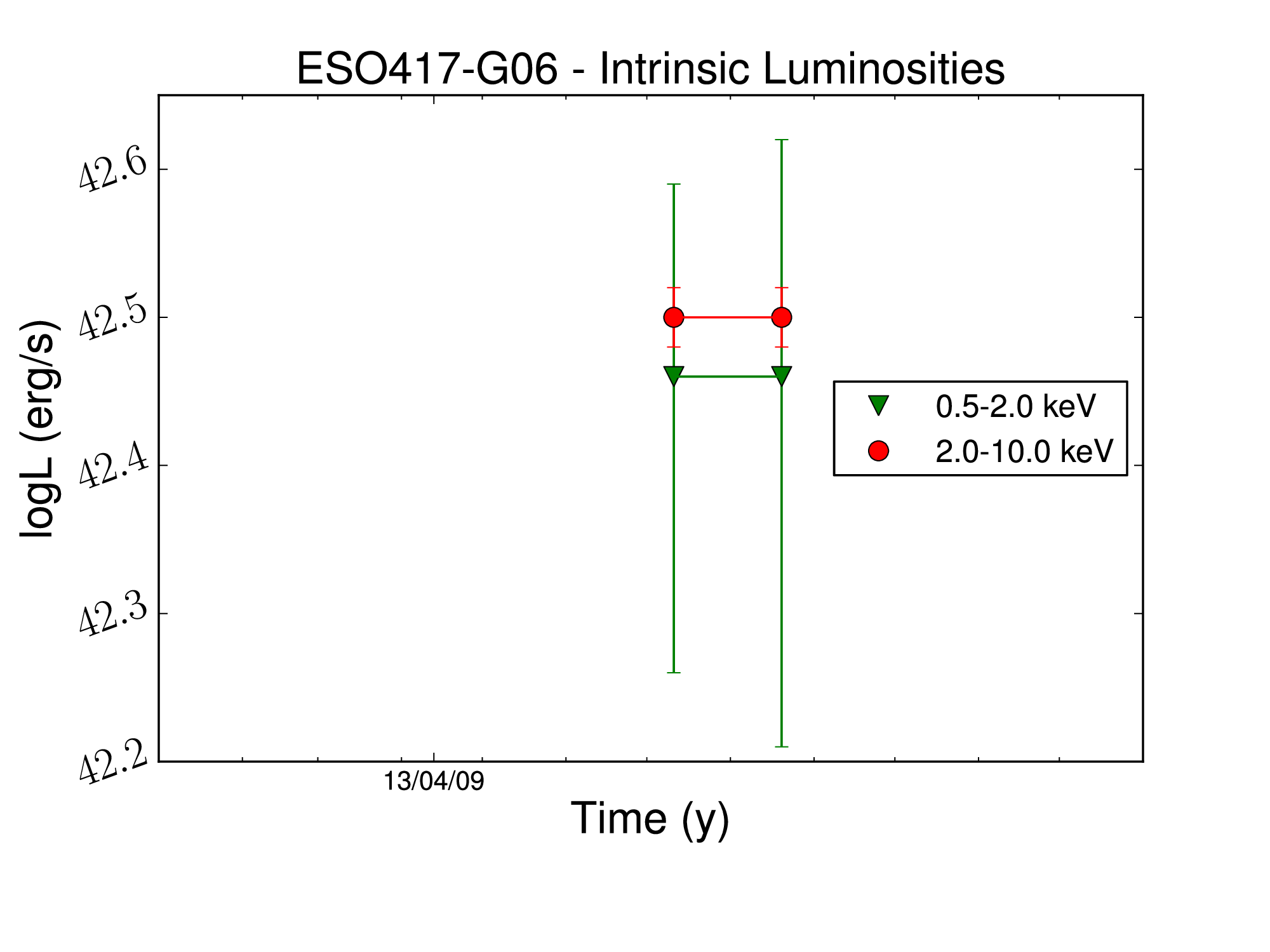}}
\subfloat{\includegraphics[width=0.30\textwidth]{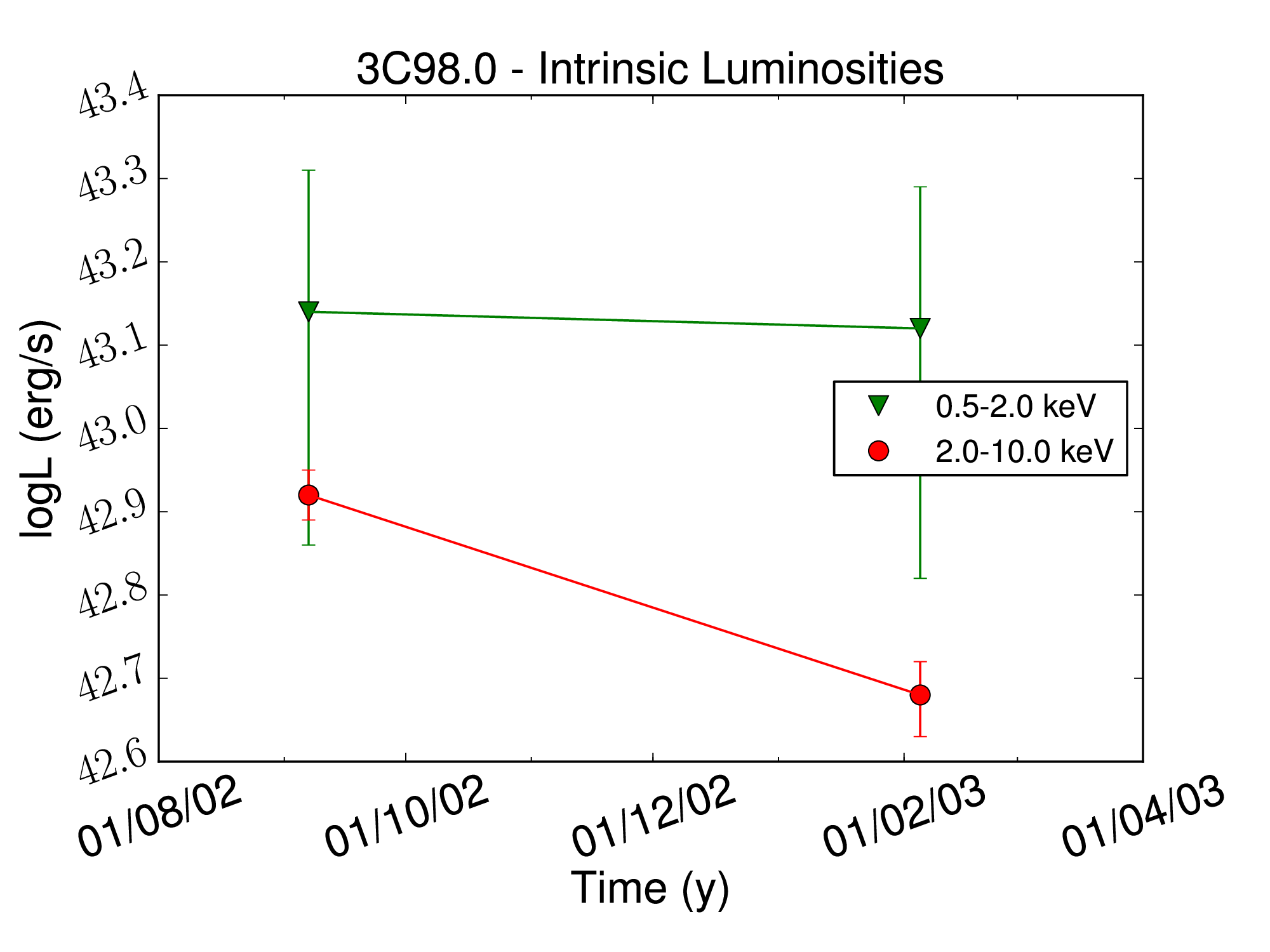}}

\subfloat{\includegraphics[width=0.30\textwidth]{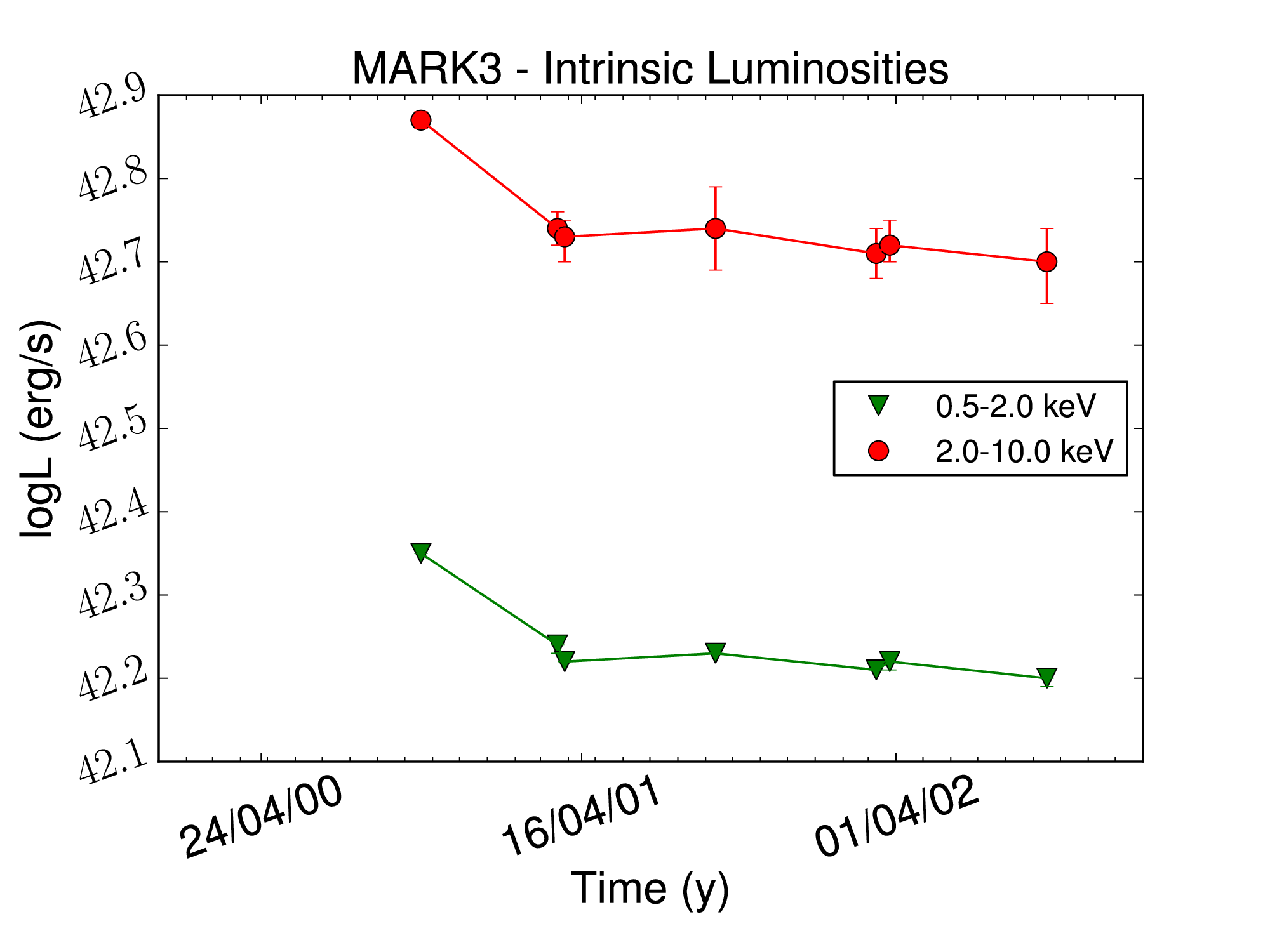}}
\subfloat{\includegraphics[width=0.30\textwidth]{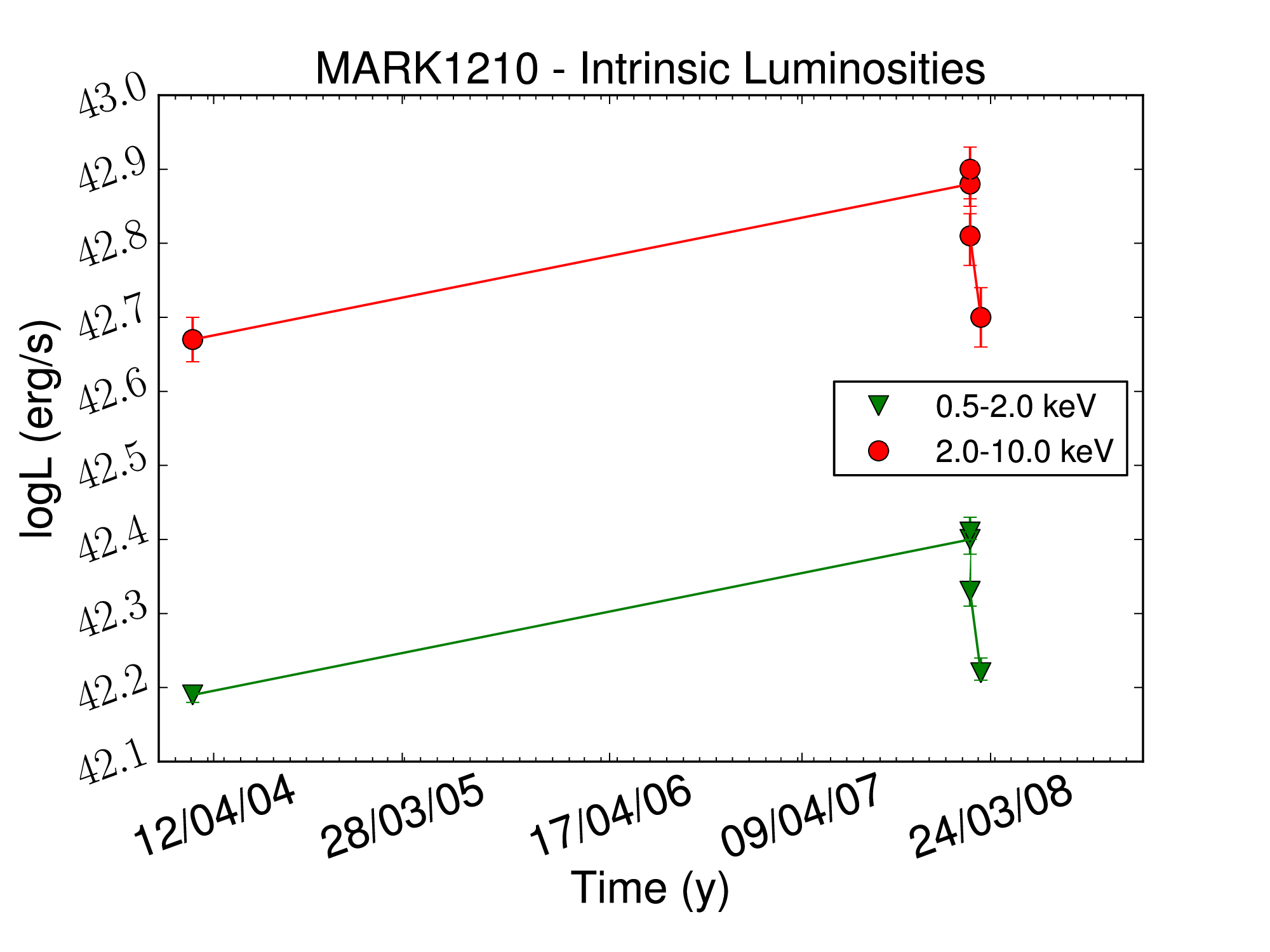}}
\subfloat{\includegraphics[width=0.30\textwidth]{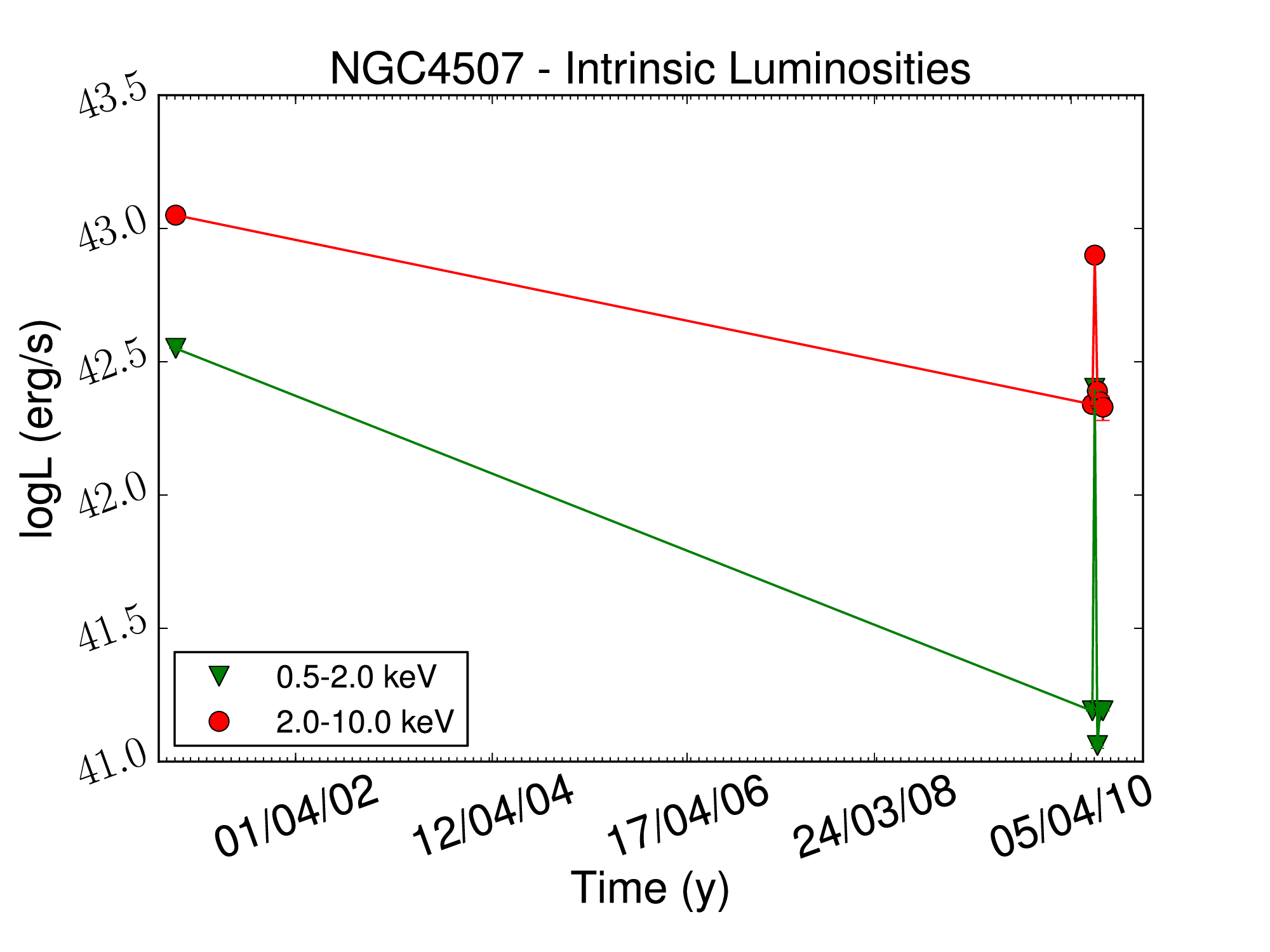}}

\subfloat{\includegraphics[width=0.30\textwidth]{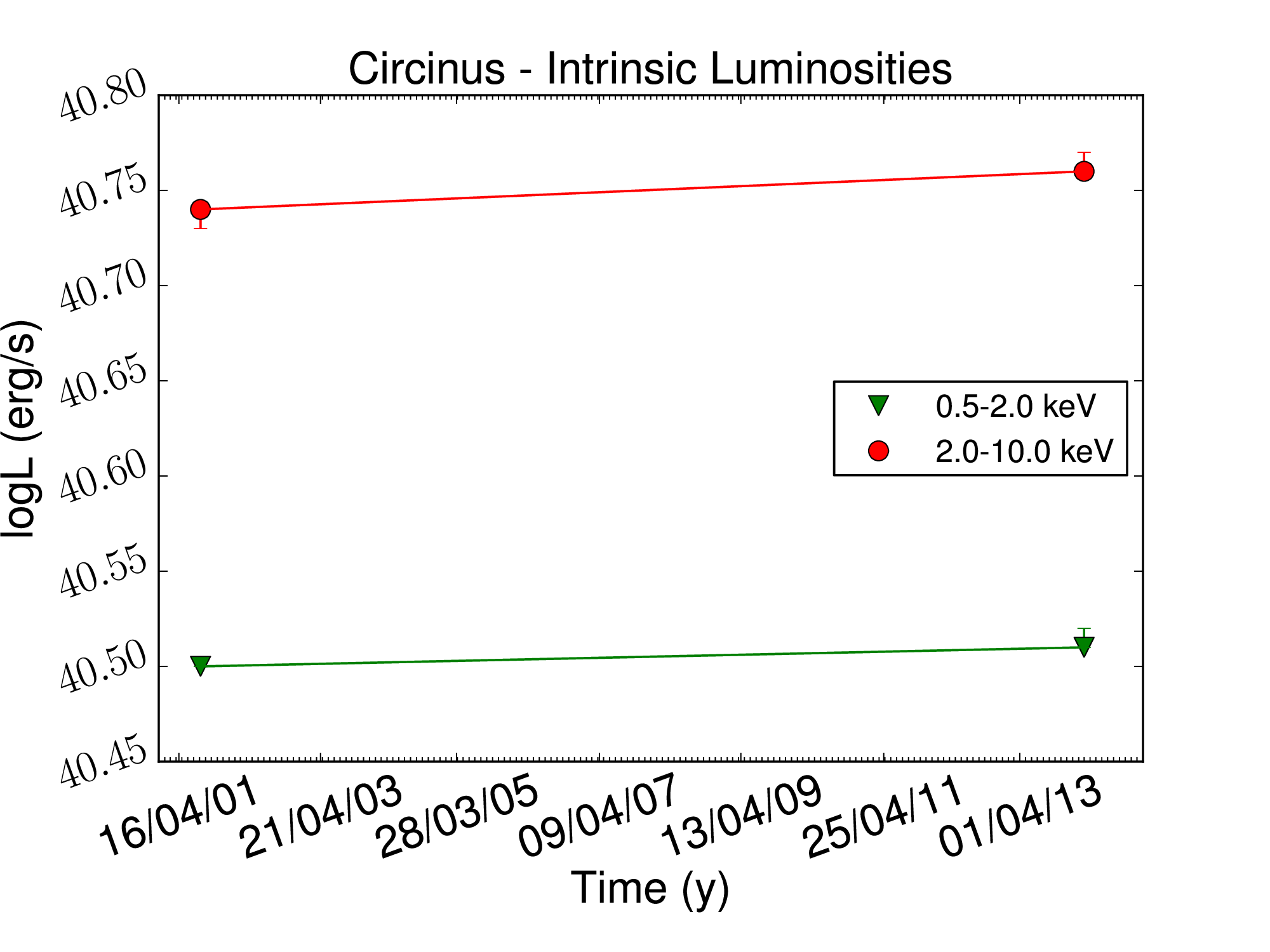}}
\subfloat{\includegraphics[width=0.30\textwidth]{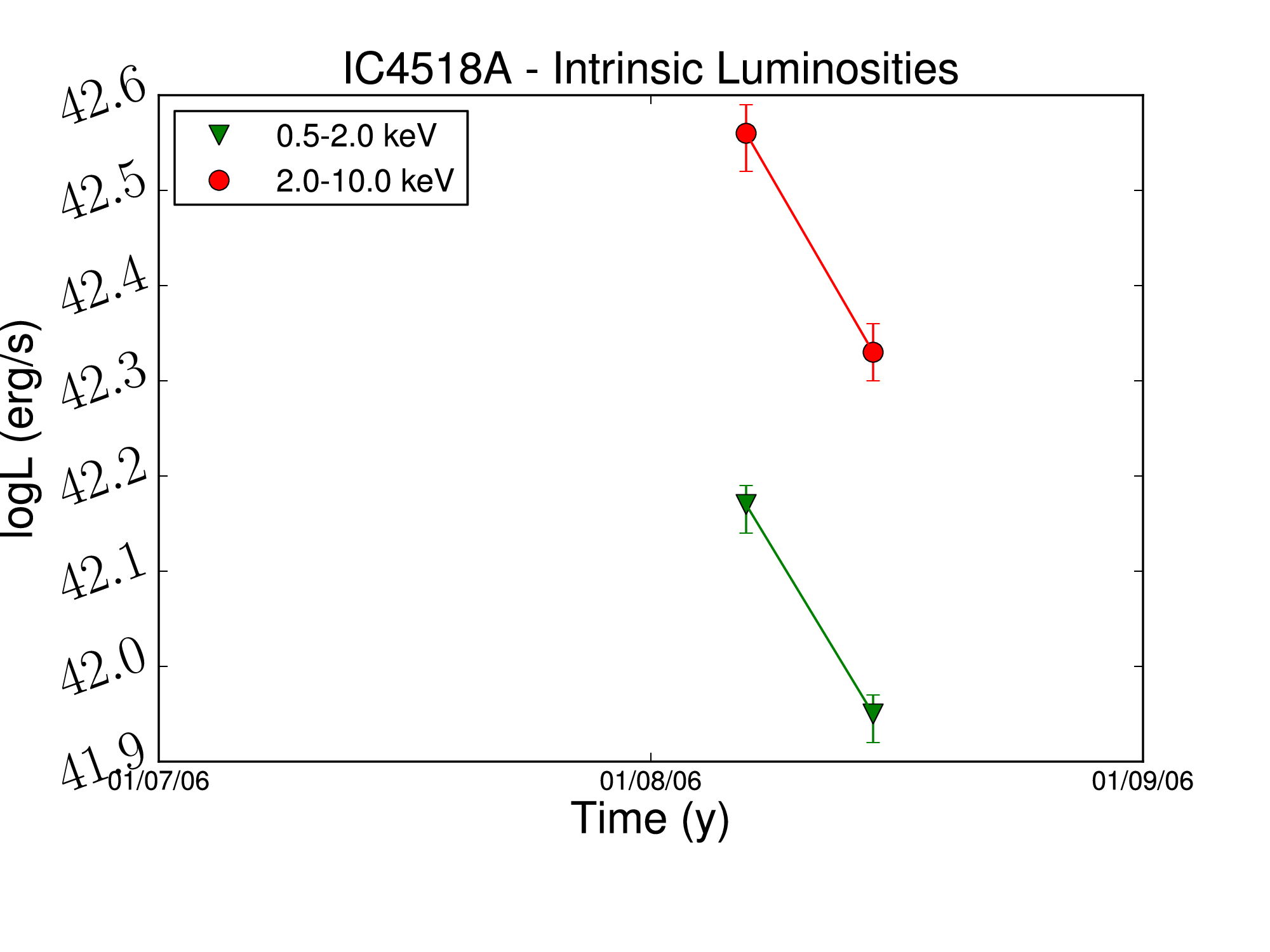}}
\subfloat{\includegraphics[width=0.30\textwidth]{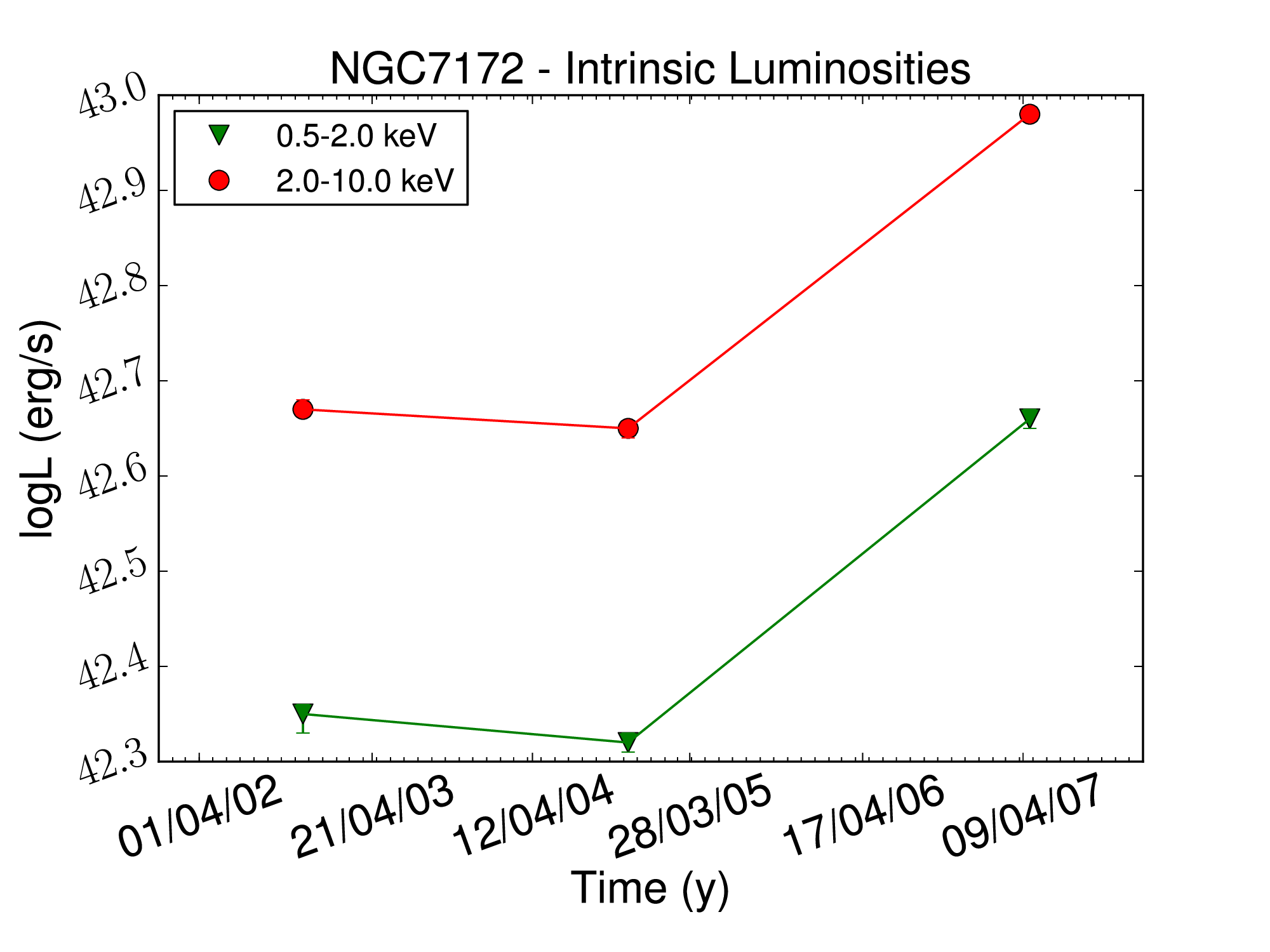}}

\subfloat{\includegraphics[width=0.30\textwidth]{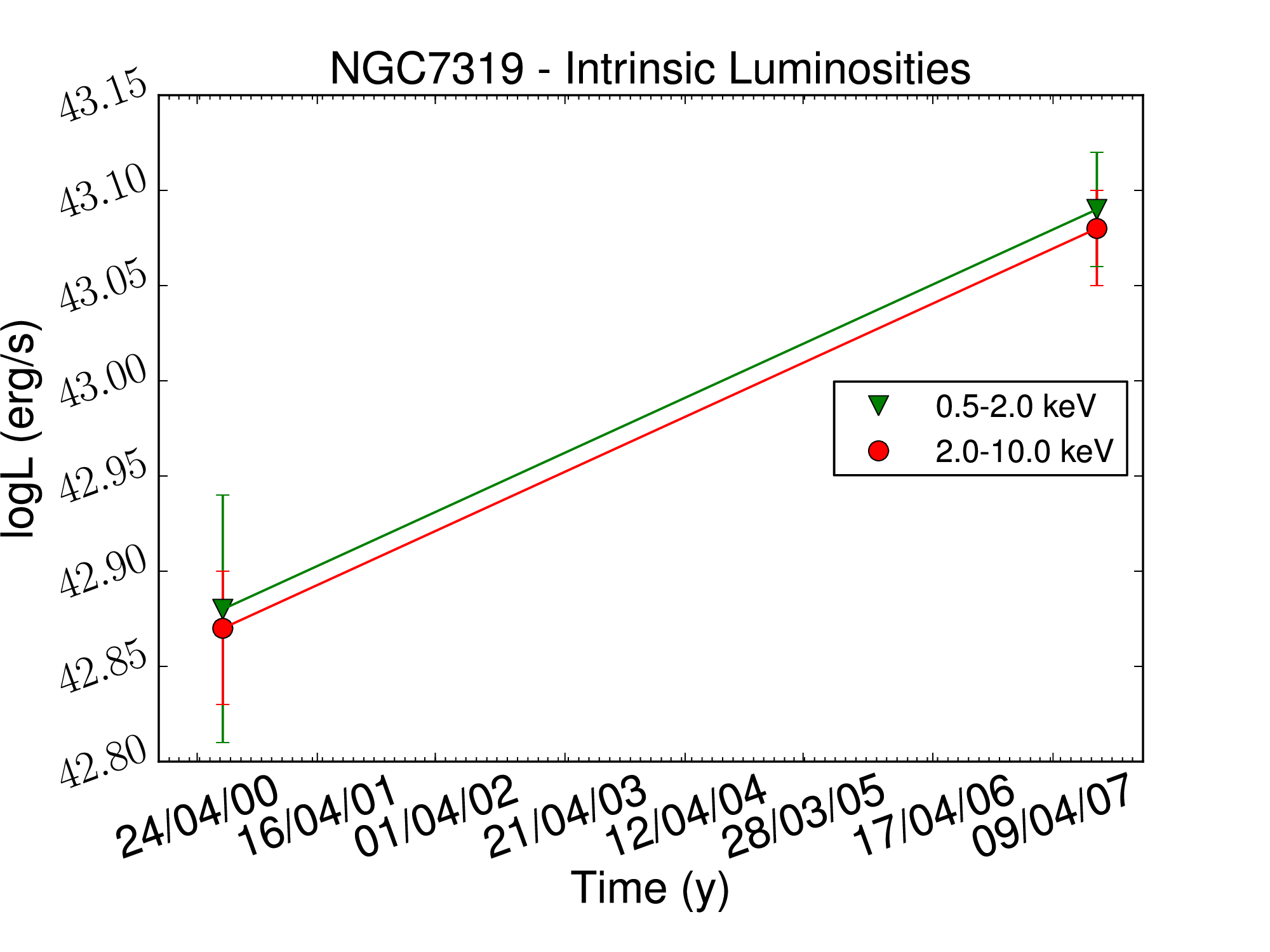}}
\caption{X-ray intrinsic luminosities calculated for the soft
  (0.5--2.0 keV, green triangles) and hard (2.0--10.0 keV, red
  circles) energies in the simultaneous fits, only for the variable
  objects.}
\label{luminXfigSey}
\end{figure*}

For details on the data and results, we refer the reader to the
following tables and figures: the observations used in the analysis
(Table \ref{obsSey}); UV luminosities with simultaneous OM data
(Col. 9 of Table \ref{obsSey} and Fig. \ref{luminUVfigSey});
individual and simultaneous best fit, and the parameters varying in
the model (Table \ref{bestfitSey} and Fig. \ref{bestfitSeyim}); X-ray
flux variations (Table \ref{lumincorrSey} and
Fig. \ref{luminXfigSey}); comparison of \emph{Chandra} and
\emph{XMM}--Newton data using the annular region (Table
\ref{annulusSey}); the simultaneous fit between these observations
(Table \ref{simultanilloSey} and Figs. \ref{ringSey} and
\ref{luminXfigSeyring}); short-term variability from the analysis of
the light curves (Table \ref{estcurvasSey} and Appendix
\ref{lightcurves}); and the \emph{Compton}-thickness analysis, where
an object was classified on the basis that at least two of the three
criteria presented in Sect \ref{thick} were met (Table \ref{ew}). We
notice that the addition of a cold reflection component is not
statistically required by the data, so we do not mention the analysis
except in one case (3C\,98.0) where the simultaneous fit was
performed.

\tiny
\begin{table*}
\begin{center}
\caption{\label{variab} Results of the variability analysis. } 
\begin{tabular}{lccccccccc} \hline \hline
Name & Type & log ($L_{soft}$) & log ($L_{hard}$) & log ($M_{BH}$) & log ($R_{Edd}$) & & Variability & & $\Delta$T$_{max}$  \\ \cline{7-9} 
 & & (0.5-2 keV) & (2-10 keV) & & & SMF0 &  SMF1 & SMF2 & (Years)  \\ 
(1) & (2) & (3)          & (4) & (5) & (6) & (7) & (8) & (9) & (10)  \\ \hline
MARK\,348 (X) &  HBLR  & 42.76 & 43.15 & 7.58 & -1.02 & ME2PL & $Norm_2$ & - & 10  \\
              &   &  69$^+_-$5\% & 68$^+_-$3\% & & & & 69$^{+17}_{-14}$\% & \\
NGC\,424 (C,X)* & HBLR  & 41.74 & 41.85 & 7.78 & -2.53 & 2ME2PL & - & -   & 0.16  \\
               &  & 0\% & 0\% & \\
MARK\,573 (C)* & HBLR  & 41.65 & 41.54 & 7.37 & -2.42 & 2ME2PL & - & - & 4  \\
              &  & 0\%  & 0\% &     &  & (+3gauss) & \\
\hspace*{1.5cm}   (X,C) &  & 41.73 & 41.41 & & & 2ME2PL & - & - & 2 \\          
              &  & 0\%  & 0\% &     &  &  & \\                                
NGC\,788 (X,C) & HBLR  & 42.11 & 42.60 & 7.43 & -1.43 & 2ME2PL & - & - &  0.33 \\
               &   & 0\%   & 0\%   &  &  & \\
ESO\,417-G06 (X) & - & 42.46 & 42.50 & 7.44 & -1.53 & MEPL & $N_{H2}$ & - &  0.08  \\
                 &    &  0\%  &  0\%  & & &      &  21$^{+5}_{-5}$\%    & \\
MARK\,1066 (X,C)* & NHBLR & 41.40 & 41.43 & 7.23 & -2.38 & ME2PL & - & - & 2   \\
                 &    &   0\% &  0\%  & & & \\               
3C\,98.0 (X) & - & 43.13 & 42.80 & 7.75 & -1.73 &  MEPL & $Norm_2$ & - & 0.41 \\
             &    &  5$^+_-$4\%  &  42$^+_-$7\% & & &       &  43$^{+41}_{-26}$\% \\
\hspace*{1cm}  (X,C) &  & 42.40 & 42.60 & & & MEPL & - & - & 5 \\
             &      &   0\% & 0\% &      &  & \\        
MARK\,3 (X)*   & HBLR  & 42.24 & 42.74  & 8.74 & -2.58 & 2ME2PL & $Norm_2$ & - & 1  \\
              &   & 29$^+_-$7\% & 32$^+_-$4\%   &  &  &        & 37$^{+16}_{-14}$\% & \\
MARK\,1210 (C) & HBLR & 42.31 & 42.79 & 7.70 & -1.50 & 2ME2PL & $Norm_2$ & $N_{H2}$ & 4  \\
               &    &  7$^+_-$5\% & 7$^+_-$1\%  & & &      &  11$^{+10}_{-6}$\%    &  20$^{+5}_{-4}$\% & \\       
IC\,2560 (X,C)* & -  & 40.57 & 41.03 & 6.46 & -2.02 &  2ME2PL & - & - & 0.16  \\
               &  & 0\%   & 0\%   &  &  &  (+1gauss) & \\
NGC\,3393 (C)* & -   & 41.64 & 41.29 & 8.10 & -3.41 & 2ME2PL & - & - & 7  \\
              &   & 0\% & 0\% &      &  & \\
\hspace*{1.38cm}  (X,C) &  & 41.44 & 41.26 &  &  & 2ME2PL & - & - & 0.66 \\ 
              &   & 0\% & 0\% &      &  & \\                                   
NGC\,4507 (X)  &  HBLR &  42.04  &  42.67  & 8.26 & -2.28 & 2ME2PL & $Norm_2$ & $N_{H2}$ & 9  \\
             &  &  96$^+_-$4\% & 81$^+_-$10\% &  &  & (+2gauss) & 51$^{+26}_{-20}$\% & 4$^{+12}_{-9}$\% & \\
\hspace*{1.38cm}  (X,C) &  & 41.96 & 42.85 &  &  & & $Norm_2$ & - & 0.41  \\   
                        &  &  45$^+_-$3\% & 38$^+_-$3\%  &  &  & & 53$^{+36}_{-27}$\% \\
NGC\,4698 (X) & - & 40.14 & 40.08 & 7.53 & -4.04 & 2PL & - & - & 9  \\
              &  & 0\%  & 0\%   & & & \\
NGC\,5194 (C)* & NHBLR & 39.53 & 39.51 & 6.73 & -3.82 & ME2PL & - & - & 11  \\
              &    &  0\%  & 0\%   & & &    \\
\hspace*{1.4cm} (X,C) &  & 39.94 & 39.39 & & & 2ME2PL & - & - & 0.6 \\          
              &  & 0\%  & 0\% &     &  &  & \\                        
MARK\,268 (X) & -  & 41.34 & 42.92 & 7.95 & -1.62 & ME2PL & - & - & 0.01  \\
          &  & 0\%   & 0\%   & & & \\
MARK\,273 (X,C)$^{CL?}$ & -  & 41.34 & 42.29 & 7.74 & -2.05 & 2ME2PL & $N_{H2}$ & - & 2  \\
                &  & 24$^+_-$2\% & 32$^+_-$6\%   & & & & 51$^{+15}_{-14}$\% & \\
Circinus (C)* & HBLR  & 39.80 & 40.60 & 7.71 & -3.71 & 2ME2PL & - & - & 9  \\
             &  & 0\%  & 0\%    &  &  & (+4gauss)  & \\ 
NGC\,5643 (X)* &  NHBLR & 40.44 & 40.87  & 6.30 & -2.02 & 2ME2PL & - & - & 6  \\
              &   & 0\%  & 0\%   &   &  & \\
MARK\,477 (X)*  & HBLR & 42.60 & 43.11 & 7.20 & -0.68 & 2ME2PL & - & - & 0.01  \\
             &  & 0\% & 0\%         &  &  &  \\
IC\,4518A (X) &  - & 42.06 & 42.45 & 7.48 & -1.63 & 2ME2PL & $Norm_2$ & - & 0.02  \\
              &   & 40$^+_-$2\% & 41$^+_-$6\%   &  &  &        & 42$^{+45}_{-30}$\%   & \\
ESO\,138-G01 (X)* & -  & 42.23 & 42.11 & 5.50 & 0.01 & ME2PL & - & - & 6  \\
             &   & 0\% & 0\% & \\        
NGC\,6300 (C) & - & 41.32 & 41.95 & 7.18 & -2.68 & 2PL & - & - & 0.01  \\
          &    &  0\%  & 0\%   & & & \\
\hspace*{1.38cm}  (X,C) &  & 41.06 & 41.68 & & & 2PL & $Norm_2$ & $Norm_1$ & 8 \\
                     &  &  98$^+_-$50\% & 98$^+_-$16\%   & & &     & 98$^{+12}_{-77}$\%    & 93$^{+25}_{-25}$\% & \\                                                                       
\hline
\end{tabular}
\end{center}
\end{table*} 

\begin{table*}
\setcounter{table}{1}
\begin{center}
\caption{Cont. } 
\begin{tabular}{lccccccccc} \hline \hline
Name & Type & log ($L_{soft}$) & log ($L_{hard}$) & log ($M_{BH}$) & log ($R_{Edd}$) & & Variability & & $\Delta$T$_{max}$  \\ \cline{7-9} 
 & & (0.5-2 keV) & (2-10 keV) & & & SMF0 &  SMF1 & SMF2 & (Years)  \\ 
(1) & (2) & (3)          & (4) & (5) & (6) & (7) & (8) & (9) & (10)  \\ \hline         
NGC\,7172 (X) & NHBLR &     42.50 & 42.82 & 8.20 & -1.98 & ME2PL & $Norm_2$ & - & 5  \\
              &  &  51$^+_-$2\% & 51$^+_-$1\% & & & & 51$^{+5}_{-5}$\% & \\
NGC\,7212 (X,C)* & HBLR & 41.81 & 42.60 & 7.54 & -1.55 & 2ME2PL & - & - & 1  \\              
               &  & 0\% & 0\% &   &  & \\
NGC\,7319 (C)$^{CL?}$ & - &   42.99 & 42.98 & 7.43 & -1.26 & ME2PL & $Norm_2$ & $N_{H1}$ & 7  \\
              &    & 38$^+_-$8\% & 38$^+_-$5\% & & & & 39$^{+53}_{-22}$\% & 100$^{+27}_{-23}$\% & \\
\hspace*{1.38cm}  (X,C) &  & 42.58 & 42.84 &  &  & ME2PL & $Norm_2$ & - & 6 \\
                        &  & 71$^+_-$8\%  & 69$^+_-$7\%  &  &  &       & 72$^{+64}_{-46}$\% & \\                                                                 
\hline
\end{tabular}
\caption*{{\bf Notes.} (Col. 1) Name (the asterisks represent
  \emph{\emph{Compton}}--thick or changing look candidates), and the
  instrument (C: \emph{Chandra} and/or X: \emph{XMM}--Newton) in
  parenthesis; (Col. 2) (non) hidden broad line region objects only in
  the cases where there are available observations; (Cols. 3 and 4)
  logarithm of the soft (0.5--2 keV) and hard (2--10 keV) X-ray
  luminosities, where the mean was calculated for variable objects,
  and percentages in flux variations; (Col. 5) black-hole mass on
  logarithmical scale, determined using the correlation between
  stellar velocity dispersion (from HyperLeda) and black-hole mass
  \citep{tremaine2002}, or obtained from the literature otherwise
  (MARK\,1210 and NGC\,4507 from \cite{nicastro2003}; IC\,4518A from
  \cite{alonsoherrero2013}; NGC\,6300 and NGC\,5643 from
  \cite{davis2014}; IC\,2560 from \cite{balokovic2014}; MARK\,268 from
  \cite{khorunzhev2012}; and MARK\,477 from \cite{singh2011});
  (Col. 6) Eddington ratio, $L_{bol}/!  L_{Edd}$, calculated from
  \cite{eracleous2010b} using $L_{bol}=33L_{2-10 keV}$; (Col. 7) best
  fit for SMF0; (Col. 8) parameter varying in SMF1, with the
  percentage of variation; (Col. 9) parameter varying in SMF2, with
  the percentage of variation; (Col. 10) and the sampling timescale,
  corresponding to the difference between the first and the last
  observation. The percentages correspond to this $\Delta$T$_{max}$.}
\end{center}
\end{table*} 
\normalsize

\begin{itemize}
\item \underline{MARK\,348}: SMF1 with variations in $Norm_2$ (69\%)
  represents the data best. These variations were found within a
  nine-year period, which implies intrinsic flux variations of 69\%
  (68\%) in the soft (hard) energy band. We classify it as a
  \emph{\emph{Compton}}-thin candidate.
\item \underline{NGC\,424}: Two \emph{XMM}--Newton data sets are
  available. SMF0 results in $\chi^2_r$=2.20, and SMF1 does not
  improve the fit; this is most probably because the spectra from 2008
  shows a more complex structure compared to 2000, preventing a proper
  simultaneous spectral fitting. Thus, we do not perform the
  simultaneous spectral fit between the two \emph{XMM}--Newton data
  sets. The contribution from the annular region is negligible, thus
  the spectral analysis can be jointly performed using
  \emph{XMM}--Newton and \emph{Chandra} data together. SMF0 is the
  best representation of the data. Short-term variations from the
  \emph{XMM}--Newton light curve are not found. We classify it as a
  \emph{\emph{Compton}}-thick candidate.
\item \underline{MARK\,573}: The \emph{Chandra} data do not show
  variations (SMF0 was used) within a four-year period. When compared
  with \emph{XMM}--Newton data, the annular region contributes with
  24\% to the \emph{Chandra} data. Again, SMF0 results in the best
  representation of the data. Three additional Gaussian lines are
  needed to fit the data at 1.20 keV (Ne X), 1.36 keV (Mg XI), and 2.4
  keV (S XIV). Two \emph{Chandra} light curves are analyzed, and
  variations are not detected. We classify it as a
  \emph{\emph{Compton}}-thick candidate.
\item \underline{NGC\,788}: One observation per instrument is
  available. The emission from the annular region is negligible so we
  jointly fit \emph{Chandra} and \emph{XMM}--Newton data. SMF0 was
  used, thus no variations are found in a two years period. We
  classify it as a \emph{\emph{Compton}}-thin candidate.
\item \underline{ESO\,417-G06}: SMF1 with $N_{H2}$ (21\%) because the
  parameter varying represents the data best. These variations were
  obtained within about a one-month period, corresponding to no flux
  intrinsic variations. We classify it as a \emph{\emph{Compton}}-thin
  candidate.
\item \underline{MARK\,1066}: Only one observation per instrument is
  available. The annular region contributes with 8\% to \emph{Chandra}
  data. The simultaneous fit without allowing any parameter to vary
  (i.e., SMF0) results in a good fit of the data. We classify it as a
  \emph{\emph{Compton}}-thick candidate.
\item \underline{3C\,98.0}: This is the only object where the
  unabsorbed PEXRAV component improves the fit. The values of the
  spectral parameters in this fit are $Norm_1=70.22^{81.21}_{58.82}
  \times 10^{-4} Photons \hspace*{0.1cm} keV^{-1} cm^{-2} s^{-1}$,
  $N_{H2}=9.68^{11.31}_{8.20} \times 10^{22} cm^{-2}$,
  $\Gamma=1.30^{1.54}_{1.07}$, $Norm_{pex}=0.10^{0.15}_{0.07} \times
  10^{-4} Photons \hspace*{0.1cm} keV^{-1} cm^{-2} s^{-1}$,
  $Norm_2=5.51^{8.72}_{3.55} \times 10^{-4} Photons \hspace*{0.1cm}
  keV^{-1} cm^{-2} s^{-1}$ (\emph{XMM}--Newton obsID. 0064600101),
  $3.03^{4.86}_{1.93} \times 10^{-4} Photons \hspace*{0.1cm} keV^{-1}
  cm^{-2} s^{-1}$ (\emph{XMM}--Newton obsID. 0064600301), and
  $\chi/d.o.f=109.30/126$. Thus, the best representation of the data
  requires $Norm_2$ to vary between the two \emph{XMM}--Newton data
  sets, while the reflection component remains constant. This spectral
  fit with $Norm_2$ varying agrees with the one using the MEPL model
  (Table \ref{bestfitSey}).  The percentages of the variations are
  compatible between the two SMF1 and also the luminosities. For
  simplicity, we report the results of the MEPL model in the
  following.  The simultaneous fit of the \emph{XMM}--Newton data
  needs SMF1 with $Norm_2$ (43\%) varing over a period of about half a
  year. This implies an intrinsic flux variation of 5\% (42\%) at soft
  (hard) energies. The annular region contributes with 8\% to the
  \emph{Chandra} data, and SMF0 was used when comparing \emph{Chandra}
  and \emph{XMM}--Newton data, i.e., variations were not found within
  a five-year period.  Short-term variations are not detected from the
  \emph{Chandra} data. UV data from the UVW1 filter did not show any
  variability. We classify it as a \emph{\emph{Compton}}-thin
  candidate.
\item \underline{MARK\,3}: The \emph{XMM}--Newton data need SMF1 with
  $Norm_2$ (37\%) as the parameter responsible for the
  variations. This corresponds to flux variations of 29\% (32\%) in
  the soft (hard) energy band in a one-year period. We classify it as
  a \emph{\emph{Compton}}-thick candidate.
\item \underline{MARK\,1210}: X-rays observations with \emph{Chandra}
  covering a period of about four years are simultaneously fitted,
  resulting in SMF2 with $N_{H2}$ (20\%) and $Norm_2$ (43\%) as the
  parameters varying in this model. This corresponds to intrinsic flux
  variations of 40\% (41\%) at soft (hard) energies.  We classify the
  object as a \emph{\emph{Compton}}-thin candidate.
\item \underline{NGC\,3079}: One observation per instrument is
  available. The annular region contributes with 79\% to
  \emph{Chandra} data. The estimated number counts in the nuclear
  component of the \emph{XMM}--Newton spectrum is 235 counts, so we do
  not perform a simultaneous fitting. This object will not be used to
  discuss long-term variations. We classify it as a
  \emph{\emph{Compton}}-thin candidate. We refer the reader to
  Appendix \ref{indivnotes} for the discussion of this source.
\item \underline{IC\,2560}: Only one observation per instrument is
  available. When comparing the data, the annular region contributes
  with 11\% to the \emph{Chandra} data. No variations were observed
  within two months, i.e., SMF0 was used for the simultaneous fit. An
  additional Gaussian line was needed in the fit at 1.85 keV (Si
  XIII).  A \emph{XMM}--Newton and a \emph{Chandra} light curve were
  analyzed. We notice that the \emph{XMM}--Newton light curve showed a
  positive value of $\sigma_{NXS}^2$ at 2.5$\sigma$ of confidence
  level, close to our limit (see Sect. \ref{short}).  We classify it
  as a \emph{\emph{Compton}}-thick candidate.
\item \underline{NGC\,3393}: \emph{Chandra} data are fitted with SMF0,
  resulting in no variations in a seven years period. When comparing
  with \emph{XMM}--Newton data, the annular region contributes with
  17\%, and SMF0 is needed to fit the data within a one-year
  period. Short-term variations are not found from one \emph{Chandra}
  light curve. We classify it as a \emph{\emph{Compton}}-thick
  candidate.
\item \underline{NGC\,4507}: SMF2 was used to fit the
  \emph{XMM}--Newton data, with $Norm_2$ (36\%) and $N_{H2}$ (21\%)
  varying in a nine-year period. This corresponds to a flux variation
  of 96\% (81\%) in the soft (hard) energy band. Two additional
  Gaussian lines at 1.36 (Mg XI) and 1.85 (Si XIII) keV are needed to
  fit the data. The annular region contributes with 13\% to the
  \emph{Chandra} data. When comparing \emph{Chandra} and
  \emph{XMM}--Newton data, the best fit resulted in SMF1 with $Norm_2$
  (53\%) varying over nine years. Short-term variations are found from
  neither \emph{Chandra} nor \emph{XMM}--Newton light curves. We
  classify it as a \emph{\emph{Compton}}-thin candidate.
\item \underline{NGC\,4698}: SMF0 was used in the simultaneous fit,
  resulting in no variations in a nine-year period. UV data in the
  UVM2 filter is available, where the object does not show changes. We
  classify it as a \emph{\emph{Compton}}-thin candidate.
\item \underline{NGC\,5194}: The simultaneous fit results in no
  variations (i.e., SMF0 was used) within an 11-year period. The
  annular region contributes with 91\% to the \emph{Chandra}
  data. When comparing data from \emph{XMM}--Newton and
  \emph{Chandra}, SMF0 results in the best representation of the
  data. Six \emph{Chandra} light curves were analyzed in three energy
  bands, but variations are not reported. UV data are available in
  three filters, one showing variations (UVW1) and the remaining two
  not (UVW2, UVM2). We classify it as a \emph{\emph{Compton}}-thick
  candidate.
\item \underline{MARK\,268}: The \emph{XMM}--Newton observations are
  separated by two days. SMF0 was used to fit the data. UV data are
  available in two filters (UVW1 and UVM2); none of them show variability. 
We classify it as a
  \emph{\emph{Compton}}-thin candidate.
\item \underline{MARK\,273}: Only one observation per instrument can
  be used for the variability analysis. The annular region contributes
  with 31\% to the \emph{Chandra} data. Variations in $N_{H2}$ (51\%)
  were needed in the SMF1. This corresponds to a luminosity variation
  of 24\% (32\%) in the soft (hard) energy band over a two-year
  period. UV data are available in two epochs, with no variations
  observed. The analysis of the \emph{Chandra} light curve results in
  no short-term variations.  \emph{\emph{Compton}}-thick and
  \emph{\emph{Compton}}-thin classifications were obtained for
  different observations, so we classify it as a changing-look
  candidate (see Table \ref{ew}).
\item \underline{Circinus}: \emph{Chandra} and \emph{XMM}--Newton data
  are available at different epochs. The \emph{Chandra} data analysis
  results in SMF0 (i.e., no variations) in a nine-year period, while
  the \emph{XMM}--Newton data set needs SMF2 with $Norm_1$ (34\%) and
  $Norm_2$ (31\%) varying within a 13-year period. However, the
  \emph{XMM}--Newton data did not show any flux variations. The
  spectra are quite complex, so two (at 1.85 (Si XIII) and 2.4 (S XIV)
  keV) and four (at 1.2 (Ne X), 1.36 (Mg XI), 1.85 (Si XIII), and 2.4
  (S XIV) keV) additional Gaussian lines are required for the
  \emph{XMM}--Newton and \emph{Chandra} fits, respectively. The
  annular region contributes with 28\% to the \emph{Chandra}
  data. However, the comparison between the data sets was not carried
  out owing to the complexity of the spectra. Short-term variations
  are not found from a \emph{Chandra} light curve. We classify it as a
  \emph{Compton}-thick candidate. We notice that the variations
  obtained from \emph{XMM}--Newton data will not be used for further
  discussion, because this variability seems to be caused by
  extranuclear sources (see \ref{circi} for details), and therefore
  this nucleus is considered as non-variable.
\item \underline{NGC\,5643}: The \emph{XMM}--Newton data were fitted
  with the SMF0; i.e., variations were not observed within a six-year
  period. We classify it as a \emph{\emph{Compton}}-thick candidate.
\item \underline{MARK\,477}: The two observations are separated by two
  days. SMF0 was used, so no variations are reported. At UV
  frequencies variations are not found. We classify the source as a
  \emph{\emph{Compton}}-thick candidate.
\item \underline{IC\,4518A}: The \emph{XMM}--Newton data need SMF1
  with $Norm_2$ (42\%) varying. The variations are found in an
  eight-day period, and correspond to a flux variation of 40\% (41\%)
  in the soft (hard) energy band. We classify it as a
  \emph{\emph{Compton}}-thin candidate.
\item \underline{ESO\,138-G01}: No variations are found (i.e., SMF0
  was used) within a five-year period. We classify it as a
  \emph{\emph{Compton}}-thick candidate.
\item \underline{NGC\,6300}: The \emph{Chandra} observations are
  separated by four days. SMF0 results in the best fit; i.e.,
  variations are not found. The annular region contributes with 5\% to
  the \emph{Chandra} data. When comparing \emph{Chandra} and
  \emph{XMM}--Newton data, SMF2 was used, with $Norm_1$ (98\%) and
  $Norm_2$ (98\%) varying over an eight-year period. We classify it as
  a \emph{\emph{Compton}}-thin candidate.

\begin{figure*}
\centering
\subfloat{\includegraphics[width=0.45\textwidth]{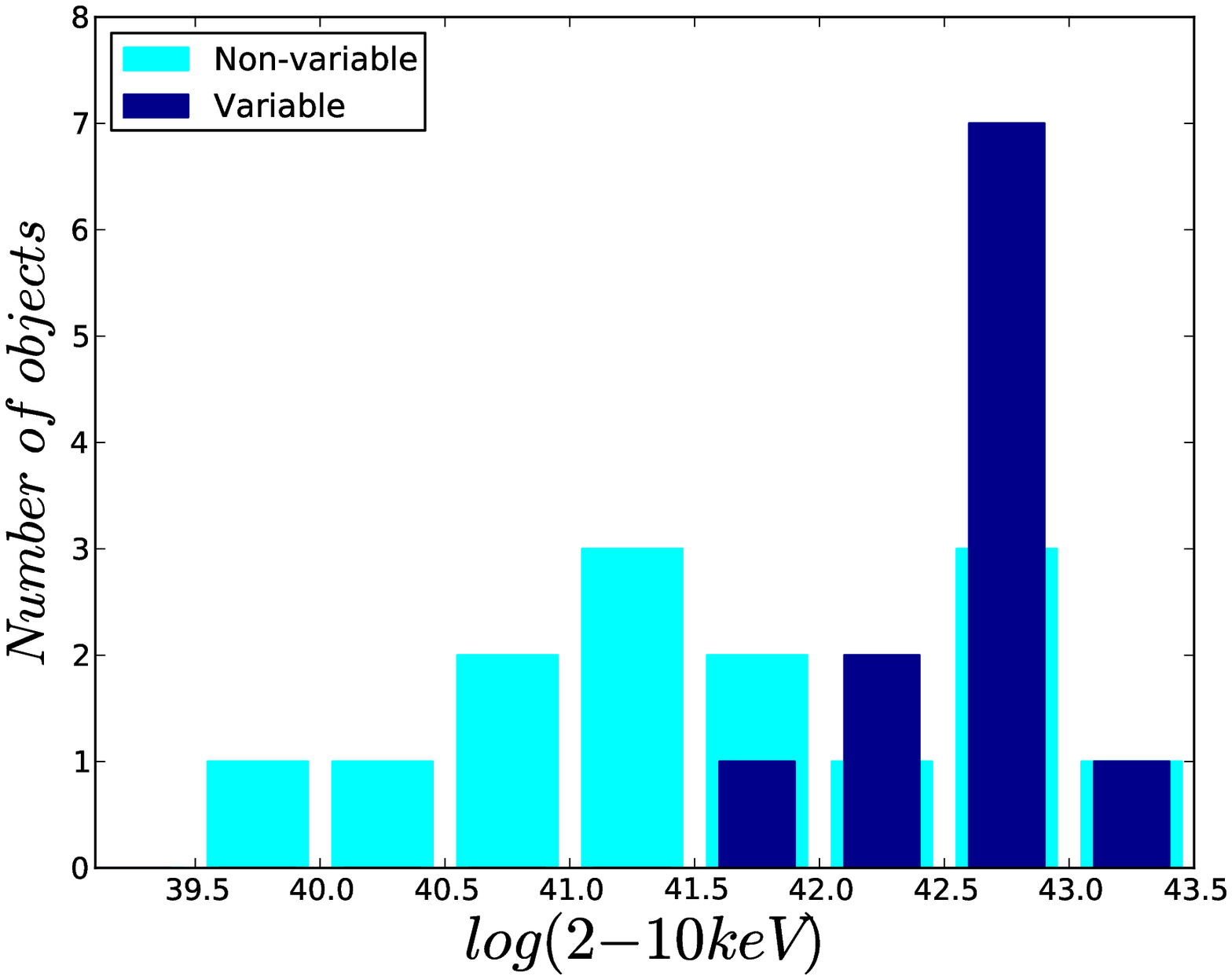}}
\subfloat{\includegraphics[width=0.45\textwidth]{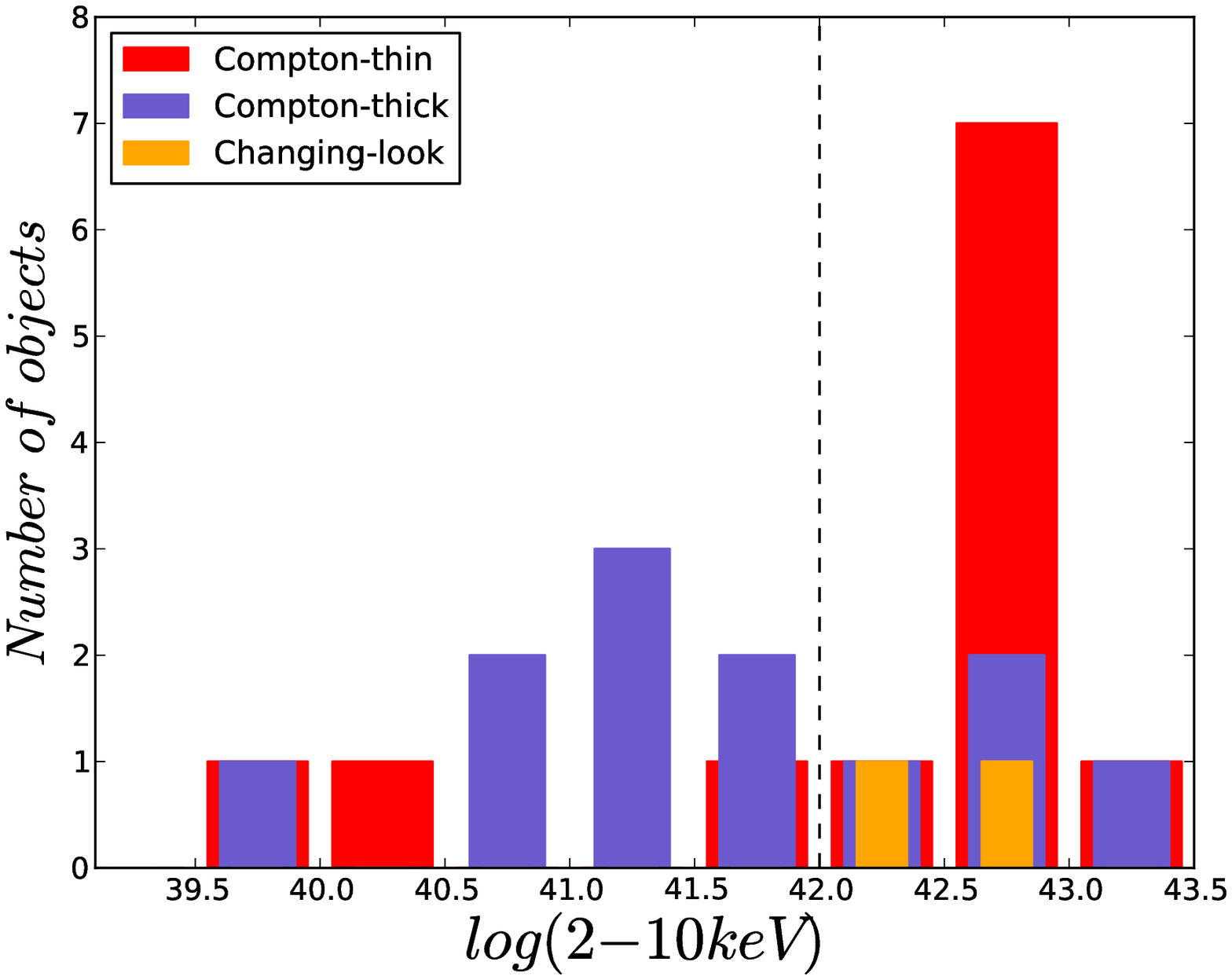}}
\caption{Histograms of: (Left): the luminosities for the variable
  (dark blue) and non-variable (light blue) galaxies in the sample;
  and (Right): the luminosities for the galaxies in the sample divided
  into \emph{\emph{Compton}}-thick (purple),
  \emph{\emph{Compton}}-thin (red), and changing-look (orange)
  candidates. The dashed line represents the value for the selection
  of faint (below) and bright (above) Seyfert 2s. }
\label{histolumin}
\end{figure*}

\item \underline{NGC\,7172}: SMF1 is the best representation of the
  \emph{XMM}--Newton data, with $Norm_2$ (54\%) varying over a
  three-year period. This implies an intrinsic flux variation of 54\%
  (53\%) at soft (hard) energies.  We classify it as a
  \emph{\emph{Compton}}-thin candidate.
\item \underline{NGC\,7212}: One observation per instrument is
  available. The annular region contributes with 16\% to the
  \emph{Chandra} data. When comparing both data sets, SMF0 is needed;
  i.e., variations are not found. We classify this source as a
  \emph{\emph{Compton}}-thick candidate.
\item \underline{NGC\,7319}: The best representation of the data used
  SMF2 with $N_{H1}$ (passed from $N_{H1}=6.5 \times 10^{21} cm^{-2}$
  to $N_{H1}=N_{Gal}$) and $Norm_2$ (39\%) varying in a seven-year
  period. Intrinsic flux variations of 38\% in both the soft and hard
  energy bands are obtained. The annular region contributes with 17\%
  to the \emph{Chandra} data. When comparing \emph{XMM}--Newton and
  \emph{Chandra} data, SMF1 with $Norm_2$ (54\%) varying is required,
  implying flux variations of 71\% (69\%) at soft (hard) energies over
  six years. Short-term variations were not detected. We classify it
  as a changing-look candidate because \emph{\emph{Compton}}-thick and
  \emph{\emph{Compton}}-thin classifications were obtained for
  different observations (see Table \ref{ew}).
\end{itemize}

\subsection{\label{shape}Spectral characteristics}

The sample of 26 optically classified Seyfert 2 galaxies
presented in this work show a variety of spectral shapes. None of them
are well-fitted with the ME or the PL models alone. Composite models
are required in all cases.

The models we used in previous works \citep[to represent the spectra
  of LINERs,][]{omaira2009a,lore2013,lore2014} describe the spectra of
12 galaxies well (MARK\,348, ESO\,417-G06, MARK\,1066, 3C\,98.0,
NGC\,3079, NGC\,4698, NGC\,5194, MARK\,268, ESO\,138-G01, NGC\,6300,
NGC\,7172, and NGC\,7319).  Three models are required (2PL, MEPL, and
ME2PL) for the spectral fits.  Among the 15 objects in our sample
observed in polarized light (see Table \ref{properties}), one galaxy
in this group has a HBLR and four a NHBLR.

On the other hand, 14 objects (NGC\,424, MARK\,573, NGC\,788, MARK\,3,
MARK\,1210, IC\,2560, NGC\,3393, NGC\,4507, MARK\,273, Circinus,
NGC\,5643, MARK\,477, IC\,4518A, and NGC\,7212) show a more complex
structure at energies below and around 2 keV, which cannot be fitted
with a single thermal component.  These nuclei need the 2ME2PL model
to fit the data. Besides, four of the objects need additional Gaussian
lines to properly fit the data.  Nine galaxies in this group have a
HBLR and one a NHBLR.

The addition of a cold reflection component to the best-fit model is
not statistically required by the data, except in obsID 0064600101
(\emph{XMM}--Newton) of 3C98.0. It is worth noting that even if a
model including this component is physically more meaningful, the lack
of data at harder energies prevents us from setting the best values
required by the model, and therefore a single power law is enough for
studying nuclear variations. On the other hand, we find that the cold
reflection component remains constant for 3C\,98.0 in SMF1. If this is
the general scenario (see Sect. \ref{disCT}), the lack of this
component in the models will not introduce biases into the variability
analysis.

A thermal component at soft energies is needed to fit the data in 24
out of the 26 sources; in 14 cases, two MEKAL are needed. It is worth
recalling that even if a MEKAL model fits the data well, because of its
spectral resolution,
photoionized models would be required to properly describe the data
(see Sect. \ref{indiv}). The values of the temperatures are in the
range $kT_1$ = [0.04--0.26] keV (only when the 2ME2PL model is fitted)
with a mean value of 0.12$^+_-$0.03 keV, and $kT_2$ = [0.13-1.00] keV
with a mean value of 0.60$^+_-$0.14 keV. The values of the spectral
index (which is the same at soft and hard energies, when two are
required) is in the range $\Gamma$ = [0.61--3.23], with a mean value
of 1.56$^+_-$0.40, and the absorbing column densities at hard energies
$N_{H2}$ = [5.15--152.21] $\times 10^{22} cm^{-2}$, with a mean value
of 34.69$^+_-$15.30 $\times 10^{22} cm^{-2}$.

\subsection{\label{spectral}Long-term X-ray spectral variability}

>From the 26 galaxies in our sample, we compared data at different
epochs from the same instrument in 19 cases.  Among these, seven
objects were observed with \emph{Chandra}, 13 with \emph{XMM}--Newton,
and in one case (namely Circinus) observations at different epochs
with both instruments were available.

\emph{Chandra} and \emph{XMM}--Newton data are available for the same
object in 15 cases (see Table \ref{obsSey}).  We did not compare these
data sets for NGC\,3079 because the number counts of the nuclear
contribution of \emph{XMM}--Newton spectrum (after decontaminating
from the annular region) is not enough for a reliable spectral fit.
Given that NGC\,3079 has one observation per instrument that cannot be
compared, this object will not be used to discuss long-term
variations.  Additionally, the \emph{Chandra} and \emph{XMM}--Newton
spectra of Circinus are very different, most probably because
extranuclear sources are included in the \emph{XMM}--Newton aperture
radius, thus preventing us from properly comparing both.  For the
remaining 13 objects, the simultaneous analysis was carried out (Table
\ref{simultanilloSey}), where the extranuclear emission were negligible in two cases
(NGC\,424 and NGC\,788).  Four of these sources showed spectral
variations.

In total, 25 (out of 26) nuclei have been analyzed to study long-term
X-ray spectral variations, with 11 of them (excluding
Circinus\footnote{We exclude the variations found with
  \emph{XMM}--Newton data because they are most probably due to
  extranuclear sources, while variations with \emph{Chandra} data are
  not reported.}) showing variability.  In Fig. \ref{histolumin}
(left) we present a histogram of the luminosities of the variable and
non-variable sources. A K-S test results in p=0.006, so we can reject
the hypothesis that the sample came from the same normal distribution.
The spectral changes are mainly due to variations in the nuclear power
(i.e., $Norm_2$), which is observed in nine nuclei (MARK\,348,
3C\,98.0, MARK\,3, MARK\,1210, NGC\,4507, IC\,4518A, NGC\,6300,
NGC\,7172, and NGC\,7319). Changes in the column density (i.e.,
$N_{H2}$) are also present in four cases (ESO\,417-G06, MARK\,273,
MARK\,1210, and NGC\,4507 -- in the last two accompained by changes in
$Norm_2$). Changes at soft energies are found in two objects:
NGC\,7319 ($N_{H1}$ together with $Norm_2$) and NGC\,6300 ($Norm_1$
together with $Norm_2$). This means that from the 11 sources showing
variations, most of them (nine out of 11) show variations in the
nuclear continuum (i.e., $Norm_2$), while variations due to
absorptions are less common (four in total, in two objects accompained
by variations in $Norm_2$).

\begin{figure*}
\centering
\subfloat{\includegraphics[width=0.34\textwidth]{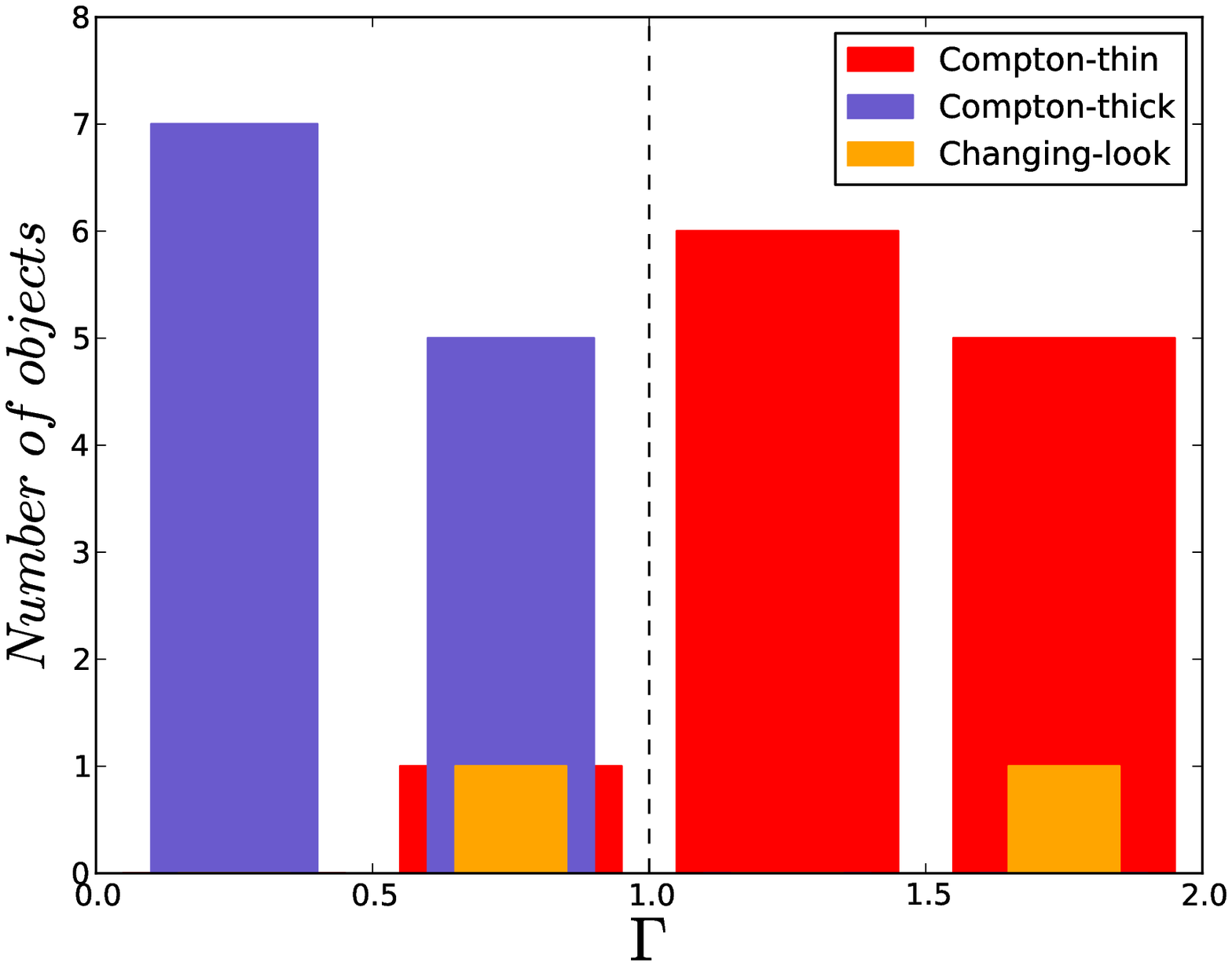}}
\subfloat{\includegraphics[width=0.34\textwidth]{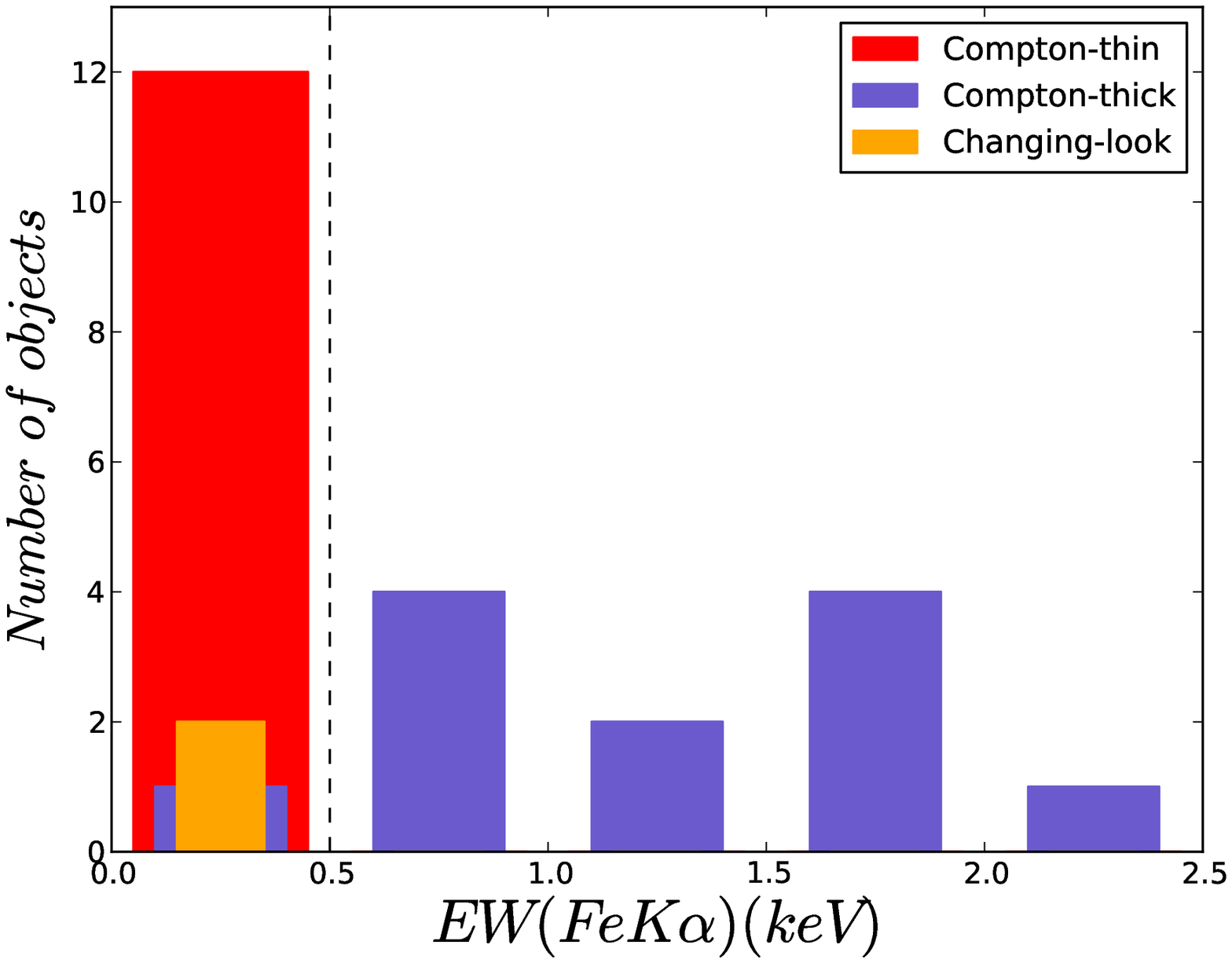}}
\subfloat{\includegraphics[width=0.34\textwidth]{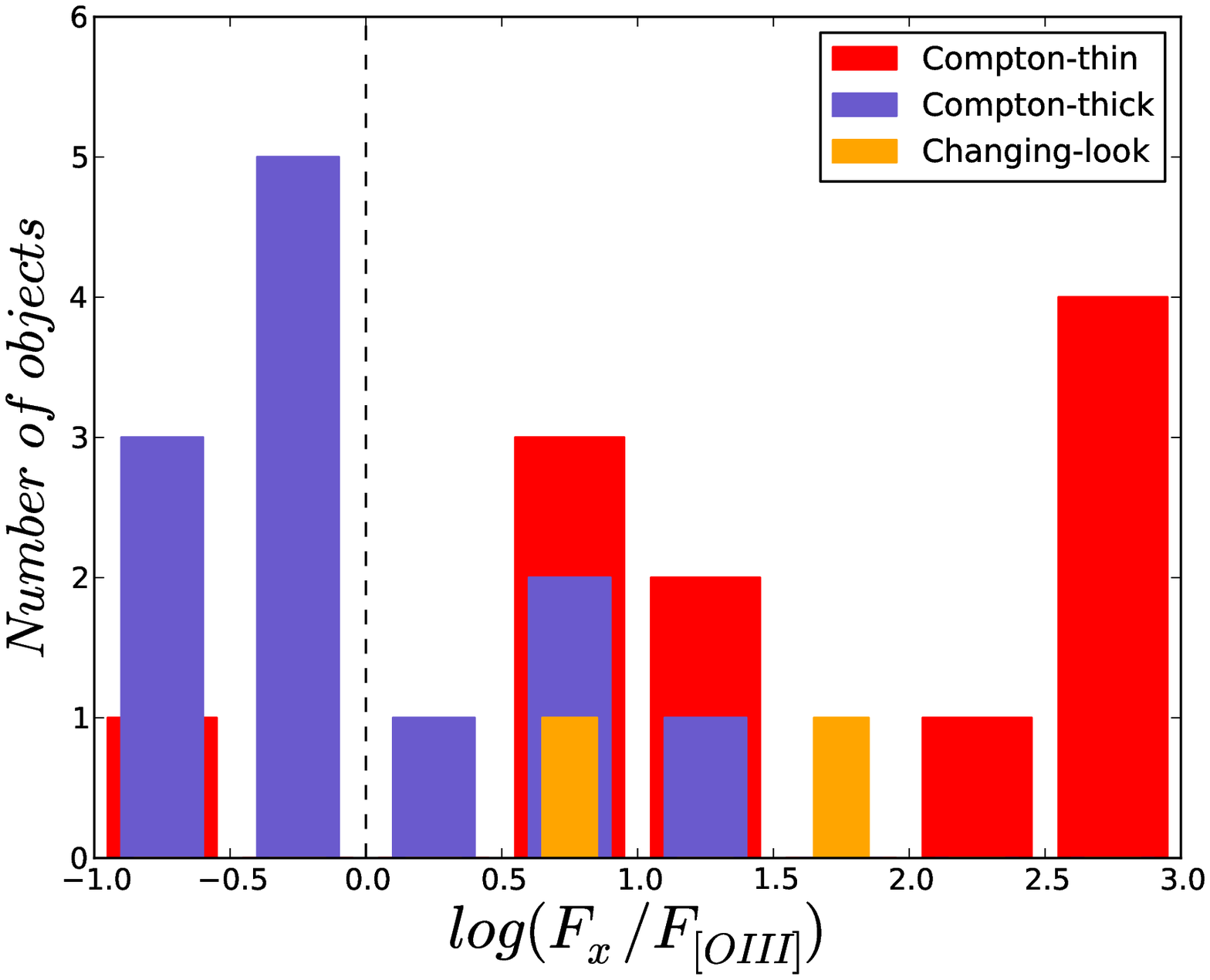}}
\caption{Histograms of (Left): the slope of the power law, $\Gamma$
  from Table \ref{ew}; (Middle): equivalent width of the iron line,
  EW(FeK$\alpha$); and (Right): the X-ray to [O III] flux ratios. In
  all cases the sample is divided into \emph{\emph{Compton}}-thick
  (purple), \emph{\emph{Compton}}-thin (red), and changing-look
  (orange) candidates. The dashed lines represents the values for the
  selection of \emph{\emph{Compton}}-thick (below) and
  \emph{\emph{Compton}}-thin (above) candidates. }
\label{histothick}
\end{figure*}

\subsubsection{HBLR vs. NHBLR}

>From the 15 objects in the sample with available observations in
polarized light (see Table \ref{properties}), ten are HBLR objects and
five NHBLR.  Nine out of the ten HBLR objects need the 2ME2PL model
for the spectral fits (except MARK\,348). The mean values of the
parameters in the simultaneous fits are reported in Table
\ref{resultparam}.  From the ten HBLR, four (MARK\,348, MARK\,3,
MARK\,1210, and NGC\,4507) show variations in $Norm_2$, in two sources
accompained by variations in $N_{H2}$.  One (NGC\,7172) out of the
four NHBLR sources shows variations in $Norm_2$.

Therefore, although the number of objects in this subsample is not
enough to be conclusive, it seems that there is no difference in
either the proportion of variable objects or in the pattern of the
variations.

\subsubsection{\emph{\emph{Compton}}-thick vs. \emph{\emph{Compton}}-thin}

We select \emph{\emph{Compton}}-thick candidates when at least two out
of the three indicators were met (see Sect. \ref{thick}). These
indicators are obtained from X-ray (EW(FeK$\alpha$) and $\Gamma$) and
the [O III] line ($F_x/F_{[O III]}$) data. In Fig. \ref{histothick} we
represent the histogram of these values for the whole sample, where
the mean was calculated when multiple observations were available
(from Table \ref{ew}). One \emph{\emph{Compton}}-thin candidate has
$\Gamma<1$ (NGC\,4698), one \emph{\emph{Compton}}-thick candidate has
EW(FeK$\alpha$)$<$0.5 keV (MARK\,477), one \emph{\emph{Compton}}-thin
candidate has log($F_x/F_{[O III]}$)$<$ 0 (NGC\,3079), and four
\emph{\emph{Compton}}-thick candidates have log($F_x/F_{[O III]}$)$>$
0 (NGC\,424, IC\,2560, ESO\,138-G01, and NGC\,7212; see discussion in
Sect. \ref{disCT}).

>From the 26 nuclei, 12 are classified as \emph{\emph{Compton}}-thick
candidates (NGC\,424, MARK\,573, MARK\,3, MARK\,1066, IC\,2560,
NGC\,3393, NGC\,5194, Circinus, NGC\,5643, MARK\,477, ESO\,138-G01,
and NGC\,7212), 12 as \emph{\emph{Compton}}-thin candidates
(MARK\,348, NGC\,788, ESO\,417-G06, 3C\,98.0, MARK\,1210, NGC\,3079,
NGC\,4507, NGC\,4698, MARK\,268, IC\,4518A, NGC\,6300, and NGC\,7172),
and two as changing-look candidates (MARK\,273, and NGC\,7319).  The
mean values of the spectral parameters in these subgroups are reported
in Table \ref{resultparam}, where \emph{\emph{Compton}}-thin
candidates are more luminous and less obscured and have steeper
spectral indices than \emph{\emph{Compton}}-thick candidates. The
spectral index of \emph{\emph{Compton}}-thick candidates was estimated
using $\Gamma_{soft} \neq \Gamma_{hard}$ (see details in
Sect. \ref{thick}) and the values are reported in Table \ref{ew}.

Only one (out of the 12) \emph{\emph{Compton}}-thick candidates shows
variations (MARK\,3), in $Norm_2$.  Eight (out of 11)
\emph{\emph{Compton}}-thin candidates show changes, with these
variations related mainly to $Norm_2$ (seven cases, in three sources
accompained by variations in $N_{H2}$ or $Norm_1$) and only in one
case to $N_{H2}$ alone.  The two changing-look candidates show X-ray
long-term variations, MARK\,273 varies $N_{H2}$, and NGC\,7319 needs
variations in $N_{H1}$ plus $Norm_2$.

Therefore, the number of variable \emph{\emph{Compton}}-thin and
changing look candidates is notably higher than that of
\emph{\emph{Compton}}-thick candidates.

\subsubsection{Bright vs. faint nuclei}

In Fig. \ref{histolumin} (right), we present the histogram of the
luminositites of the AGN in the sample as reported in Table
\ref{variab}, for \emph{\emph{Compton}}-thick (purple),
\emph{\emph{Compton}}-thin (red), and changing-look (orange)
candidates. A bimodal distribution can be appreciatted (K-S test,
p=0.030), with the difference around log(L(2--10 keV))$\sim$42. Based
on this histogram we separate the objects into faint (with log(L(2--10
keV))$<$42) and bright (log(L(2--10 keV))$>$42) Seyfert 2s.

>From these, 15 sources are bright, including four
\emph{\emph{Compton}}-thick (one variable, MARK\,3), two changing-look
(both variable, MARK\,273, and NGC\,7319), and nine
\emph{\emph{Compton}}-thin (seven variable, MARK\,348, ESO\,417-G06,
3C\,98.0, MARK\,1210, NGC\,4507, IC\,4518A, and NGC\,7172).  The
remaining 11 objects are faint Seyfert 2s, including three
\emph{\emph{Compton}}-thin (one shows variations, NGC\,6300) and eight
\emph{\emph{Compton}}-thick (none varies).

In total, 10 (out of 15) bright nuclei, and one (out of 10) faint
nuclei show variations. Therefore, brighter sources include more
variable sources and less \emph{\emph{Compton}}-thick candidates, a
trend that can be derived by comparing left- and righthand panels in
Fig. \ref{histolumin}. Moreover, we note that NGC\,6300 (i.e., the
only faint source that varies) has log(L(2--10 keV))=41.95, very close
to the established luminosity limit.  The mean values of the spectral
parameters of these subgroups are reported in Table \ref{resultparam},
where faint objects show a steeper power law index than bright
objects.

\begin{table}
\caption{\label{resultparam} Mean values of the spectral parameters for the subgroups.}
\begin{tabular}{l|cccc} \hline
\hline
Group & $\Gamma$ & $N_{H2}$ & log(L(2--10 keV)) & \\ \hline
All & 1.56$^+_-$0.40 & 34.69$^+_-$15.30 & 42.56$^+_-$0.89 \\ 
HBLR & 1.34$^+_-$0.43 & 39.22$^+_-$18.62 & 42.72$^+_-$0.80 \\
NHBLR & 1.58$^+_-$0.48 & 40.17$^+_-$20.23 & 41.40$^+_-$1.04 \\ 
\emph{\emph{Compton}}-thick & 0.57$^+_-$0.29$^1$  & 43.95$^+_-$19.53 & 42.33$^+_-$1.01 \\
\emph{\emph{Compton}}-thin & 1.43$^+_-$0.32 & 20.31$^+_-$14.39 & 42.73$^+_-$1.12 \\
Changing-look & 1.68$^+_-$0.49 & 45.99$^+_-$1.24 & 42.76$^+_-$0.49 \\ 
Bright & 1.44$^+_-$0.40 & 32.11$^+_-$20.12 & 42.78$^+_-$0.29 \\
Faint & 1.69$^+_-$0.61 & 34.53$^+_-$21.20 & 41.38$^+_-$0.82 \\
\hline
\end{tabular}
\caption*{ (Col. 1) Group, (Col. 2) values of $\Gamma$, (Col. 3)
  column density in units of $10^{22} cm^{-2}$, and (Col. 4) intrinsic
  luminosity in the 2--10 keV energy band. }

\vspace*{-0.4cm}

$^1$ This value is calculated from the simultaneous values reported in Table \ref{ew}. 
\end{table}

\subsection{\label{lightcurve}Short-term X-ray variability}

Observations with a net exposure time $>$ 30 ksec are used to study
short-term variations. This requirement leaves us with ten sources for
the analysis (see Table \ref{estcurvasSey}). Three of them (IC\,2560,
NGC\,5194, and MARK\,573) show positive values of $\sigma^2_{NXS}$,
but below 3$\sigma$ of confidence level in all cases. Therefore we
cannot claim short-term variations in any of the objects in our
sample. Upper limits of $\sigma^2_{NXS}$ have been estimated for all
the other cases.

\subsection{\label{flux}Long-term UV flux variability}

\emph{XMM}--Newton data at different epochs were used to study
long-term X-ray spectral variations in 13 sources. In nine of them
data from the OM cannot be used because the source is outside the
detector or because the same filter is not available at different
epochs. In contrast, two objects (MARK\,273 and NGC\,5194) have OM
data while the sources were out of the pn detector, so these data were
also used to search for variations at UV frequencies.  Thus, UV data
for variability studies are available for six galaxies (3C\,98.0,
NGC\,4698, NGC\,5194, MARK\,268, MARK\,273, and MARK\,477). Only
NGC\,5194 shows variations above 3$\sigma$ of the confidence level in
one filter (UVW1).

We also searched in the literature for UV variations for the sources
in the sample, but this information was available only for MARK\,477
(see Appendix \ref{indivnotes}). Comparing the analyses at X-rays and
UV, two out of the six sources do vary at X-rays but not at UV
frequencies (3C98.0 and MARK\,273), and one (NGC\,5194) does not show
variations in X-rays but it does at UV. The remaining three objects do
not vary neither in X-rays nor at UV frequencies.

%
%______________________________________________________________

\section{\label{discusion}Discussion}

\subsection{X-ray spectral variability}

A long-term X-ray variability analysis was performed for 25 out of the
26 nuclei in our sample of Seyfert 2 galaxies\footnote{We recall
  that NGC\,3079 will not be used for the discussion of variability,
  see Sect. \ref{ind}.}.  From these, 11 sources are variable at
X-rays. Among the remaining 14 nuclei where variations are not
detected, 11 are \emph{\emph{Compton}}-thick candidates, and therefore
variations are not expected \citep[e.g.,][and references
  therein]{matt2013}. This agrees well with our results, where only
one out of the 12 \emph{\emph{Compton}}-thick candidates shows
variations. We refer the reader to Sect. \ref{disCT} for a complete
discussion about \emph{\emph{Compton}}-thick candidates.  The other
three nuclei where variations are not detected are
\emph{\emph{Compton}}-thin candidates (NGC\,788, NGC\,4698, and
MARK\,268). The lack of variations may be due to the short timescale
between observations for MARK\,268 (two days).  The timescales between
observations for the other two sources are on the order of years, so,
in principle, variations could be detected.  New data would therefore
be required before confirming the non-variable nature of these
sources.

In this section the discussion is focused on the different patterns of
variability obtained for the 11 variable nuclei, including eight
\emph{\emph{Compton}}-thin, two changing-look, and one
\emph{\emph{Compton}}-thick candidates. We notice that this is the
first time that transitions from a \emph{\emph{Compton}}-thin to a
\emph{\emph{Compton}}-thick (or vice versa) appearance have been
reported for MARK\,273 and NGC\,7319, which should be added to the
short list of known changing-look Seyfert 2s, such as NGC\,2992
\citep{gilli2000}, MARK\,1210 \citep{guainazzi2002}, NGC\,6300
\citep{guainazzi2002a}, NGC\,7674 \citep{bianchi2005a}, and NGC\,7582
\citep{bianchi2009}.

\subsubsection{\label{soft}Variations at soft energies}

We found that most of the objects in our sample do not vary at soft
X-ray energies, indicating that the mechanism responsible for the soft
emission should be located far from the nucleus.  Indeed, using
artificial neural networks, \cite{omaira2014} compared the spectra of
different classes of AGN and starburst galaxies and find that 
Seyferts 2 have a high contribution from processes that are related star
formation, which may be related to emission coming from the host
galaxy.

Notwithstanding, two sources show variations at soft energies ($<$2
keV), each showing a different variability pattern, but in both cases
these variations are accompanied by variations in the normalization of
the hard power law; NGC\,6300 shows variations in the normalization at
soft energies, $Norm_1$, when comparing data from \emph{XMM}--Newton
and \emph{Chandra}; and NGC\,7319 showed variations in the absorber at
soft energies, $N_{H1}$, when comparing two \emph{Chandra}
observations.  It is worth noting that the soft X-ray fluxes are on
the order of $10^{-13} erg \hspace*{0.1cm} cm^{-2} s^{-1}$ in the two
nuclei, which is typical of Seyfert galaxies \citep{guainazzi2005a},
so these variations are not related to low-count number statistics.
However, variations at soft energies in these sources have not been
reported before.  Up to now, such variations have only been found for
two Seyfert 2s.  \cite{paggi2012} found variations at soft X-rays
in the Seyfert 2 MARK\,573 when comparing four \emph{Chandra}
observations. This nucleus is also included in the present sample, but
variations are not found here, mainly because we did not use two of
the observations included in the work of \cite{paggi2012} since they
were affected by a pileup fraction higher than 10\%.
\cite{guainazzi2012} speculate that variations at soft X-ray energies
in MARK\,3 may be present when comparing \emph{XMM}--Newton and
\emph{Swift} data, but confirmation is still required. They argue that
these variations are most probably due to cross-calibration
uncertainties between the instruments, but if true, soft X-ray
variations could be related to the innermost part of the narrow-line
region.

On the other hand, the variability patterns found in this work have
also been reported for other types of AGN. Variations in the
absorbers, as seen in NGC\,7319, were found by \cite{omaira2011b}, who
used \emph{Suzaku} data to study the  LINER 2 NGC\,4102. They
argue that the variations at soft energies are due to an absorbing
material located within the torus and perpendicular to the plane of
the disk. Variability timescales can be used to estimate the lower
limits of the cloud velocity \citep[e.g.,][]{risaliti2007}. However,
the timescales between our observations were obtained randomly, so the
variability timescale of the eclipse can be shorter.  In the case of
NGC\,7319, variations are obtained within a timescale of seven years,
which is too long to estimate the distance at which the cloud is
located. It is worth noting that we classified this object as a
changing-look candidate.  Besides, we found that NGC\,6300 varied the
normalizations at soft and hard energies. Using the same method as
explained in this work, \cite{lore2013} find the same variability
pattern in the LINER 2 NGC\,4552, indicating that these
variations may be intrinsic to the emitting material.

\subsubsection{\label{nh2}Absorber variations} 

Variations in the circumnuclear absorbers are thought to be very
common in Seyfert galaxies.  In fact, these variations are usually
observed in Seyferts 1-1.9 (e.g., NGC\,1365,
\citealt{risaliti2007}; NGC\,4151, \citealt{puccetti2007}; MARK\,766,
\citealt{risaliti2011}), where it has been shown that the changes are
most probably related to the broad line region (BLR), although it has
been suggested that multiple absorbers may be present in an AGN,
located at different scales \citep{braito2013}.  However, it is not so
clear whether variations due to absorbers are common for optically
classified Seyfert 2s, for which this kind of variation has only
been reported in a few cases (e.g., MARK\,348, \citealt{marchese2014};
NGC\,4507, \citealt{braito2013} and \citealt{marinucci2013};
MARK\,1210, \citealt{risaliti2010}).

>From the 11 variable sources in our sample, variations due to
absorbers at hard energies are detected in four nuclei. In two of
them, MARK\,1210 and NGC\,4507, variations in $N_{H2}$ are accompained
by variations in the nuclear continuum, $Norm_2$.  The variability
pattern reported for these objects agrees with previous results
presented by \cite{risaliti2010} and \cite{braito2013}, who argue that
the physical properties of the absorber are consistent with these
variations occurring in the BLR.  Following prescriptions in
\cite{risaliti2010} and using the BH masses (Table \ref{variab}) and
variability timescales of one and ten days for MARK\,1210 and
NGC\,4507, respectively, we estimate the cloud velocities to be higher
than $10^3 km \hspace*{0.1cm} s^{-1}$ in both cases, thus also
locating the absorbers at the BLR.

On the other hand, ESO\,417-G06 and MARK\,273 showed variations only
in $N_{H2}$. \cite{trippe2011} report variations of a factor about two
in the count rate of ESO\,417-G06 from the 22-month survey of
\emph{Swift}, and \cite{balestra2005} fit the \emph{XMM}--Newton and
\emph{Chandra} spectra of MARK\,273 studied in this work and note that
different column densities were required to fit the data well (its
values in good agreement with ours), indicating variations due to
absorption.  The timescale between observations for ESO\,417-G06 is 40
days and two years for MARK\,273. Therefore, we cannot estimate the
cloud velocity for MARK\,273 because the timescale is too
large. Assuming the variability timescale of ESO\,417-G06 (40 days)
and following prescriptions in \cite{risaliti2010}, we estimate a
cloud velocity $>$ 60 $km \hspace*{0.1cm} s^{-1}$, so too low to
restrict the location of the cloud.  Since this estimate is a lower
limit of the cloud velocity, a monitoring campaign of these sources
would be needed to constrain their variability timescales, in order to
properly constrain the locus of the absorbers.

\subsubsection{\label{norm2}Flux variations}

The most frequently varying parameter in our sample is $Norm_2$, which
is related to the nuclear continuum. These kinds of variations are
observed in nine out of the 11 X-ray variable sources -- sometimes
accompanied by variations in other parameters (see Sects. \ref{soft}
and \ref{nh2}). Therefore the most natural explanation for the
observed variations in Seyfert 2 galaxies is that the nuclear
power is changing with time. We recall that variations are not due to
changes in the power law index, $\Gamma$, but related to its
normalization.  It has been shown that hard X-ray variability is usual
in Seyfert 2 galaxies \citep[e.g.,][]{turner1997,
  trippe2011,marchese2014}. In fact, this kind of variation has
already been reported in the literature for objects included in the
present work from intrinsic flux variations indicating changes in the
nuclear continuum \citep{isobe2005} or because they needed to set free
the normalization of the power law for a proper fit to the data
\citep{lamassa2011}.  Also at higher energies, \cite{soldi2013}
studied the long-term variability of 110 AGN selected from the BAT
58-month survey and argue in favor of a variable nuclear continuum
plus a constant reflection component. Their result is independent of
the classification of the objects, which includes Seyferts, NLSy1s,
radio galaxies, and quasars.

Flux variations are indeed a property of AGN, and they have been
reported at different frequencies for Seyfert 2s, such as in
radio \citep{nagar2002,neff1983} or infrared
\citep{sharples1984,honig2012}.  In the present work we used data from
the OM onboard \emph{XMM}--Newton to study UV variability. These data
are available at different epochs for six objects in our sample, but
only NGC\,5194 shows variations in the UVW1 filter. This is a
\emph{\emph{Compton}}-thick candidate that does not vary in X-rays, so
variations at UV frequencies from the nuclear component are not
expected.  It has been shown that the UV/optical spectra of 
Seyferts 2 include scattered AGN light, and it can sometimes be produced
by young starbursts, including supernovae explosions
\cite[e.g.,][]{gonzalezdelgado2004}. In fact, supernovae explosions in
NGC\,5194 have been reported in 1945, 1994, 2005, and 2011
\citep{vandyk2011}, which could account for the observed variations in
the UV.

 None of the remaining five nuclei show variations at UV frequencies,
 although there are two nuclei that are variable in X-rays (3C\,98.0
 and MARK\,273). The lack of UV variations could be explained because
 X-ray and UV variations might not happen simultaneously
 \citep[e.g.,][]{lore2014} or because we are not directly observing
 the nucleus. \cite{munozmarin2009} studied 15 Seyfert galaxies with
 \emph{HST} data (including types 1 and 2) and found that most type 2
 nuclei appear resolved or absent at UV frequencies, concluding that
 the UV emission in Seyfert 2s does not come from the nucleus. Thus,
 the lack of UV variations in Seyfert 2s is most probably because
 we are not directly observing the nucleus at UV.

\subsection{\label{disCT}\emph{Compton}-thickness}

\cite{brightman2011} show that at column densities $\sim 4 \times
10^{24} cm^{-2}$, the observed flux below 10 keV is half that of the
intrinsic flux at harder energies (see also \citealt{ghisellini1994}).
This indicates that in \emph{\emph{Compton}}-thick objects, the
primary continuum is so absorbed in the 2-10 keV energy band that the
emission is optically thick to Compton scattering, and the spectrum is
reflection-dominated. For this reason, we have distinguished between
\emph{\emph{Compton}}-thin and \emph{\emph{Compton}}-thick candidates
(see Sects. \ref{thick} and \ref{spectral}).

However, the task of classifying \emph{\emph{Compton}}-thick objects
with X-ray data comprising energies up to $\sim$ 10 keV is hard because the peak
of the primary emission is above 10 keV. Instead, three different
indicators involving X-ray and [O III] emission line data are used for
their selection (see Sect. \ref{thick}, for details). While the three
criteria are met in most cases, our results have shown that the X-ray
to [O III] line flux ratio, log($F_x/F_{[O III]}$) is the most
unsuitable indicator (see Fig. \ref{histothick}). This agrees with
\cite{brightman2011a}, who argue that this parameter can be inaccurate
for classifying \emph{\emph{Compton}}-thick sources because of the
uncertainty in the reddening correction of the [O III] line
flux. Moreover, in Fig. \ref{histothick} (right) there are four
objects with log($F_x/F_{[O III]})>2.5$, which is higher than the
values found by other authors
\citep{bassani1999,cappi2006,panessa2006}, what may be due to a
underestimation of the [O III] line flux. Although the [O III] line
is a good luminosity indicator, the reddening correction might depend
on the geometry of the narrow line region, leading to an
underestimation of its flux if we do not take it into account and
leading to very high values of $F_x/F_{O III}$.

In the present work, 12 nuclei are classified as
\emph{\emph{Compton}}-thick candidates.  Among them, variations are
found only in MARK\,3, which was previously classified as a
\emph{\emph{Compton}}-thick candidate
\citep{bassani1999,goulding2012}, with a column density of $1.1 \times
10^{24} cm^{-2}$ measured by \emph{BeppoSAX} \citep{cappi1999}.  In
fact, variations in MARK\,3 have already been reported by
\cite{guainazzi2012}, who studied its variability using
\emph{XMM}--Newton, \emph{Suzaku}, and \emph{Swift} data, and found
variations on timescales of months.  We found that the changes in
MARK\,3 are related to $Norm_2$, i.e., intrinsic to the source. The
most likely explanation for these variations could therefore be that
part of the emission is still transmitted below 10 keV, so variations
can be observed.

Interestingly, we found that most of the \emph{\emph{Compton}}-thick
candidates are non-variable and tend to be fainter than
\emph{\emph{Compton}}-thin and changing-look candidates, which show
X-ray variations (see Fig. \ref{histolumin}). This can be explained
because the intrinsic luminosity is underestimated if the primary
continuum is suppressed at energies below 10 keV, in agreement with
the results of \cite{brightman2011}.  In fact, the only
\emph{\emph{Compton}}-thick candidate that shows variations in X-rays
is included as a bright Seyfert 2.  It could be that variations
are not observed because the spectra of \emph{\emph{Compton}}-thick
sources are dominated by the reflection component. If so, this
component might be located farther away from the central source, so it
remains constant. This scenario agrees with the results we have
obtained for the only source where a reflection component was
statistically required by the data (namely 3C\,98.0).  These results
are also in good agreement with those found by other authors, who did
not find X-ray variability for objects classified as
\emph{\emph{Compton}}-thick (e.g., NGC\,424 and NGC\,5194,
\citealt{lamassa2011}; Circinus, \citealt{arevalo2014b}; NGC\,5643,
\citealt{matt2013}).

As noted above, if the reflection component does not vary, it might
indicate that the reflection of the primary continuum occurs at large
distances from the SMBH. The same result was obtained by
\cite{risaliti2002}, who studied Seyfert 2s with \emph{BeppoSAX}
and found that the cold reflection component is compatible with being
non-variable. They argue that if the reflection originates in the
accretion disk, the reflection and the transmitted components must be
closely related, but if the distance of the reflector to the SMBH is
greater than the light crossing time of the intrinsic variations, the
reflected component must remain constant. Therefore a reflector
located far away from the SMBH is supported by our results, maybe in
the torus or in the host galaxy.

\subsection{Caveats and limitations of the analysis}

The models used in this work to characterize the spectra of 
Seyfert 2 galaxies are a simplification of the true physical scenario
occurring in these nuclei. In particular, the 2--10 keV energy band --
where variations are mostly found -- is represented by an absorbed
power law continuum, which could be an oversimplification of the real
scenario.

Spectral variability analyses of seven sources studied in this work
have been reported previously. Since at least some of these works
study individual sources, the models used in their analyses might be
more complex than ours (see Appendix \ref{indivnotes}, for
details). This comparison shows that our results are almost always
compatible with those reported in the literature (MARK\,1210,
\citealt{matt2009} and \citealt{risaliti2010}; NGC\,4507,
\citealt{matt2004}, \citealt{marinucci2013}, and \citealt{braito2013};
MARK\,273, \citealt{balestra2005}; Circinus, \citealt{arevalo2014b};
NGC\,6300, \citealt{guainazzi2002a}; and NGC\,7172,
\citealt{lamassa2011}). However, we cannot discard variations due to
components that we did not fit in the models. For instance,
\cite{marchese2014} analyzed the \emph{XMM}--Newton and \emph{Suzaku}
data of MARK\,348 (also included in the present work), and report
variations due to a neutral plus an ionized absorbers, together with a
change in the ionization parameter of the ionized absorber. Their
analysis is based on the residuals of the spectral fitting, where they
include as many components as required, and the variability analysis
is performed by testing different scenarios, including a variable
continuum plus a constant reflection component
($\chi^2/d.o.f$=567.7/407), a variable continuum plus a variable
reflection component ($\chi^2/d.o.f$=551.1/406, but variations are not
observed), variations due to absorptions, and changes in the
ionization state ($\chi^2/d.o.f$=551.6/407).  We notice that our
spectral fit of MARK\,348 with $Norm_2$ varies results in a very good
fit ($\chi^2/d.o.f$=1520.5/1368) when comparing the two
\emph{XMM}--Newton data sets, and residuals are mostly at energies
below $\sim$ 2.5 keV (see Fig. \ref{bestfitSeyim}).  Therefore, the
presence of complex variations like these in at least some sources in
our sample cannot be completely discarded.

\section{\label{conclusion}Conclusions}

Using \emph{Chandra} and \emph{XMM}--Newton public archives we
performed a spectral, flux, short-, and long-term variability
analysis of 26 optically selected Seyfert 2 galaxies. The
main results of this study can be summarized as follows:

\begin{enumerate}
\item Long-term variability was found in 11 out of the 25 analyzed
  nuclei, which are more frequent among the brightest sources
  (log(L(2-10 keV)) $>10^{42} erg \hspace*{0.1cm} s^{-1}$). From the
  11 variable sources, eight are \emph{\emph{Compton}}-thin
  candidates, two are changing-look, and only one (namely MARK\,3) is
  a \emph{\emph{Compton}}-thick candidate. No difference in the
  variability is found among the HBLR and NHBLR objects. We report two
  changing-look candidates for the first time: MARK\,273 and
  NGC\,7319.
\item Short-term variability has not been detected in any of the sources. Nor UV variability.
\item The main driver of the observed variations is due to the power
  of the central engine manifested through variations in the
  normalization of the power\ law at high energies. At soft energies
  variations are rare, and column density variations have only been
  observed in four cases.
\end{enumerate}

Our results are compatible with a scenario where a constant reflection
component located far away from the nucleus and a variable nuclear
continuum take place. Within this scenario,
\emph{\emph{Compton}}-thick objects are dominated by reflection and do
not show any X-ray spectral or flux variations. This implies that
their luminosities are suppressed at hard X-rays, making them fainter
sources than \emph{\emph{Compton}}-thin objects. In contrast, most of
the \emph{\emph{Compton}}-thin or changing-look candidates are
variable, showing different patterns of variability. These changes are
mainly due to variations in the nuclear continuum.  However,
variations of the absorber or at soft energies are also found in some
cases, with many of them accompanied by variations of the nuclear
continuum. These variations are mainly due to clouds intersecting our
line of sight.

\begin{acknowledgements}

We acknowledge the referee, M. Guainazzi, for his comments and
suggestions that helped to improve the paper, and the AGN group at the
IAA for helpful comments during this work. This work was financed by
MINECO grant AYA 2010-15169, AYA 2013-42227-P, Junta de Andaluc\'{i}a
TIC114 and Proyecto de Excelencia de la Junta de Andaluc\'{i}a
P08-TIC-03531. LHG acknowledges financial support from the Ministerio
de Econom\'{i}a y Competitividad through the Spanish grant FPI
BES-2011-043319. This research made use of data obtained from the
\emph{Chandra} Data Archive provided by the \emph{Chandra} X-ray
Center (CXC). This research made use of data obtained from the
\emph{XMM}-Newton Data Archive provided by the \emph{XMM}-Newton
Science Archive (XSA). This research made use of the NASA/IPAC
extragalactic database (NED), which is operated by the Jet Propulsion
Laboratory under contract with the National Aeronautics and Space
Administration. We acknowledge the usage of the HyperLeda data base
(http://leda.univ-lyon1.fr).

\end{acknowledgements}

\bibliographystyle{aa}
\bibliography{seyfert}

\newpage

\appendix

\onecolumn

\section{\label{tables}Tables}

\tiny
\renewcommand{\arraystretch}{1.4}
\begin{longtable}{lcccccccc}
\caption[Observational details]{\label{obsSey} Observational details.} \\   \hline \hline
 Name & Instrument & ObsID  & Date & R  & Net Exptime & Counts &   log($L_{UV}$) & Filter \\
 &     &           &                    & ($\arcsec$)  & (ksec) &  & (erg/s) &  \\
 (1) & (2) & (3) & (4) & (5) & (6) & (7) & (8) & (9)  \\
\hline 
\endfirsthead
\caption{(Cont.)} \\
\hline \hline
 Name & Instrument & ObsID  & Date & R  & Net Exptime & Counts &   log($L_{UV}$) & Filter \\
 &     &           &                    & ($\arcsec$)  & (ksec) &  & (erg/s) &  \\
 (1) & (2) & (3) & (4) & (5) & (6) & (7) & (8) & (9)  \\
 \hline  
\endhead
%\multicolumn{11}{r}{continued on next page\ldots}\\
\endfoot
\\
\endlastfoot
MARK\,348 & \emph{XMM}--Newton & 0067540201  &  2002-07-18 & 25  &  18.5 &  39552   & - \\
 & \emph{XMM}--Newton & 0701180101  &  2013-01-04 &  25 &  7.2 &  5681   & - \\ \hline
NGC\,424 & \emph{XMM}--Newton & 0002942301$^c$  &  2001-12-10 &  20  &  4.5 & 1777 &  - \\
 & \emph{XMM}--Newton & 0550950101  &  2008-12-07 &  20  & 127.5  & 33452   & - \\
 & \emph{Chandra} & 3146$^c$ & 2002-02-04 &  2  &    9.2 & 1266 &  - \\ \hline
MARK\,573 &  \emph{Chandra} &  7745$^c$ &  2006-11-18 & 2  &  38.1 & 3181 &  - \\
 &  \emph{Chandra} & 13124 &  2010-09-17 & 2  &  52.4 &  3456 &  -  \\   
 & \emph{XMM}--Newton & 0200430701$^c$ &  2004-01-15 & 20 &  9.0 &  3605   & 42.75$^+_-$0.01 & UVW1 \\
\end{longtable}

\tiny
\renewcommand{\arraystretch}{1.4}
\begin{longtable}{lccccccccc}
\caption[Spectral fittings]{\label{bestfitSey}  Final compilation of the best-fit models for the sample, including the individual best-fit model for each observation, and the simultaneous best-fit model with the varying parameters. }  \\  \hline \hline
Analysis & ObsID & Model & $N_{H1}$ & $N_{H2}$ & kT & $\Gamma$ & $Norm_1$ & $Norm_2$ & $\chi^2/d.o.f$  \\
  & & &  &  & keV & & ($10^{-4}$) & ($10^{-4}$) & F-test \\
(1) & (2) & (3) & (4) & (5) & (6) & (7) & (8) & (9) & (10)  \\
\hline 
\endfirsthead
\caption[]{(Cont.)} \\
\hline \hline
Analysis & ObsID & Model & $N_{H1}$ & $N_{H2}$ & kT & $\Gamma$ & $Norm_1$ & $Norm_2$ & $\chi^2/d.o.f$  \\
  & & &  &  & keV & & ($10^{-4}$) & ($10^{-4}$) & F-test \\
(1) & (2) & (3) & (4) & (5) & (6) & (7) & (8) & (9) & (10)  \\
\hline  
\endhead
%\multicolumn{11}{r}{continued on next page\ldots}\\
\endfoot
\\
\endlastfoot
\multicolumn{10}{c}{MARK\,348} \\ \hline
Ind & 0067540201* & ME2PL & 0.00$ _{ 0.00 }^{ 0.03 }$ & 13.40$ _{ 12.94 }^{ 13.61 }$ &  0.18$ _{ 0.14 }^{ 0.20 }$ & 1.50$ _{ 1.44 }^{ 1.56 }$  & 0.59$ _{ 0.55 }^{ 0.63 }$ & 80.76$ _{ 80.76 }^{ 101.65 }$ & 1281.96/1132 \\ 
Ind & 0701180101 & ME2PL & - & 12.89$ _{ 11.74 }^{ 14.12 }$ & 0.20$ _{ 0.18 }^{ 0.23 }$ & 1.42$ _{ 1.26 }^{ 1.58 }$ & 0.37$ _{ 0.37 }^{ 0.49 }$ & 23.81$ _{ 17.39 }^{ 32.79 }$ & 219.05/227 \\ \hline
SMF1 & 0067540201 & ME2PL & - & 13.29$_{12.90}^{13.70}$ & 0.19$_{0.18}^{0.21}$ & 1.50$_{1.44}^{1.56}$ & 0.54$_{0.50}^{0.57}$ & 88.49$_{79.91}^{98.95}$ & 1520.54/1368 \\
     & 0701180101 &       & & & & & & 27.17$_{24.37}^{30.49}$ & 0 \\ 
\multicolumn{10}{c}{NGC\,424} \\ \hline
Ind & 0002942301* & 2ME2PL & - & 34.89$ _{ 23.92 }^{ 51.68 }$ &  0.07$ _{ 0.06 }^{ 0.09 }$(0.65$ _{ 0.58 }^{ 0.72 }$) & 1.49$ _{ 1.07 }^{ 1.91 }$ & 0.48$ _{ 0.48 }^{ 0.85 }$ & 4.21$ _{ 1.44 }^{ 10.69 }$ & 66.80/54 \\
Ind & 0550950101 & 2ME2PL & - & 45.55$ _{ 41.29 }^{ 51.01 }$ & 0.10$ _{ 0.09 }^{ 0.11 }$ (0.71$ _{ 0.70 }^{ 0.72 }$) & 2.03$ _{ 1.93 }^{ 2.10 }$ & 0.74$ _{ 0.74 }^{ 0.83 }$ & 11.47$ _{ 9.26 }^{ 13.75 }$ & 1165.90/532 \\
\end{longtable}

\begin{longtable}{ccccccc}
\caption[]{\label{lumincorrSey} X-ray luminosities.}  \\  \hline \hline
\multicolumn{7}{c}{ \hspace*{5.7cm} Individual \hspace*{3.5cm} Simultaneous } \\  \cmidrule(r){4-5} \cmidrule{6-7}
Name & Satellite & ObsID & log(L(0.5-2 keV))  & log(L(2-10 keV)) & log(L(0.5-2 keV))  & log(L(2-10 keV)) \\ 
(1) & (2) & (3) & (4) & (5) & (6) & (7)  \\ 
\hline 
\endfirsthead
\caption[]{ (Cont.)} \\
\hline \hline
\multicolumn{7}{c}{ \hspace*{5.7cm} Individual \hspace*{3.5cm} Simultaneous } \\ \cmidrule(r){4-5} \cmidrule{6-7}
Name & Satellite & ObsID & log(L(0.5-2 keV))  & log(L(2-10 keV)) & log(L(0.5-2 keV))  & log(L(2-10 keV)) \\ 
(1) & (2) & (3) & (4) & (5) & (6) & (7)  \\ 
\hline 
\endhead
%\multicolumn{11}{r}{continued on next page\ldots}\\
\endfoot
\\
\endlastfoot
MARK\,348 & \emph{XMM}--Newton & 0067540201 & 43.01$_{42.98}^{43.03}$  &     43.41$_{43.40}^{43.41}$  &  43.01$_{42.99}^{43.02}$  &     43.40$_{43.40}^{43.41}$  \\
          & \emph{XMM}--Newton & 0701180101 & 42.47$_{42.44}^{42.50}$  &     42.89$_{42.88}^{42.91}$  &  42.50$_{42.49}^{42.52}$  &     42.90$_{42.89}^{42.91}$  \\ \hline
NGC\,424 & \emph{XMM}--Newton & 0002942301 & 41.70$_{41.68}^{41.72}$  &     42.01$_{41.93}^{42.07}$   & 41.74$ _{     41.72  }^{   41.76 }$  &  41.87$ _{     41.81 }^{    41.92}$ \\
         & \emph{XMM}--Newton & 0550950101 & 41.95$_{41.94}^{41.95}$  &     42.00$_{41.99}^{42.01}$   & \\
         & \emph{Chandra} (2$\arcsec$) & 3146 & 42.02$_{41.97}^{42.06}$  &     41.91$_{41.83}^{41.97}$   & 41.74$ _{     41.72 }^{    41.76 }$  &  41.84$ _{     41.79 }^{    41.89}$ \\
         & \emph{Chandra} (20$\arcsec$) & 3146 &  42.00$_{41.95}^{42.04}$  &     41.94$_{41.86}^{42.01}$  & \\ \hline
MARK\,573 & \emph{Chandra} (2$\arcsec$) & 7745 & 42.02$_{41.98}^{42.06}$  &     41.69$_{41.59}^{41.77}$   & 41.65$_{41.56}^{41.72}$  &     41.54$_{41.48}^{41.59}$  \\
          & \emph{Chandra} (20$\arcsec$) & 7745 & 42.19$_{42.17}^{42.21}$  &     41.87$_{41.76}^{41.96}$   & \\
\end{longtable}

\begin{longtable}{lcccccccccc}
\caption[]{\label{annulusSey} Results for the best fit of the annular region (ring) in \emph{Chandra} data, and the best fit obtained for the nucleus of \emph{XMM}--Newton data when the contribution from the annular region was removed. }  \\  \hline \hline
Name (obsID) & Region & Model & $N_{H1}^a$ & $N_{H2}^a$ & kT & $\Gamma$ & $\chi^2_r$ & $log(L_{soft})$ & $log(L_{hard})$ & Cont. \\
 & & & & & (keV) & & & (0.5-2 keV) & (2-10 keV) & \%  \\
(1) & (2) & (3)          & (4) & (5) & (6) & (7) & (8) & (9) & (10) & (11) \\ 
\hline 
\endfirsthead
\caption[]{(Cont.)} \\
\hline \hline
Name (obsID) & Region & Model & $N_{H1}^a$ & $N_{H2}^a$ & kT & $\Gamma$ & $\chi^2_r$ & $log(L_{soft})$ & $log(L_{hard})$ & Cont. \\
 & & & & & (keV) & & & (0.5-2 keV) & (2-10 keV) & \%  \\
(1) & (2) & (3)          & (4) & (5) & (6) & (7) & (8) & (9) & (10) & (11) \\ \hline  
\hline 
\endhead
%\multicolumn{11}{r}{continued on next page\ldots}\\
\endfoot
\\
\endlastfoot
MARK\,573 (7745) & Ring* & ME2PL & - & 75.90$ _{ 26.64 }^{ 160.34 }$ & 0.67$ _{ 0.57 }^{ 0.75 }$ & 3.86$ _{ 3.64 }^{ - }$ & 1.97 & 43.12 & 42.01 & 24 \\
MARK\,573 (0200430701) & Nucleus** & 2ME2PL & - & 9.82$_{5.83}^{18.25}$ & 0.15$_{0.14}^{0.18}$ (0.76$_{0.68}^{0.86}$) & 3.01$_{2.61}^{3.37}$ & 0.86 & 41.88 & 41.34 & - \\ \hline
MARK\,1066 (4075) & Ring* & PL & 0.21$_{0.00}^{0.35}$ & - & - & 3.99$_{2.68}^{4.00}$ & 1.40 & 39.65 & 39.26 & 8 \\
MARK\,1066 (0201770201) & Nucleus** & ME2PL & 0.08$_{0.00}^{0.16}$ & 53.37$_{28.95}^{110.46}$ & 0.76$_{0.61}^{0.87}$ & 2.00$_{1.50}^{2.56}$ & 0.78 & 41.61 & 41.67 & - \\ \hline
3C\,98.0 (10234) & Ring* & ME & 15.78$_{ 8.21 }^{ 27.17 }$ & - & 3.99$ _{ 1.24 }^{ 0 }$ & - & 4.30 & 41.58 & 41.65 & 8 \\
3C\,98.0 (0064600301) & Nucleus** & MEPL & 0.72$_{0.52}^{0.94}$ & 6.98$_{4.45}^{9.76}$ & 0.17$_{0.11}^{0.21}$ & 0.99$_{0.43}^{1.61}$ & 0.72 & 42.93 & 42.67 & - \\ \hline
NGC\,3079 (2038) & Ring* & MEPL & 0.01$_{0.00}^{0.24}$ & 0.07$_{0.00}^{0.16}$ & 0.65$_{0.61}^{0.69}$ & 2.16$_{1.76}^{2.75}$ & 0.91 & 39.92 & 39.63 & 78 \\
NGC\,3079 (0110930201) & Nucleus** & MEPL & 0.59$_{0.00}^{1.82}$ & - & 0.23$_{0.20}^{0.28}$ & 1.24$_{0.57}^{4.96}$ & 1.04 & 40.36 & 41.92 & - \\
\end{longtable}

\newpage

\begin{longtable}{lccccccccr}
\caption[]{\label{simultanilloSey} Simultaneous fittings taking into account the contribution from the annular region given in Table \ref{annulusSey}.}  \\  \hline \hline
 ObsID & $N_{H1}$ & $N_{H2}$ & kT & $\Gamma$ & $Norm_1$ & $Norm_2$ & $\chi^2/d.o.f$ & $log(L_{soft})$ & $log(L_{hard})$ \\
  & ($10^{22} cm^{-2}$) & ($10^{22} cm^{-2}$) & (keV) & & ($10^{-4}$) & ($10^{-4}$) & & (0.5-2 keV) & (2-10 keV)  \\
(1) & (2) & (3) & (4) & (5) & (6) & (7) & (8) & (9) & (10)  \\
\hline 
\endfirsthead
\caption[]{\textbf{(Cont.)}} \\
\hline \hline
 ObsID & $N_{H1}$ & $N_{H2}$ & kT & $\Gamma$ & $Norm_1$ & $Norm_2$ & $\chi^2/d.o.f$ & $log(L_{soft})$ & $log(L_{hard})$ \\
  & ($10^{22} cm^{-2}$) & ($10^{22} cm^{-2}$) & (keV) & & ($10^{-4}$) & ($10^{-4}$) & & (0.5-2 keV) & (2-10 keV)  \\
(1) & (2) & (3) & (4) & (5) & (6) & (7) & (8) & (9) & (10)  \\ 
\hline 
\endhead
%\multicolumn{11}{r}{continued on next page\ldots}\\
\endfoot
\\
\endlastfoot
\multicolumn{10}{c}{NGC\,424} \\  \hline
0002942301 & - & 24.49$ _{ 18.65 }^{ 31.94 }$ & 0.09 $_{ 0.07 }^{ 0.10 }$ (0.67 $_{ 0.61 }^{ 0.73 }$) & 1.82$ _{ 1.46 }^{ 2.15 }$ & 0.72 $_{ 0.58 }^{0.88 }$ & 6.70$ _{ 3.11 }^{ 13.11 }$ & 138.97/103 & 41.74$ _{     41.72  }^{   41.76 }$  &  41.87$ _{     41.81 }^{    41.92}$ \\
3146 & & & & & & & & 41.74$ _{     41.72 }^{    41.76 }$  &  41.84$ _{     41.79 }^{    41.89}$ \\
\multicolumn{10}{c}{NGC\,788} \\  \hline
0601740201 & - & 46.61 $_{ 42.14 }^{ 51.35 }$ & 0.11 $_{ 0.09 }^{ 0.12 }$ (0.71$ _{ 0.67 }^{ 0.75 }$) & 1.28 $_{ 1.06 }^{ 1.51 }$ & 0.35 $_{ 0.29 }^{ 0.41 }$ & 12.43 $_{ 7.55 }^{ 20.51 }$ & 262.36/205 & 42.11 $_{    42.04   }^{   42.17 }$   & 42.63  $_{   42.60  }^{    42.66}$ \\
11680 & & & & & & & & 42.11 $_{    42.04  }^{    42.17 }$  &  42.59 $_{    42.56  }^{    42.62}$ \\

\multicolumn{10}{c}{MARK\,573} \\  \hline
0200430701 & - &  11.40$_{ 0.00 }^{ 53.26 }$ & 0.12 $_{ 0.11 }^{ 0.15 }$ (0.71$ _{ 0.67 }^{ 0.76 }$) & 2.67 $_{ 2.32 }^{ 2.91 }$ &  0.02$_{0.01  }^{ 0.02 }$ &  0.51$_{ 0.43 }^{ 0.59 }$ & 198.75/171 & 41.75 $_{   41.74   }^{  41.76  }$   &  41.41 $_{  41.31   }^{  41.48  }$ \\
\end{longtable}

\begin{figure}[H]
\centering
\subfloat{\includegraphics[width=0.30\textwidth]{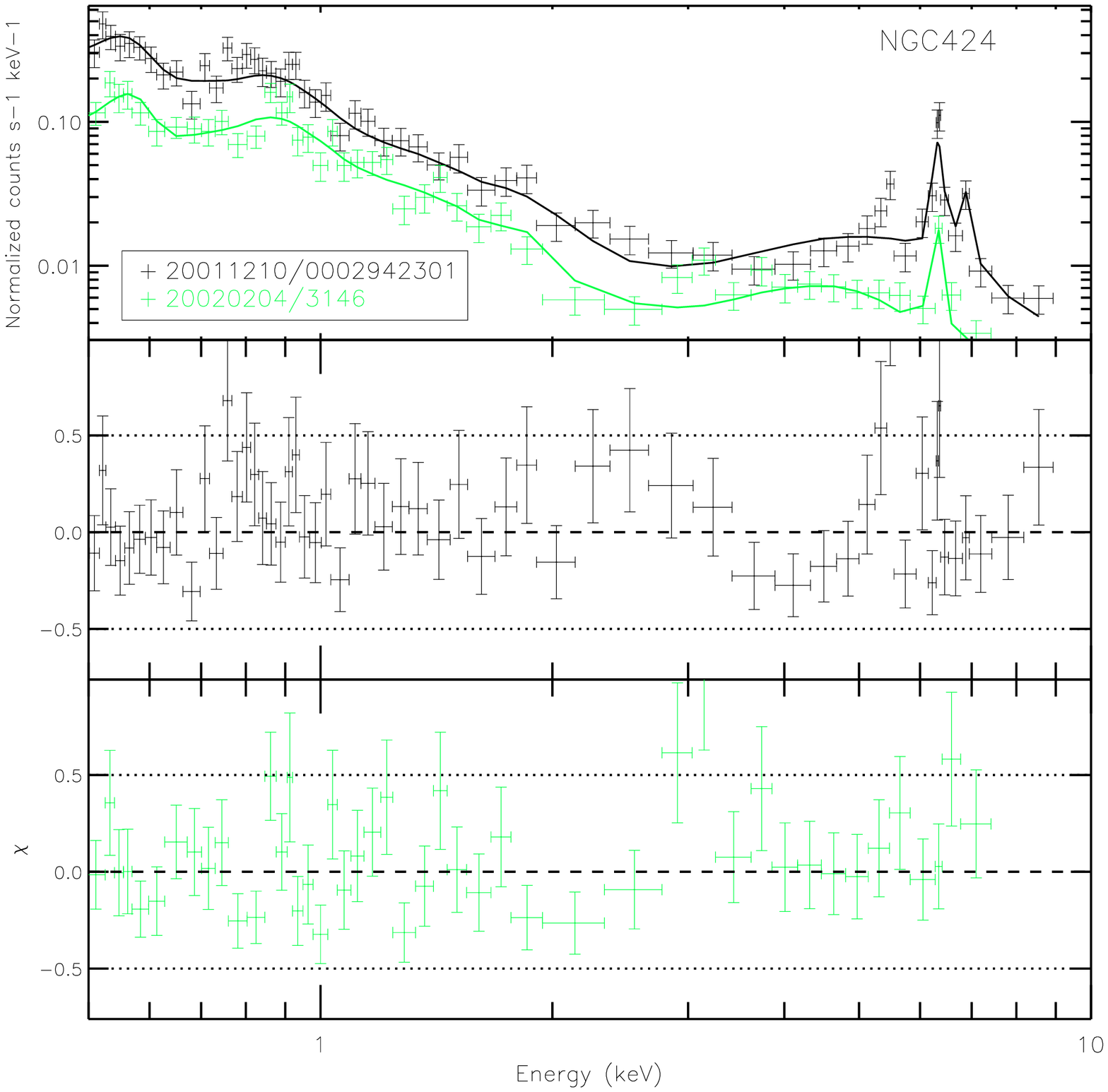}}
\subfloat{\includegraphics[width=0.30\textwidth]{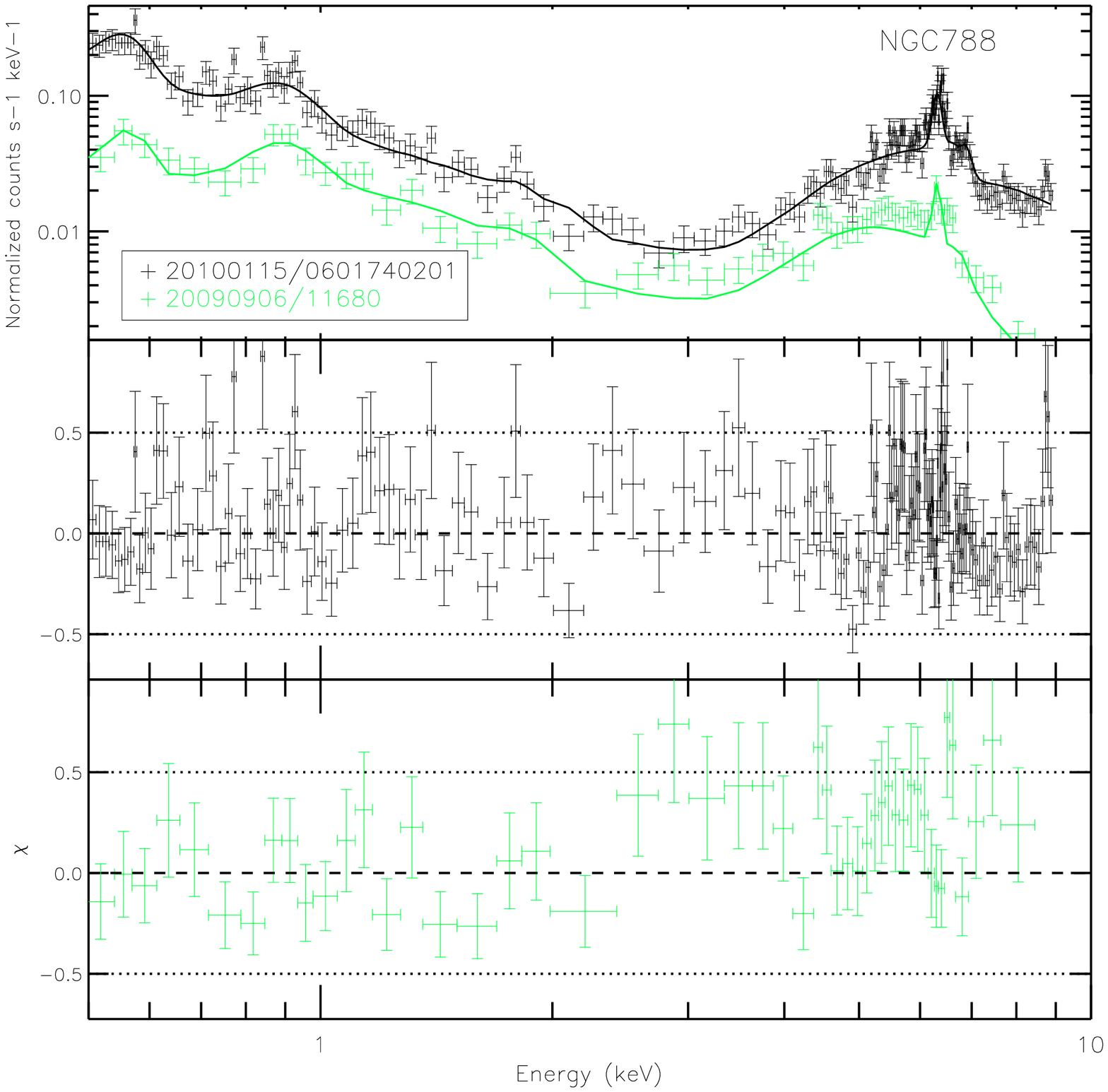}}
\subfloat{\includegraphics[width=0.30\textwidth]{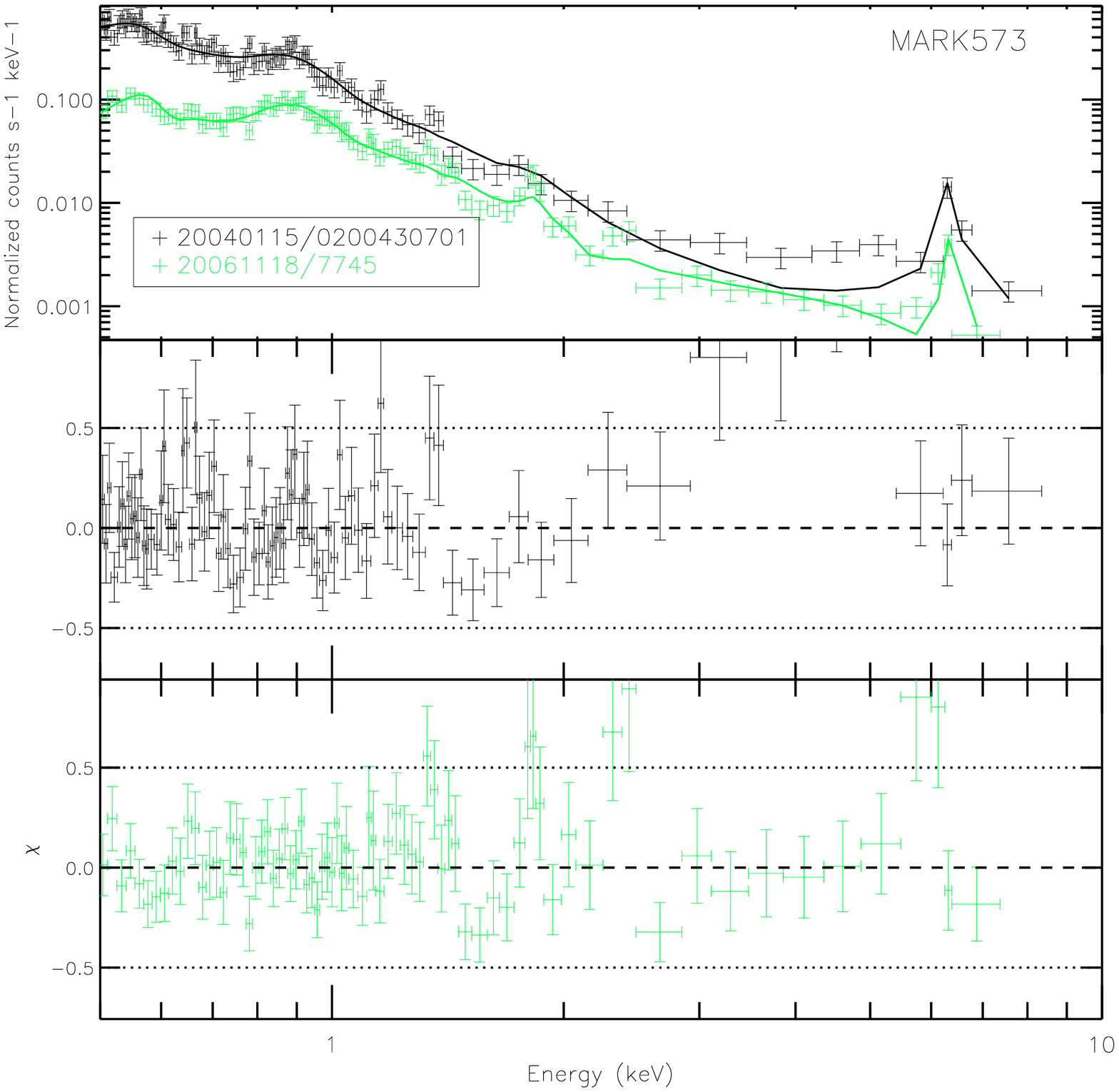}}

\subfloat{\includegraphics[width=0.30\textwidth]{MARK1066_MARK1066_ME2PLring_1.ps}}
\subfloat{\includegraphics[width=0.30\textwidth]{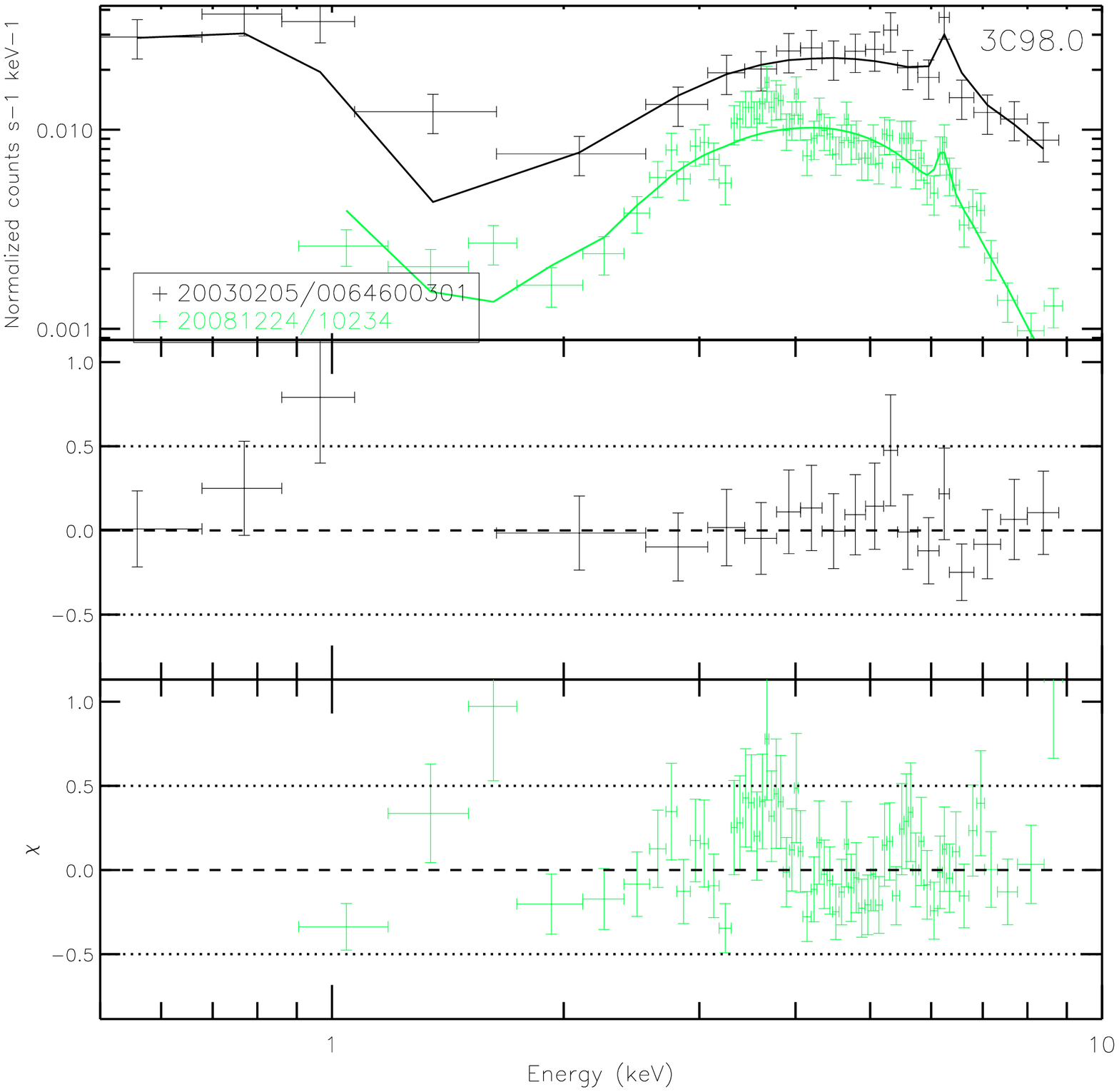}}
\subfloat{\includegraphics[width=0.30\textwidth]{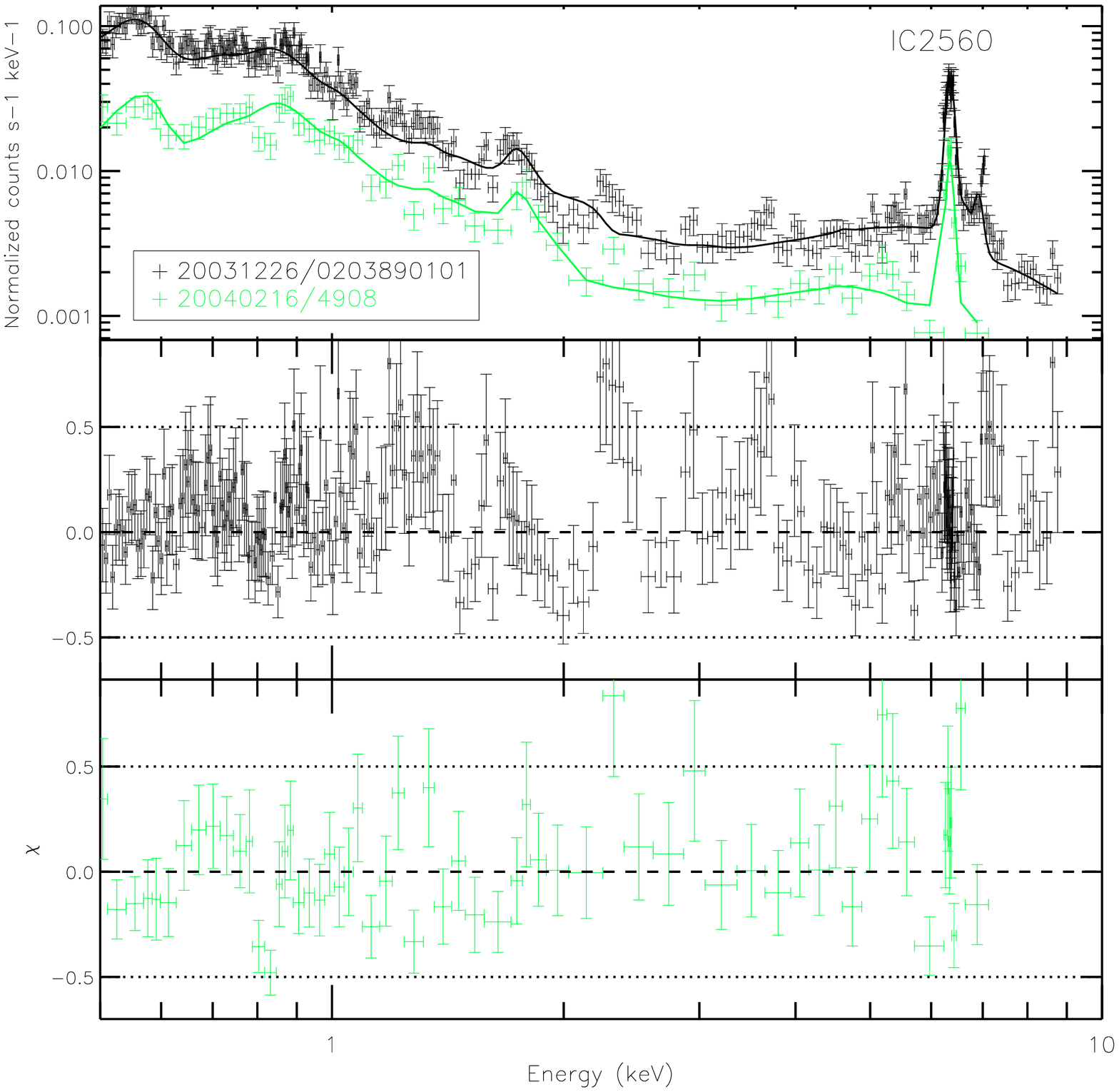}}

\subfloat{\includegraphics[width=0.30\textwidth]{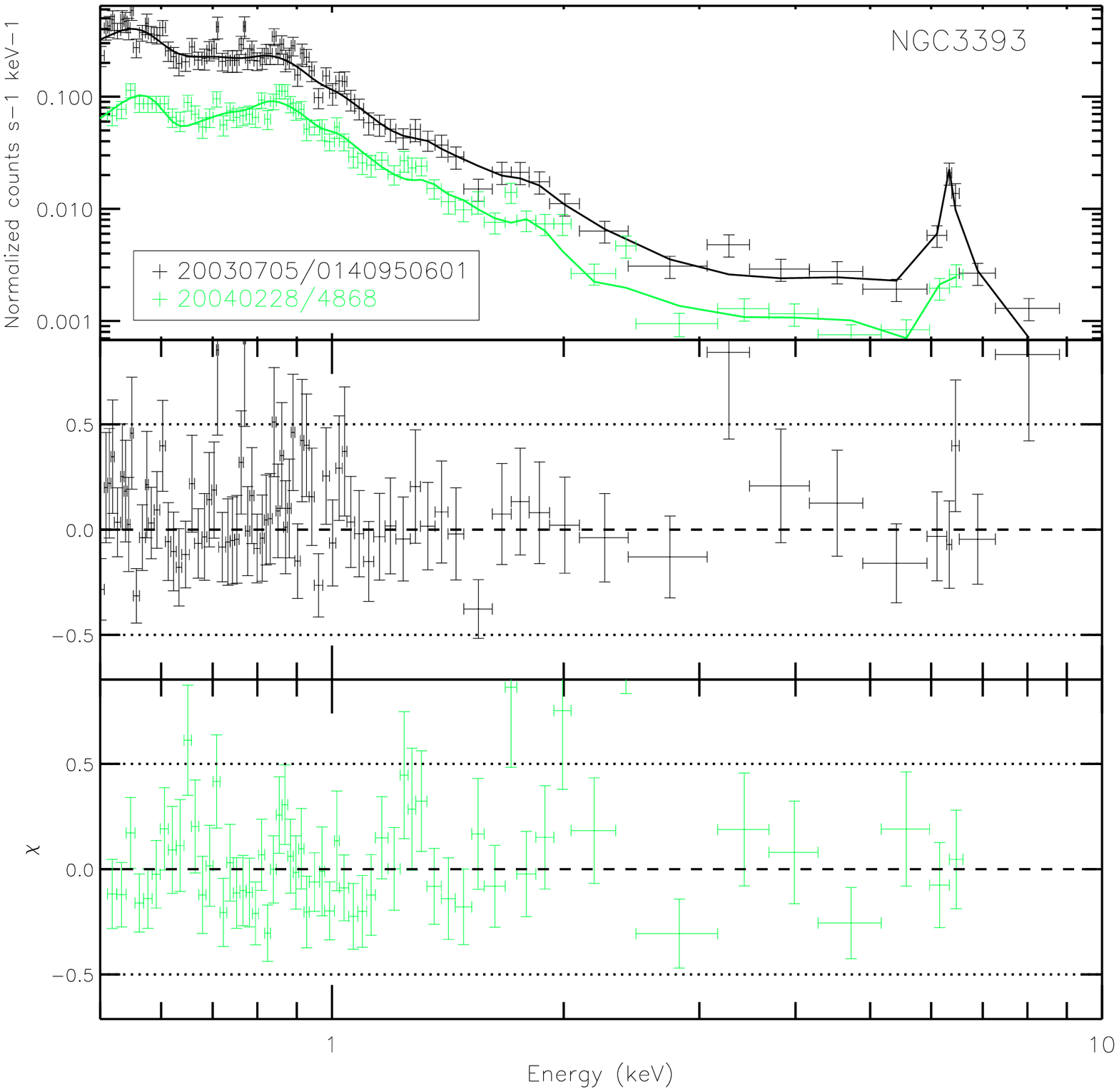}}
\subfloat{\includegraphics[width=0.30\textwidth]{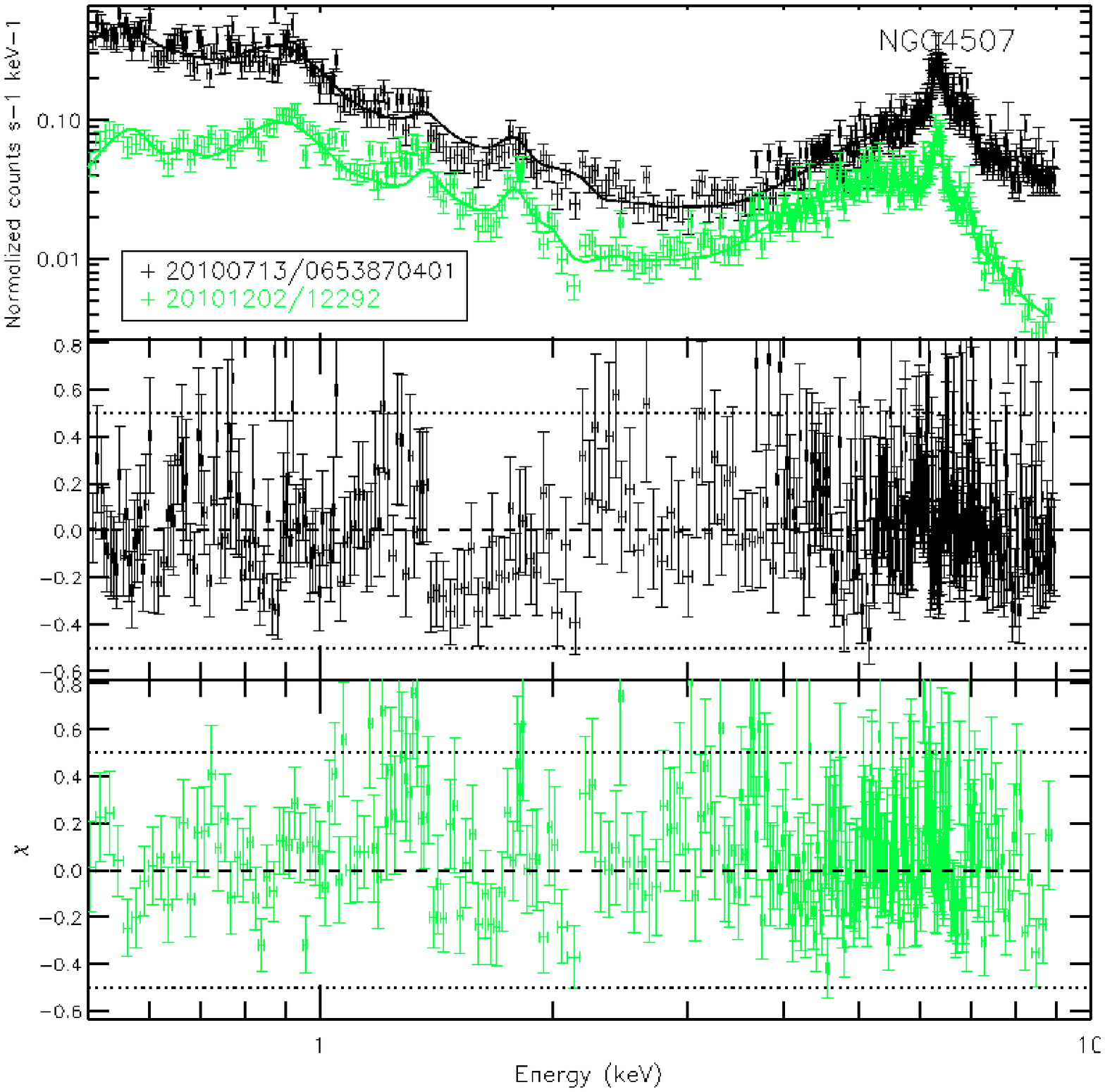}}
\subfloat{\includegraphics[width=0.30\textwidth]{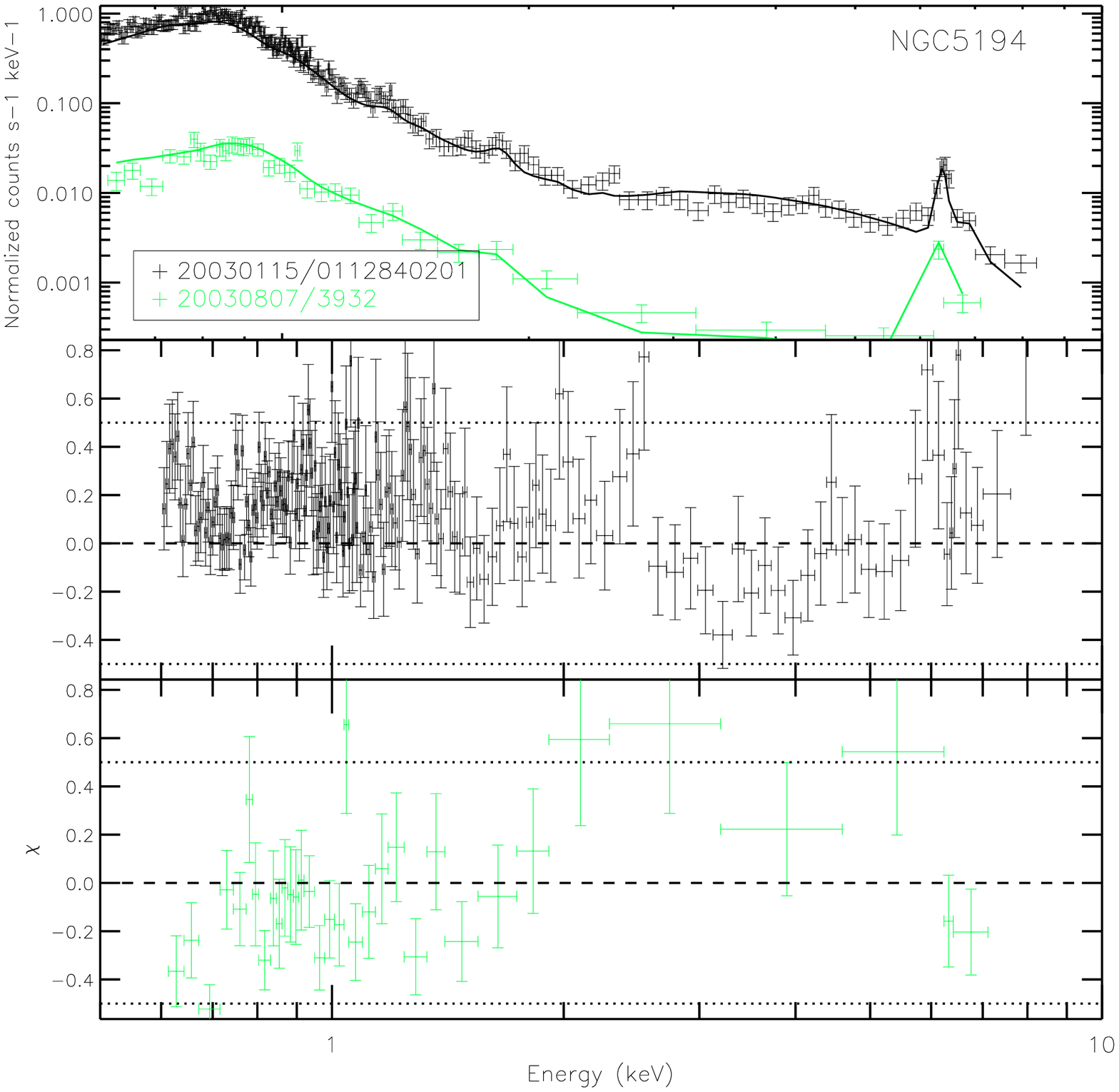}}

\subfloat{\includegraphics[width=0.30\textwidth]{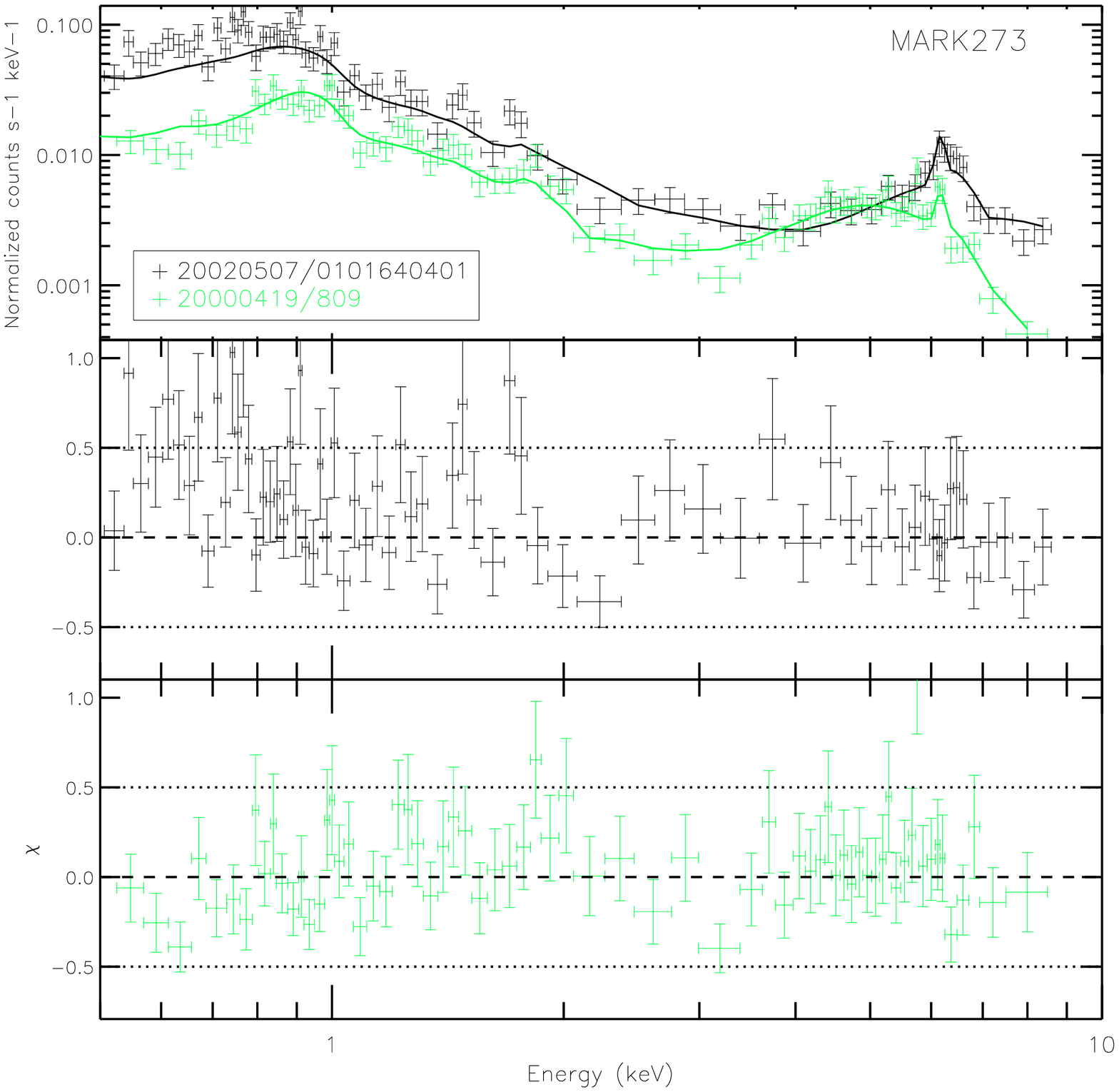}}
\subfloat{\includegraphics[width=0.30\textwidth]{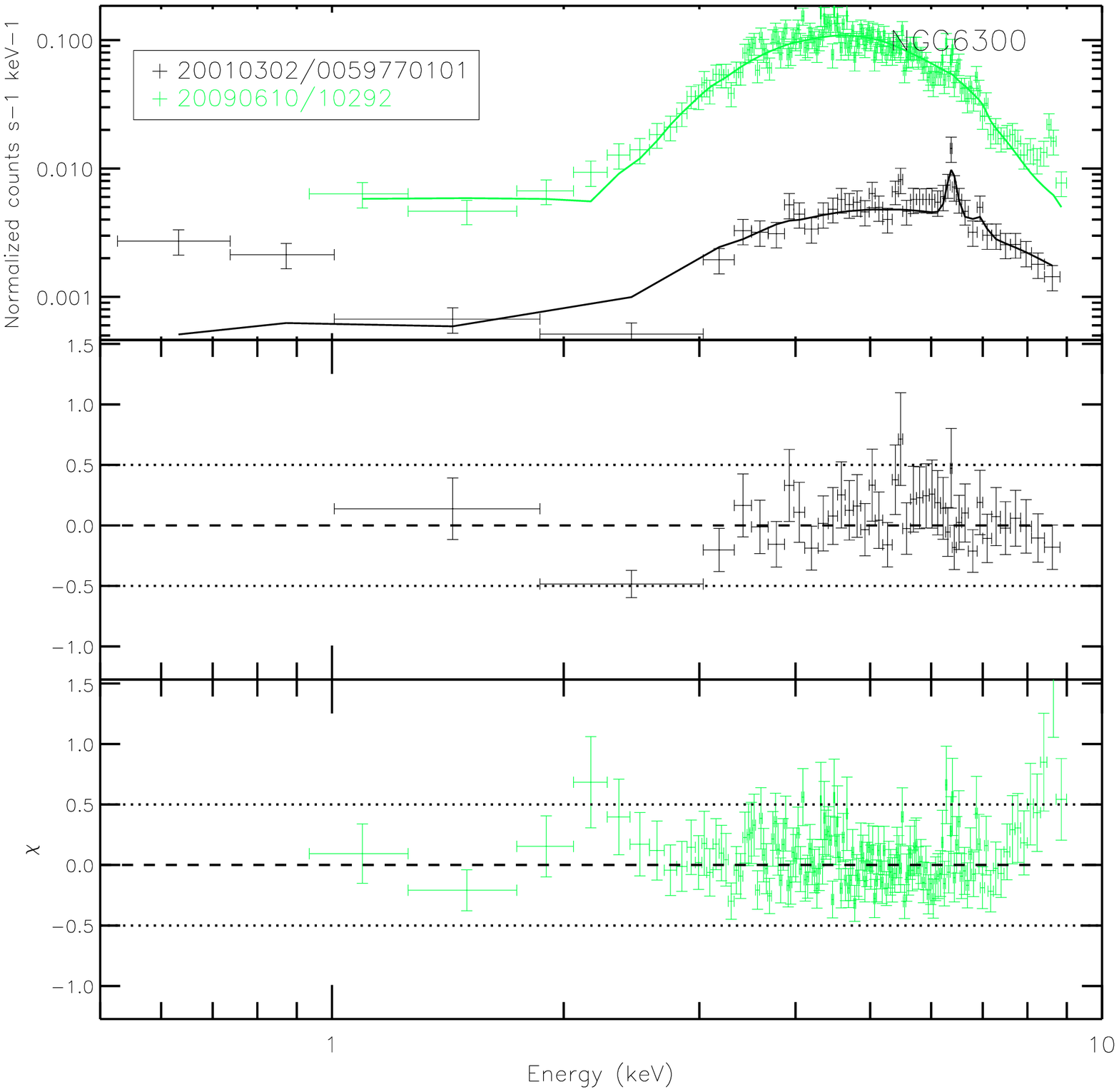}}
\subfloat{\includegraphics[width=0.30\textwidth]{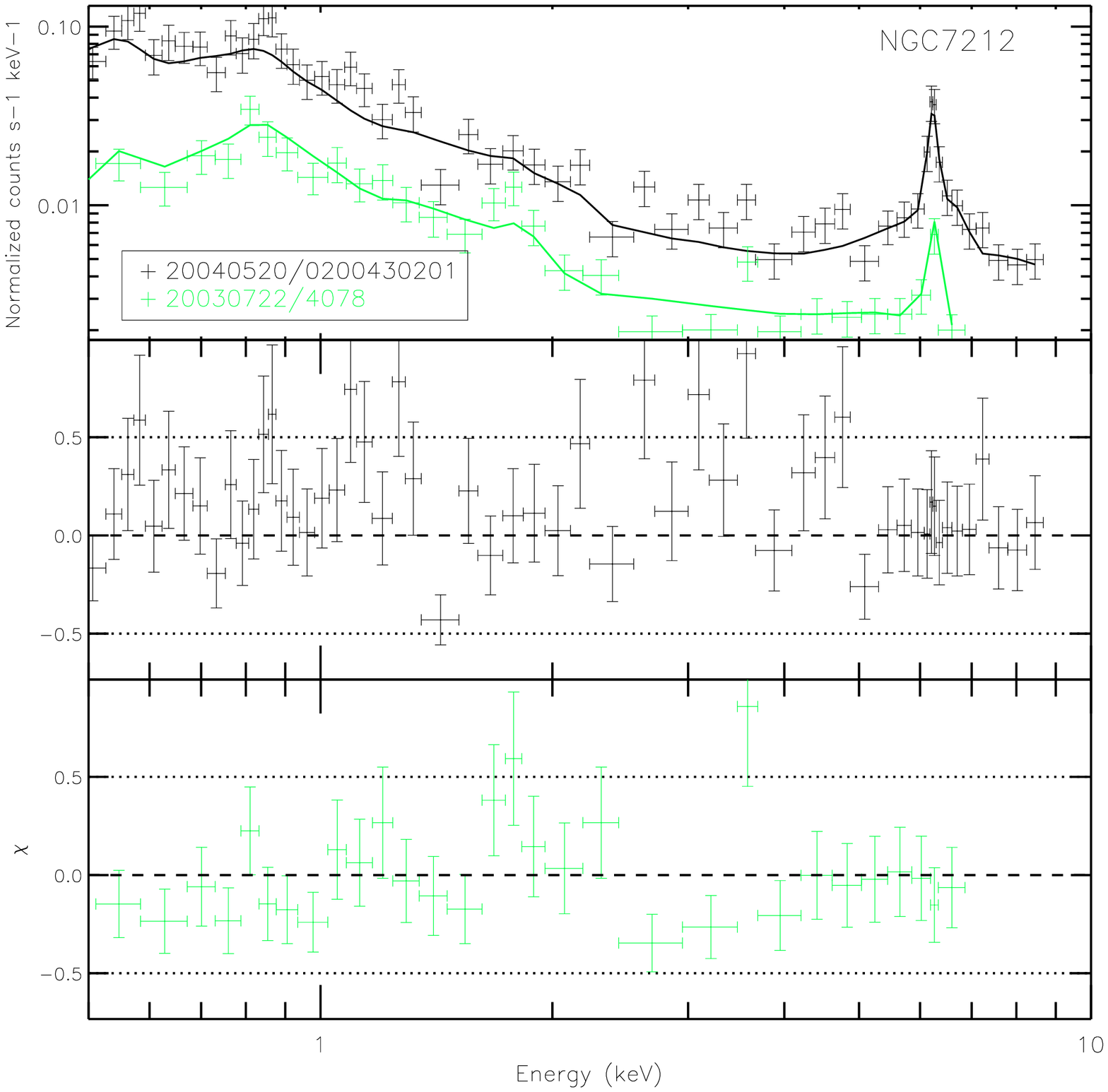}}
\caption{For each object, (top): simultaneous fit comparing \emph{Chandra} and \emph{XMM}--Newton spectra; (from second row on): residuals in units of $\sigma$. The legends contain the date (in the format yyyymmdd) and the obsID. The observations used for comparisons are marked with $c$ in Table \ref{obsSey}.}
\label{ringSey}
\end{figure}

\begin{figure}
\setcounter{figure}{0}
\centering
\subfloat{\includegraphics[width=0.30\textwidth]{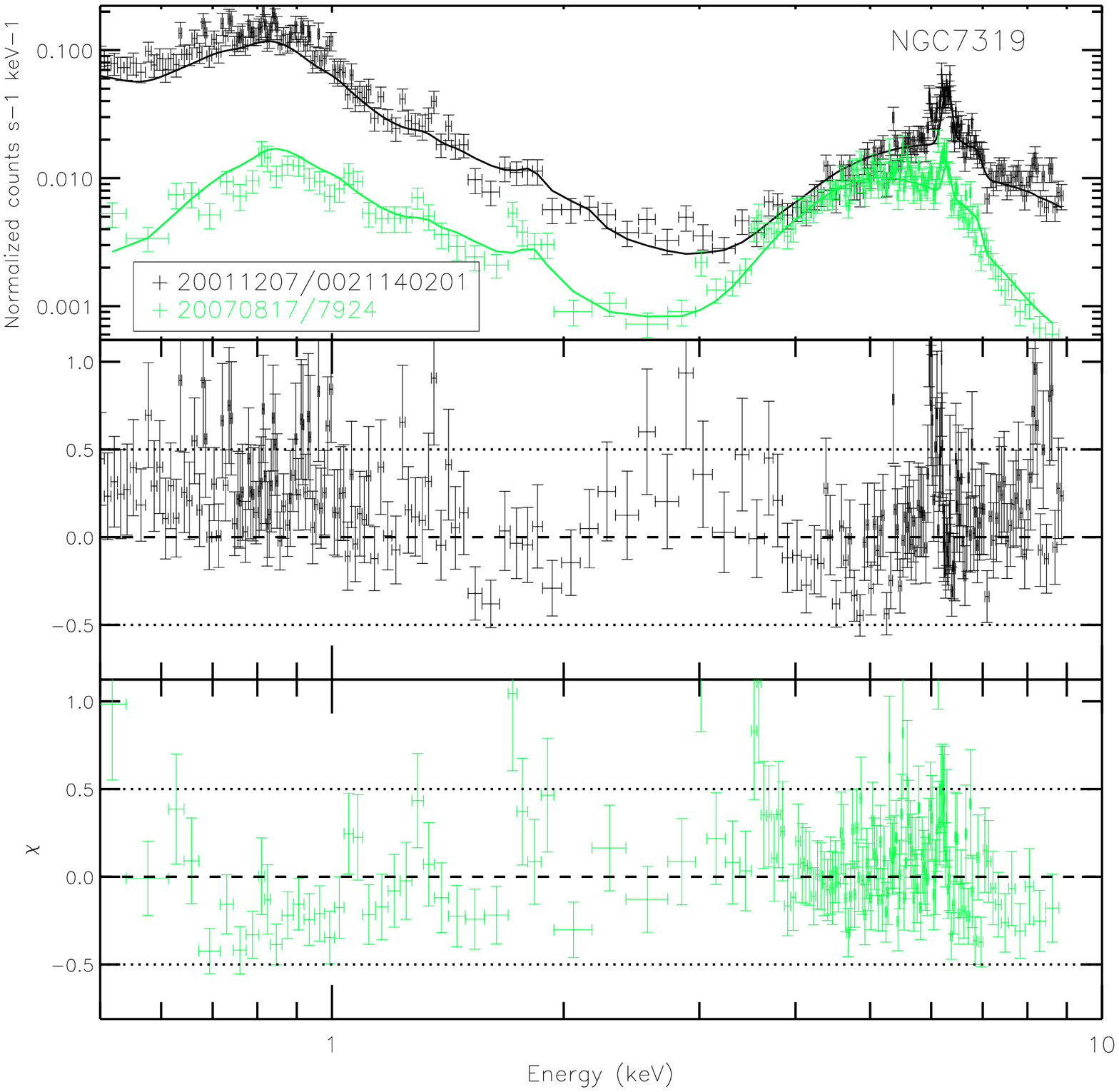}}
\caption{Cont.}
\end{figure}

\begin{figure*}
\centering
\subfloat{\includegraphics[width=0.30\textwidth]{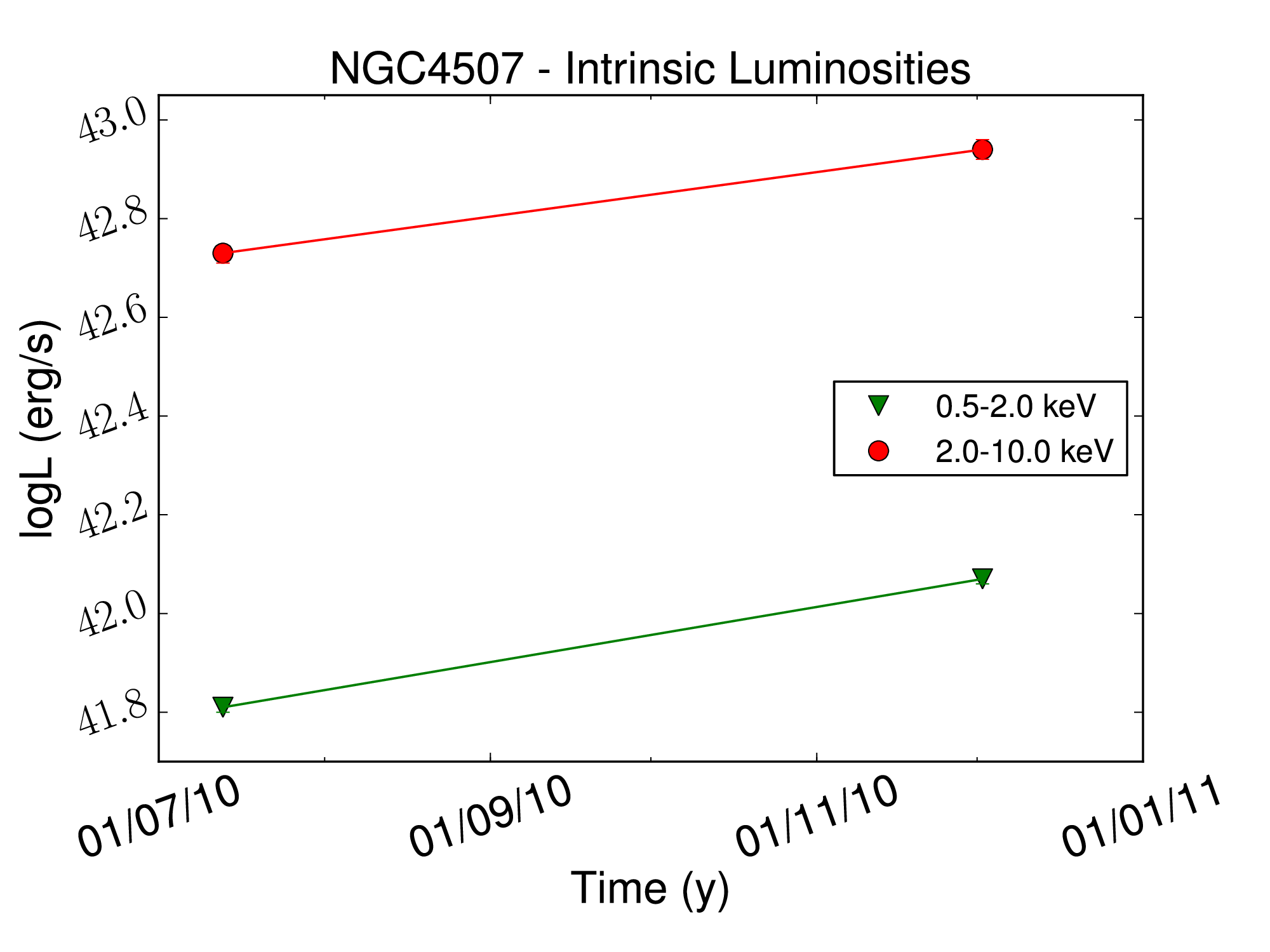}}
\subfloat{\includegraphics[width=0.30\textwidth]{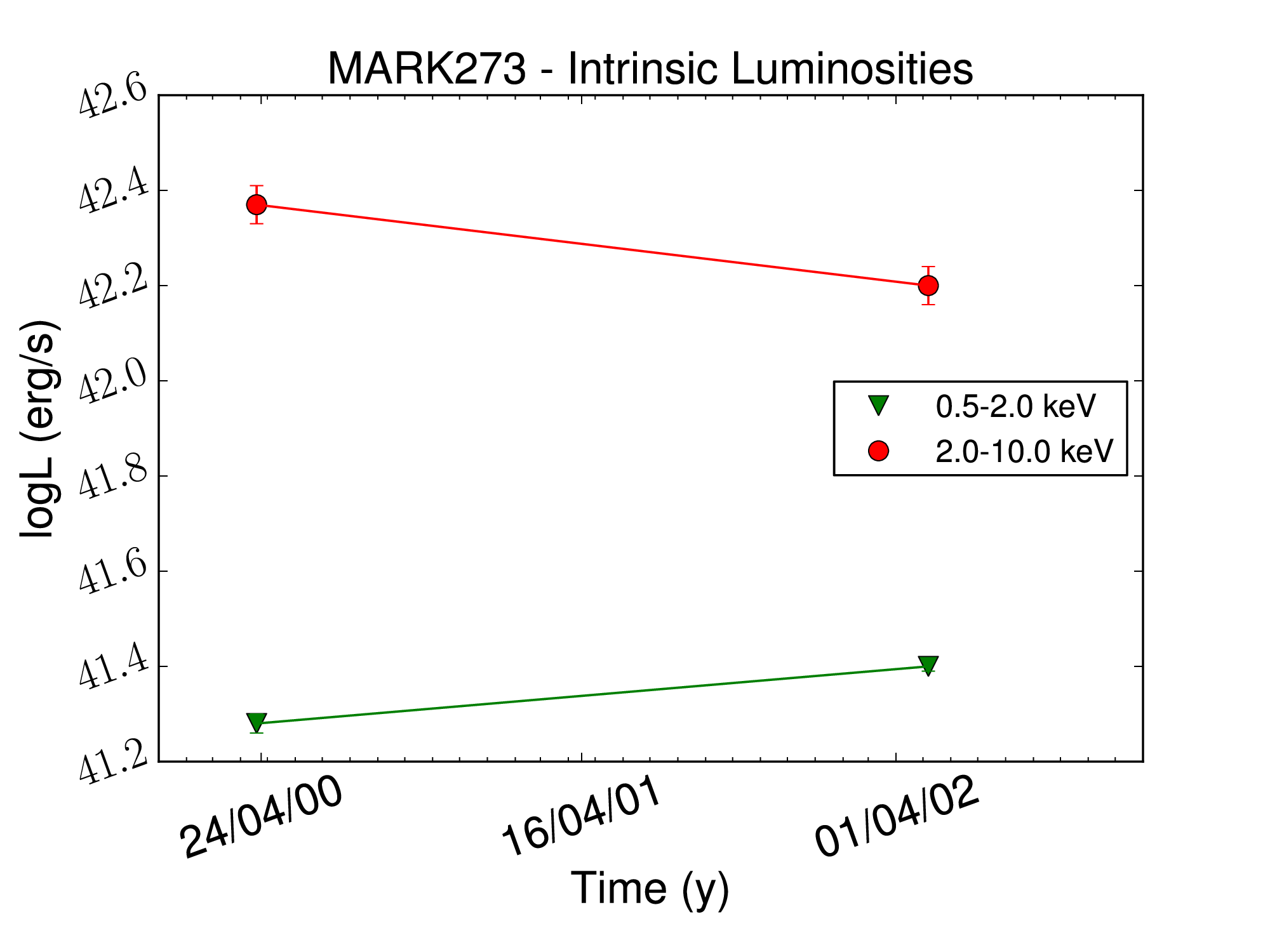}}

\subfloat{\includegraphics[width=0.30\textwidth]{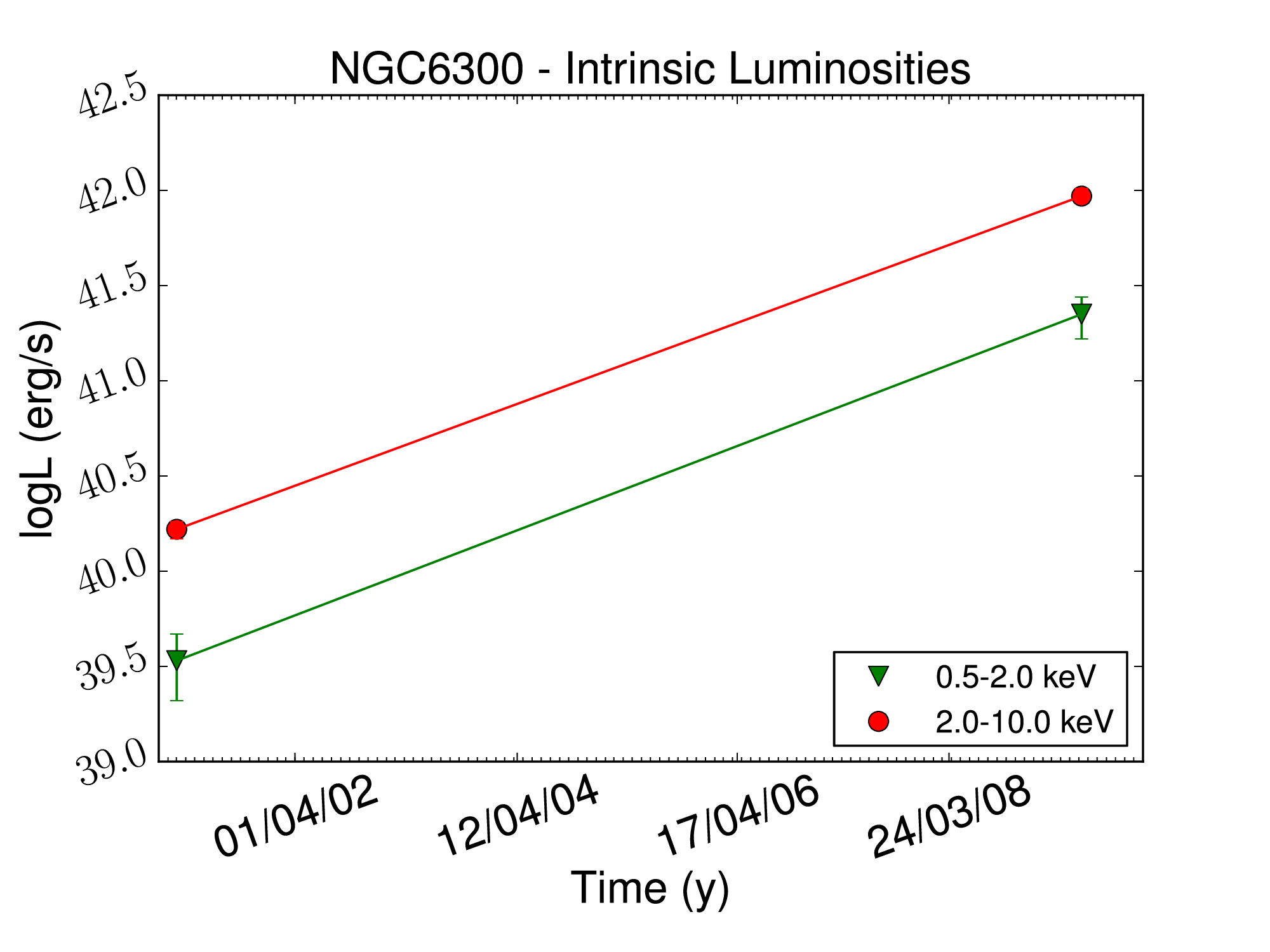}}
\subfloat{\includegraphics[width=0.30\textwidth]{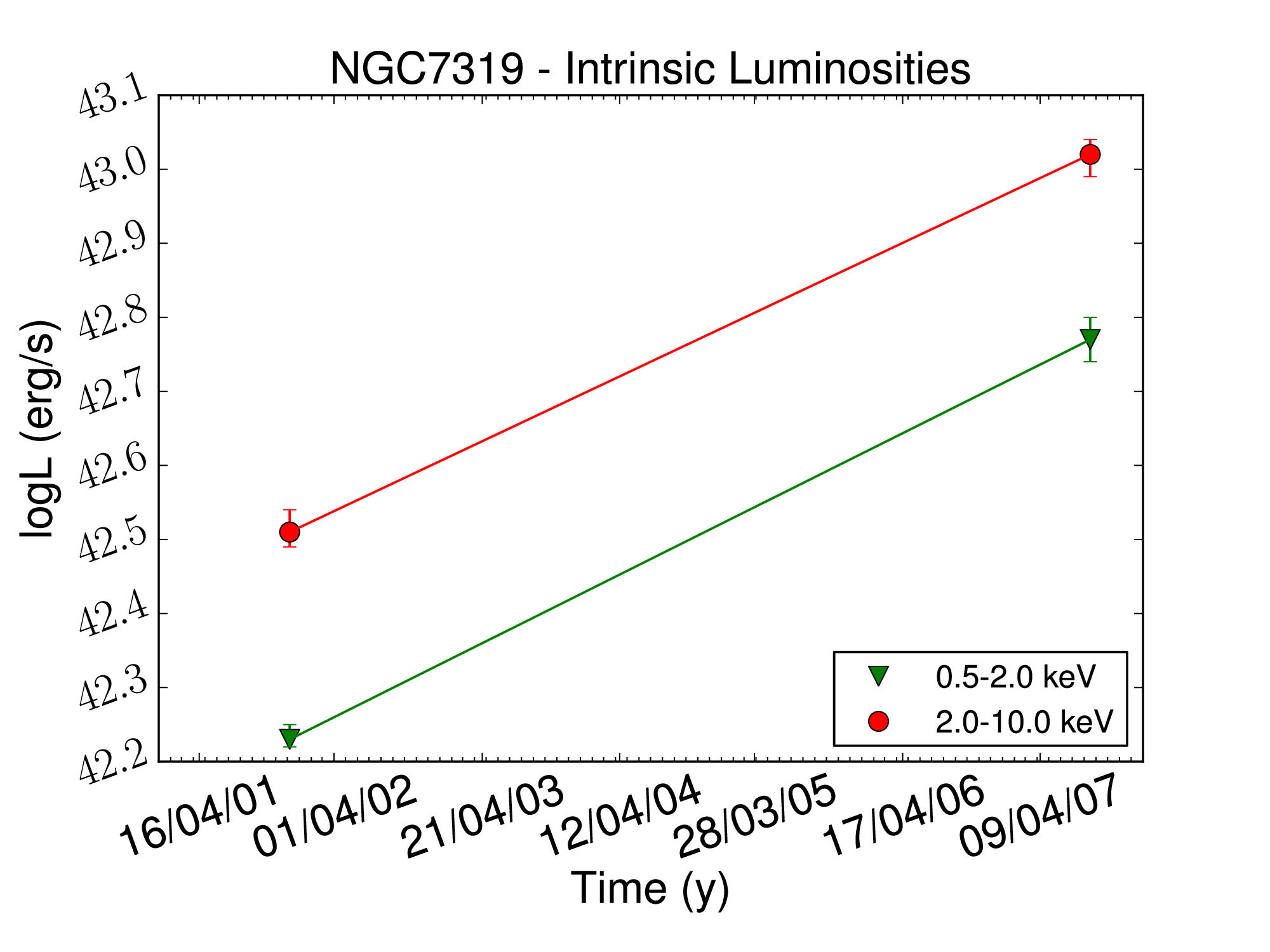}}
\caption{X-ray intrinsic luminosities calculated for the soft (0.5--2.0 keV, green triangles) and hard (2.0--10.0 keV, red circles) energies in the simultaneous fitting, only for the variable objects, when \emph{Chandra} and \emph{XMM}--Newton data are compared.}
\label{luminXfigSeyring}
\end{figure*}

\begin{longtable}{lccccccc}
\caption[]{\label{estcurvasSey} Statistics of the light curves.}  \\  \hline \hline
Name &  ObsID & Energy & $\chi^2/d.o.f$  & Prob.(\%) & $\sigma_{NXS}^2$  &  $<\sigma^2_{NXS}>$ \\
(1) & (2) & (3) & (4) & (5) & (6) & (7)   \\ 
\hline 
\endfirsthead
\caption[]{(Cont.)} \\
\hline \hline
Name &  ObsID & Energy & $\chi^2/d.o.f$  & Prob.(\%) & $\sigma_{NXS}^2$  &  $<\sigma^2_{NXS}>$  \\
(1) & (2) & (3) & (4) & (5) & (6) & (7)  \\ 
\hline 
\endhead
%\multicolumn{11}{r}{continued on next page\ldots}\\
\endfoot
\\
\endlastfoot
NGC\,424 &  0550950101    & 0.5-10 (1) & 48.7/40 & 84 & $<$0.0026 & $<$0.0020 \\
         &             & 0.5-10 (2) & 13.0/30 & 1 & $<$0.0031 \\
          &            & 0.5-2 (1) &  38.8/40 & 48 & $<$0.0034 & $<$0.0027  \\
         &            & 0.5-2 (2) &  17.3/30 & 3 & $<$0.0041  \\
          &            & 2-10 (1) & 33.0/40 & 22 & $<$0.0102 & $<$0.0077 \\
          &            & 2-10 (2) & 31.8/30 & 62 & $<$0.0116 & \\
MARK\,573 & 7745   & 0.5-10 & 44.4/38 & 88 & 0.0041$^+_-$0.0037 \\
          &            & 0.5-2 & 34.9/38 & 39 & $<$0.0096 \\
          &            & 2-10 &  76.0/38 & 100 & $<$0.0710 \\
          &  13124  & 0.5-10 &  56.0/40 & 95 & $<$0.0108 \\
          &            & 0.5-2 & 46.9/40 & 79 & $<$0.0122 \\
          &            & 2-10 &  50.7/40 & 88 & $<$0.0900 \\
3C\,98.0 & 10234 & 0.5-10 & 17.2/31 & 2 & $<$0.0157 & \\
         &       & 0.5-2 & 32.3/31 & 60 & $<$0.2035 & \\
         &       & 2-10 & 17.9/31 & 3 & $<$0.0169 & \\
IC\,2560 & 0203890101 & 0.5-10 & 62.0/40 & 99 & 0.0108 $^+_-$ 0.0043 & \\
         &            & 0.5-2 & 44.5/40 & 71 & $<$0.0156 \\
         &            & 2-10 & 49.1/40 & 85 & $<$0.0268 \\
         & 4908 & 0.5-10 & 48.1/40 & 82 & $<$0.0204 \\
         &            & 0.5-2 & 60.7/40 & 99 & 0.0172$^+_-$ 0.0140 \\
         &            & 2-10 & 29.3/40 & 11 & $<$0.0581 \\
NGC\,3393 & 12290 & 0.5-10 & 30.2/40 & 13 & $<$0.0109 \\
          &            & 0.5-2 & 31.3/40 & 16 & $<$0.0127 \\
          &            & 2-10 & 42.2/40 & 62 & $<$0.0724 \\
NGC\,4507 & 0006220201   & 0.5-10 &  35.4/30 & 77 & $<$0.0007 \\    
          &            & 0.5-2 & 25.7/30 & 31 & $<$0.0031 \\          
          &            & 2-10 &  36.4/30 & 81 & $<$0.0009 \\
          & 12292 & 0.5-10 & 39.2/39 & 54 & $<$0.0026 \\
          &            & 0.5-2 & 28.9/39 & 12 & $<$0.0079 \\
          &            &  2-10 & 47.6/39 & 84 & $<$0.0039 \\
NGC\,5194 & 3932 & 0.5-10 & 50.7/40 & 88 &   $<$0.0311    \\
          &      & 0.5-2 & 50.0/40 & 87 &   $<$0.0364    \\
          &      & 2-10  & 42.0/40 & 62 &  $<$0.2008      \\
          & 13813 & 0.5-10 (1) & 58.7/40 & 97 & $<$0.0568 & 0.0209$^+_-$0.0190 \\
          &       & 0.5-10 (2) & 36.2/40 & 46 & $<$0.0379 \\
          &       & 0.5-10 (3) & 32.8/40 & 22 & $<$0.0366 \\
          &       & 0.5-10 (4) & 58.8/40 & 97 &    0.0335$^+_-$0.0185 \\
          &       & 0.5-2 (1) & 84.8/40 & 100 & 0.0572$^+_-$0.0330 & 0.0373$^+_-$0.0289 \\
          &       & 0.5-2 (2) & 36.5/40 & 37 & $<$0.0454 \\
          &       & 0.5-2 (3) & 31.8/40 & 18 & $<$0.0435 \\
          &       & 0.5-2 (4) & 60.0/40 & 98 & 0.0236$^+_-$0.0217 \\
          &       & 2-10 (1) & 28.1/40 & 18 & $<$0.2318 &  $<$0.1218 \\
          &       & 2-10 (2) & 33.8/40 & 26 & $<$0.2203 \\
          &       & 2-10 (3) & 30.6/40 & 14 & $<$0.2473 \\
          &       & 2-10 (4) & 28.9/40 & 10 & $<$0.2716 \\
          & 13812 & 0.5-10 (1) & 48.9/40 & 84 & $<$0.0401 & $<$0.0227 \\
          &       & 0.5-10 (2) & 37.9/40 & 43 & $<$0.0382 \\
          &       & 0.5-10 (3) & 44.6/40 & 72 & $<$0.0398 \\
          &       & 0.5-2 (1) &  44.8/40 & 72 & $<$0.0485 & $<$0.0273 \\
          &       & 0.5-2 (2) & 40.9/40 & 57 & $<$0.0459 \\
          &       & 0.5-2 (3) & 40.5/40 & 55 & $<$0.0474 \\
          &       & 2-10 (1) & 38.1/40 & 45 & $<$0.2280 & $<$0.1423 \\
          &       & 2-10 (2) & 26.9/40 & 6 & $<$0.2355 \\
          &       & 2-10 (3) & 24.1/40 & 2 & $<$0.2737 \\
          & 13814 & 0.5-10 (1) & 54.5/40 & 94 & $<$0.0440 & $<$0.0208  \\
          &       & 0.5-10 (2) & 36.9/40 & 39 & $<$0.0400 \\
          &       & 0.5-10 (3) & 31.6/40 & 17 & $<$0.0403 \\
          &       & 0.5-10 (4) & 60.1/40 & 98 & $<$0.0422 \\
          &       & 0.5-2 (1) & 44.5/40 & 71 & $<$0.0525 & 0.0196$^+_-$0.0170 \\
          &       & 0.5-2 (2) & 49.0/40 & 84 & $<$0.0497 \\
          &       & 0.5-2 (3) & 32.7/40 & 21 & $<$0.0491 \\
          &       & 0.5-2 (4) & 70.3/40 & 100 & 0.0310$^+_-$0.0222 \\
          &       & 2-10 (1) & 23.9/40 & 2 & $<$0.3524 & $<$0.1471  \\
          &       & 2-10 (2) & 22.7/40 & 1 & $<$0.2235 \\
          &       & 2-10 (3) & 28.4/40 & 8 & $<$0.2553 \\
          &       & 2-10 (4) & 24.8/40 & 3 & $<$0.3271 \\
          & 13815 & 0.5-10 & 27.9/40 & 7 & $<$0.0351 \\
          &       & 0.5-2 & 25.1/40 & 3 & $<$0.0418 \\
          &       & 2-10 & 19.4/40 & 1 & $<$0.2777 \\
          & 13816 & 0.5-10 & 40.1/40 & 53 & $<$0.0391 \\
          &       & 0.5-2 & 40.5/40 & 55 & $<$0.0463 \\
          &       & 2-10 & 28.9/40 & 10 & $<$0.2729 \\ 
MARK\,273 & 809 & 0.5-10 & 71.8/40 & 100 & $<$0.0155 \\
          &       & 0.5-2 & 49.1/40 & 85 & $<$0.0287 \\
          &       & 2-10 & 60.2/40 & 98 & $<$0.0337 \\
Circinus & 9140 &    0.5-10 & 48.0/40 & 82 & $<$0.0019 \\             
         &            & 0.5-2 & 32.6/40 & 21 & $<$0.0075 \\
         &            &  2-10 &  45.5/40 & 74 & $<$0.0025 \\                 
NGC\,7319 & 7924 & 0.5-10 (1) & 30.3/40 & 13 & $<$0.0135 & $<$0.0093 \\
          &      & 0.5-10 (2) & 23.9/40 & 2 & $<$0.0127 \\
          &      & 0.5-2 (1) & 48.7/40 & 84 & $<$0.0644 & $<$0.0451 \\
          &      & 0.5-2 (2) & 69.6/40 & 99 & $<$0.0632 \\
          &      & 2-10 (1) & 37.2/40 & 40 & $<$0.0170 & $<$0.0116 \\
          &      & 2-10 (2) & 29.5/40 & 11 & $<$0.0158 \\  
          & 0021140201 & 0.5-10 & 22.8/31 & 16 & $<$0.0051 \\
          &            & 0.5-2 & 24.1/31 & 19 & $<$0.0121 \\
          &            & 2-10 & 17.3/31 & 2 & $<$0.0089 \\

\hline
\caption*{{\bf Notes.} (Col. 1) name, (Col. 2) obsID, (Col. 3) energy band in keV, (Cols. 4 and 5) $\chi^2/d.o.f$ and the probability of being variable in the 0.5-10.0 keV energy band of the total light curve, (Col. 6) normalized excess variance, $\sigma_{NXS}^2$, and (Col. 8) the mean value of the normalized excess variance, $<\sigma^2_{NXS}>$, for each light curve and energy band.}
\end{longtable}

\begin{longtable}{lcccccc|cc}
\caption[]{\label{ew} Classification of \emph{Compton}-thick objects.}  \\  \hline \hline
Name &  ObsID &  $\Gamma$  & EW  & $F_x/F_{[OIII]}$ & Ref.$^1$ & CT? & Classification & $\Gamma_{hard}$  \\
     &        &  & (keV)  & & $[OIII]$ &  & &   \\
(1) & (2) & (3) & (4) & (5) & (6) & (7) & (8) & (9)     \\ 
\hline 
\endfirsthead
\caption[]{(Cont.)} \\
\hline \hline
Name &  ObsID &  $\Gamma$  & EW  & $F_x/F_{[OIII]}$ & Ref.$^1$ & CT? & Classification  & $\Gamma_{hard}$ \\
     &        &  & (keV)  & & $[OIII]$ &  & &  \\
(1) & (2) & (3) & (4) & (5) & (6) & (7) & (8)  & (9)    \\ 
\hline 
\endhead
%\multicolumn{11}{r}{continued on next page\ldots}\\
\endfoot
\\
\endlastfoot
MARK348 & 0067540201 &  1.71 $_{ 1.64 }^{ 1.80 }$ & 0.06$ _{ 0.05 }^{ 0.07 }$   & 30.00 & 1  & \xmark & \emph{Compton}-thin \\
 & 0701180101 &  1.57 $_{ 1.33 }^{ 1.79 }$ & 0.19$ _{ 0.14 }^{ 0.25 }$  & 9.06 & & \xmark &  \\ \hline
NGC424 & 0002942301 &  1.03$ _{ 0.14 }^{ 1.90 }$ & 0.99$ _{ 0.74 }^{ 1.24 }$  & 1.84 & 2 & \cmark & \emph{Compton}-thick & 0.54$^{1.65}_{0.09}$ \\
 & 0550950101 &  0.16 $_{ 0.00 }^{ 0.37 }$ & 0.87$ _{ 0.82 }^{ 0.93 }$  & 1.81 & & \cmark &  \\
& 3146 &  0.00 $_{ 0.00 }^{ 1.84 }$ & 0.55$ _{ 0.32 }^{ 0.77 }$  & 1.46 & & \cmark & \\ \hline
MARK573 & 7745 &  0.18 $_{ 0.00 }^{ 2.71 }$ & 2.17$ _{ 1.52 }^{ 2.81 }$  & 0.49 & 3 & \cmark & \emph{Compton}-thick & 0.50$^{3.23}_{0.00}$  \\ 
 & 13124 &  0.88 $_{ 0.66 }^{ 2.08 }$ & 2.05$ _{ 1.49 }^{ 2.55 }$  & 0.41 &   & \cmark & \\ \hline
NGC788 & 0601740201 &  1.59$ _{ 0.97 }^{ 2.01 }$ & 0.43$ _{ 0.35 }^{ 0.49 }$  & 341.75 & 2  & \xmark & \emph{Compton}-thin \\
& 11680 &  1.07$ _{ 0.17 }^{ 2.61 }$ & 0.15$ _{ 0.07 }^{ 0.22 }$  & 284.26 & & \xmark & \\ \hline
ESO417-G06 & 0602560201 &  1.66 $_{ 1.27 }^{ 2.00 }$ & 0.18 $_{ 0.06 }^{ 0.30 }$  & 268.01 & 4  & \xmark & \emph{Compton}-thin \\
& 0602560301 &  1.73$ _{ 1.31 }^{ 2.17 }$ & 0.37 $_{ 0.22 }^{ 0.53 }$ & 268.01 & & \xmark & \\ \hline
MARK1066 & 0201770201 &  0.46 $_{ 0.00 }^{ 1.97 }$ & 0.60$ _{ 0.30 }^{ 0.89 }$  & 0.37 &  3 & \cmark & \emph{Compton}-thick & 0.31$^{0.76}_{0.00}$  \\ \hline
3C98.0 & 0064600101 &  1.31 $_{ 0.97 }^{ 1.67 }$ & $<$ 0.07 & 10.0 &  5  & \xmark &  \emph{Compton}-thin \\
& 0064600301 &  1.41$ _{ 0.22 }^{ 2.13 }$ & $<$ 0.38 &  5.89 & & \xmark & \\
& 10234 &  0.67$ _{ 0.09 }^{ 1.30 }$ & 0.16$ _{ 0.06 }^{ 0.27 }$ & 5.01 & & \xmark & \\ \hline
MARK3  & 0111220201 &  0.05 $_{ 0.00 }^{ 0.19 }$ & 0.55$ _{ 0.52 }^{ 0.58 }$  & 0.33 & 1 & \cmark & \emph{Compton}-thick & 0.42$^{0.62}_{0.23}$  \\
 & 0009220601 &  0.36 $_{ 0.00 }^{ 0.87 }$ & 0.67$ _{ 0.56 }^{ 0.79 }$  & 0.24 & & \cmark & \\
 & 0009220701 &  0.03 $_{ 0.00 }^{ 0.53 }$ & 0.60$ _{ 0.48 }^{ 0.73 }$  & 0.21 & & \cmark &  \\
 & 0009220901 &  0.02 $_{ 0.00 }^{ 1.05 }$ & 0.49$ _{ 0.27 }^{ 0.70 }$  & 0.15  & & \cmark & \\
& 0009220401 &  0.01 $_{ 0.00 }^{ 0.43 }$ & 0.79$ _{ 0.63 }^{ 0.96 }$   & 0.24  & & \cmark &  \\
 & 0009220501 &  0.03 $_{ 0.00 }^{ 0.48 }$ & 0.63$ _{ 0.52 }^{ 0.74 }$  & 0.20  & & \cmark & \\
 & 0009221601 &  0.01 $_{ 0.00 }^{ 0.98 }$ & 1.16$ _{ 0.85 }^{ 1.46 }$  & 0.24 & & \cmark &  \\ \hline
MARK1210 & 4875 &  1.31$ _{ 0.74 }^{ 1.95 }$ & 0.13$ _{ 0.05}^{ 0.20 }$ & 2.97 & 1 & \xmark & \emph{Compton}-thin \\
& 9264 &  0.89$ _{ 0.40 }^{ 1.42 }$ & 0.06$ _{ 0.01 }^{ 0.12 }$ & 4.49 & & \xmark & \\
& 9265 &  1.41$ _{ 0.82 }^{ 2.06 }$ & 0.12$ _{ 0.05 }^{ 0.19 }$ & 6.34  & & \xmark & \\
& 9266 &  2.03$ _{ 1.46 }^{ 2.86 }$ & 0.10$ _{ 0.03 }^{ 0.18 }$ &  2.41 & & \xmark & \\
& 9268 &  1.24$ _{ 0.51 }^{ 2.07 }$ & 0.16$ _{ 0.061 }^{ 0.25 }$ & 3.57  & & \xmark & \\ \hline
NGC3079 & 0110930201 &  1.58 $_{ 0.07 }^{ 2.47 }$ & $<$ 0.26  & 0.31 & 1 & \xmark & \emph{Compton}-thin \\ \hline
IC2560 & 0203890101 &  0.22 $_{ 0.00 }^{ 0.60 }$ & 1.95$ _{ 1.79 }^{ 2.09 }$  & 7.30 & 2  & \cmark & \emph{Compton}-thick & 0.69$^{1.04}_{0.30}$ \\
& 4908 &  $<$0.77 & 1.27$ _{ 1.04 }^{ 1.49 }$  & 4.94 & & \cmark & \\ \hline
NGC3393 & 12290 &  0.55 $_{ 0.00 }^{ 3.35 }$ & 1.85$ _{ 1.45 }^{ 2.29 }$  & 0.22 & 1  & \cmark & \emph{Compton}-thick & 0.42$^{0.00}_{1.76}$   \\
& 0140950601 &  0.95 $_{ 0.14 }^{ 1.69 }$ & 1.41$ _{ 1.00 }^{ 1.78 }$  & 0.18 & & \cmark &  \\ \hline
NGC4507 & 0006220201 &  1.73$ _{ 1.59 }^{ 1.86 }$ & 0.20$ _{ 0.19 }^{ 0.22 }$  & 33.08 & 1  & \xmark & \emph{Compton}-thin \\
 & 0653870201 &  1.44$ _{ 1.13 }^{ 1.72 }$ & 0.44$ _{ 0.39 }^{ 0.47 }$  & 16.58 & & \xmark & \\
 & 0653870301 &  1.34$ _{ 1.00 }^{ 1.65 }$ & 0.38$ _{ 0.34 }^{ 0.42 }$  & 18.60  & & \xmark & \\
 & 0653870401 &  0.91$ _{ 0.44 }^{ 1.14 }$ & 0.46$ _{ 0.42 }^{ 0.50 }$ & 16.20  & & \xmark & \\
 & 0653870501 &  1.01$ _{ 0.64 }^{ 1.34 }$ & 0.46$ _{ 0.41 }^{ 0.50 }$  & 17.36  & & \xmark & \\
 & 0653870601 &  0.91$ _{ 0.00 }^{ 2.16 }$ & 0.43$ _{ 0.28 }^{ 0.59 }$  & 10.22  & & \xmark & \\
& 12292 &  0.87$ _{ 0.54 }^{ 1.21 }$ & 0.36$ _{ 0.32 }^{ 0.40 }$ & 24.52   & & \xmark & \\ \hline
NGC4698 & 0651360401 &  0.91 $_{ 0.49 }^{ 1.50 }$ & $<$ 0.46 & 9.23 & 6  & \xmark & \emph{Compton}-thin \\ \hline
NGC5194 & 13812 &  0.04 $_{ 0.00 }^{ 2.21 }$ & 2.75$ _{ 2.27 }^{ 3.26 }$  & 1.47 & 1  & \cmark & \emph{Compton}-thick & 0.57$^{1.74}_{0.00}$  \\
 & 13813 &  0.02 $_{ 0.00 }^{ 2.41 }$ & 4.16$ _{ 3.43 }^{ 4.88 }$  & 0.13 & & \cmark &  \\
 & 13814 &  0.12 $_{ 0.00 }^{ 3.25 }$ & 4.41$ _{ 3.73 }^{ 5.14 }$  & 0.17 & & \cmark &  \\
& 0112840201 &  2.16 $_{ 1.29 }^{ 3.19 }$ & 0.99$ _{ 0.75 }^{ 1.23 }$  & 0.27 & & \cmark &  \\ \hline
MARK268 & 0554500701 &  1.80 $_{ 1.11 }^{ 3.43 }$ & $<$ 0.17   & 462.73 & 7 & \xmark & \emph{Compton}-thin \\
 & 0554501101 &  1.71 $_{ 1.32 }^{ 2.18 }$ & 0.26$ _{ 0.18 }^{ 0.33 }$  & 351.01 & & \xmark &  \\ \hline
MARK273 & 0101640401 &  0.01 $_{ 0.00 }^{ 0.95 }$ & 0.87$ _{ 0.65 }^{ 1.12 }$  & 2.75 & 1  & \cmark & Changing-look? \\
& 809 &  1.69$_{ 0.78 }^{ 2.77 }$ & 0.21$ _{ 0.10 }^{ 0.32 }$  & 4.67 & & \xmark & \\ \hline
Circinus  & 365 &  0.00 $_{ 0.00 }^{ 0.28 }$ &  2.38$ _{ 2.11 }^{ 2.65 }$  & 0.39 & 1 & \cmark & \emph{Compton}-thick & 0.07$^{0.17}_{0.00}$  \\
 & 9140 &  0.12$ _{ 0.00 }^{ 0.33 }$ & 1.90$ _{ 1.83 }^{ 1.97 }$  & 0.31 & & \cmark & \\
 & 10937 &  0.00$ _{ 0.00 }^{ 0.10 }$ & 1.73$ _{ 1.63 }^{ 1.84 }$  & 0.39  & & \cmark & \\
& 0111240101 &  1.07 $_{ 1.01 }^{ 1.13 }$ & 1.54$ _{ 1.51 }^{ 1.56 }$  & 0.35  & & \cmark & \\
 & 0656580601 &  0.49 $_{ 0.41 }^{ 0.60 }$ & 1.50$ _{ 1.47 }^{ 1.54 }$  & 0.46  & & \cmark & \\ \hline
NGC5643 & 0601420101 &  0.04 $_{ 0.00 }^{ 0.61 }$ & 1.37$ _{ 1.18 }^{ 1.56 }$ & 0.29 & 1 & \cmark & \emph{Compton}-thick & 0.84$^{1.48}_{0.09}$  \\ 
 & 0140950101 &  0.01 $_{ 0.00 }^{ 0.71 }$ & 1.37$ _{ 1.04 }^{ 1.69 }$  & 0.37 &   & \cmark & \\ \hline
MARK477 & 0651100301 &   0.93$ _{ 0.36 }^{ 1.53 }$ & 0.32$ _{ 0.22 }^{ 0.43 }$ & 0.32 & 1  & \cmark & \emph{Compton}-thick & 1.02$^{1.66}_{0.59}$  \\
 & 0651100401 &  0.88 $_{ 0.30 }^{ 1.48 }$ & 0.13$ _{ 0.05 }^{ 0.21 }$ & 0.45 & & \cmark & \\ \hline
IC4518A & 0401790901 &  1.71$ _{ 1.29 }^{ 2.16 }$ & 0.33$ _{ 0.25 }^{ 0.42 }$  & - & - & \xmark & \emph{Compton}-thin \\
 & 0406410101 &  1.27 $_{ 0.94 }^{ 1.60 }$ & 0.45$ _{ 0.38 }^{ 0.53 }$  & & & \xmark &\\ \hline
ESO138-G01 & 0405380201 &  0.92 $_{ 0.52 }^{ 1.33 }$ & 0.90$ _{ 0.78 }^{ 1.01 }$  & 23.10 & 2  & \cmark & \emph{Compton}-thick & 1.04$^{1.38}_{0.73}$  \\
 & 0690580101 &  0.97 $_{ 0.58 }^{ 1.64 }$ & 1.31$ _{ 1.10 }^{ 1.48 }$  & 19.67 & & \cmark & \\ \hline
NGC6300 & 10292 &  0.57 $_{ 0.23 }^{ 0.95 }$ & $<$ 0.08  &  361.27 &  2 & \xmark & \emph{Compton}-thin \\
 & 10293 &  1.17 $_{ 0.90 }^{ 1.69 }$ & $<$ 0.08   & 444.46 & & \xmark & \\
& 0059770101 &  1.55 $_{ 1.03 }^{ 2.11 }$ & 0.23$ _{ 0.13 }^{ 0.34 }$  & 12.24 & & \xmark & \\ \hline
NGC7172 & 0147920601 &  1.61 $_{ 1.50 }^{ 1.73 }$ & 0.12$ _{ 0.09 }^{ 0.14 }$  & 853.54 & 1  & \xmark & \emph{Compton}-thin  \\
 & 0202860101 &  1.58 $_{ 1.49 }^{ 1.67 }$ & 0.09$ _{ 0.07 }^{ 0.11 }$  & 834.12 & & \xmark & \\
 & 0414580101 &  1.71 $_{ 1.66 }^{ 1.76 }$ & 0.08$ _{ 0.07 }^{ 0.09 }$  & 1742.72  & & \xmark & \\ \hline
NGC7212 & 0200430201 &  0.00 $_{ 0.00 }^{ 0.26 }$ & 0.79$ _{ 0.59 }^{ 0.99 }$  & 3.83 & 2 & \cmark &  \emph{Compton}-thick & 0.38$^{2.19}_{0.00}$   \\
& 4078 &  0.00 $_{ 0.00 }^{ 2.62 }$ & 1.00$_{ 0.61 }^{ 1.39 }$  & 3.04 & & \cmark & \\    \hline    
NGC7319 & 789 &  1.43$ _{ 0.84 }^{ 2.67 }$ & 0.23$ _{ 0.12 }^{ 0.34 }$  & 38.69 & 1  & \xmark & Changing-look? \\
 & 7924 &  1.89 $_{ 1.52 }^{ 2.39 }$ & 0.23$ _{ 0.18 }^{ 0.29 }$  & 82.73 & & \xmark & \\
& 0021140201 &  0.23 $_{ 0.00 }^{ 0.65 }$ & 0.83$ _{ 0.73 }^{ 0.93 }$  & 22.26 & & \cmark & \\ \hline
\caption*{{\bf Notes.} (Col. 1) name, (Col. 2) obsID, (Cols. 3 and 4) index of the power law and the equivalent width of the FeK$\alpha$ line from the spectral fit (PL model) in the 3--10 keV energy band, (Col. 5) ratio between the individual hard X-ray luminosity (from Table \ref{lumincorrSey}) and the extinction corrected [O III] fluxes, (Col. 6) references for the measure of $F_{[OIII]}$, (Col. 7) classification from the individual observation, (Col. 8) classification of the object, and (Col. 9) slope of the power law at hard energies for \emph{Compton}-thick candidates from the simultaneous analysis (see Sect. \ref{thick}).
References: (1) \cite{bassani1999}; (2)  \cite{gu2006}; (3) \cite{bian2007}; (4) \cite{kraemer2011}; (5) \cite{noguchi2009}; (6) \cite{panessabassani2002}; and (7) \cite{koski1978}. }
\end{longtable}

\newpage

\twocolumn

\normalsize

\section{\label{indivnotes} Notes and comparisons with previous results for individual objects}

In this appendix we discuss the general characteristics of the galaxies in our sample at different wavelenghts, as well as comparisons with previous variability studies. We recall that long-term UV variability and short-term X-ray variations were studied only for some sources (six and ten sources, see Tables \ref{properties} and \ref{estcurvasSey}, respectively), so comparisons are only made in those cases. For the remaining objects, results from other authors are mentioned, when available.

\subsection{MARK\,348}

MARK\,348, also called NGC\,262, is an interacting galaxy \citep[with NGC\,266,][]{pogge1993}.
It was optically classified as a type 2 Seyfert \citep{koski1978}, while it shows broad lines in polarized light \citep{millergoodrich1990}. It shows a spiral nuclear structure (see \emph{HST} image in Appendix \ref{multiimages}).
\emph{VLBI} observations showed a compact radio core and jets structure at radio frequencies, and revealed variations in timescales from months to years at 6 and 21 cm \citep{neff1983}.
The \emph{XMM}--Newton image shows that the soft X-ray emission is very weak in this object (see Appendix \ref{multiimages}), which was classified as a \emph{Compton}-thin object \citep[e.g.,][]{awaki2006}.

This galaxy was observed twice with \emph{XMM}--Newton in 2002 and 2013, and once with \emph{Chandra} in 2010.
Recently, \cite{marchese2014} compared \emph{XMM}--Newton and \emph{Suzaku} data from 2002 and 2008. They fitted the data with a power law component transmitted throught three abosrbers (one neutral and two ionized), obtaining intrinsic luminosities of log(L(2--10 keV)) = 43.50 and 43.51, respectively. 
They reported variations attributed to changes in the column density of the neutral and one of the ionized absorbers, together with a variation of the ionization level of the same absorber, in timescales of months. They did not report variations in $\Gamma$ and/or the continuum of the power law.
Variations in the absorbing material in timescales of weeks/months were also reported by \cite{smith2001} using \emph{RXTE} data from 1996-97, but accompained with continuum variations in timescales of $\sim$ 1 day. They obtained luminosities in the range log(L(2--10 keV)) = [42.90-43.53]. These results were in agreement with those later reported by \cite{akylas2002}, who analyzed the same observations plus 25 \emph{RXTE} observations.
Our analysis shows that variations between the two \emph{XMM}--Newton observations are due to changes in the nuclear continuum, but variations of the absorbing material are not required. These differences may be related to the different instruments involved in the analyses.

\cite{awaki2006} did not find short term variations from the analysis of the \emph{XMM}--Newton data from 2002. 

In the 14--195 keV energy band, \cite{soldi2013} estimated a variability amplitude of 25[22-28]\% using data from the \emph{Swift}/BAT 58-month survey.

\subsection{NGC\,424}

NGC\,424 was optically classified as a type 2 Seyfert galaxy \citep{smith1975}, and broad lines have been detected in polarized light \citep{moran2000}. At radio frequencies, it was observed with \emph{VLA} at 6 and 20 cm, showing an extended structure \citep{ulvestadwilson1989}. A possible mid-IR variability was reported by \cite{honig2012} between 2007 and 2009, but it could also be due to an ``observational inaccuracy''.
At X-rays, it is a \emph{Compton}-thick source \citep[][]{balokovic2014}.

It was observed twice with \emph{XMM}--Newton in 2008 and 2011, and once with \emph{Chandra} in 2002.
\cite{matt2003} studied \emph{XMM}--Newton and \emph{Chandra} data from 2001 and 2002. Both spectra were fitted with a model consisting on two power laws, a cold reflection component (PEXRAV), and narrow Gaussian lines. They reported the same luminosity for the two spectra, log(L(2-10 keV)) = 41.68, indicating no variations.
\cite{lamassa2011} studied the same data set. They found no differences between the spectra and therefore fitted the data simultaneously with a simpler model, the 2PL. They estimated an intrinsic luminosity of log(L(2-10 keV)) = 41.56[41.39-41.75]. With the same data set we did not find variations and obtained similar hard X-ray luminosities (41.85[41.79-41.92]).

We did not find short-term variations from the \emph{XMM}--Newton light curve from 2008.

\subsection{MARK\,573}

MARK\,573 (also called UCG\,1214) is a double-barred galaxy that shows dust lanes \citep[][see also Appendix \ref{multiimages}]{martini2001}.
It was optically classified as a type 2 Seyfert galaxy \citep{osterbrock1993}.
Observations at 6 cm with \emph{VLA} showed a triple radio source \citep{ulvestadwilson1984}.
A point-like source is observed at hard X-rays, while extended emission can be observed at soft X-rays, aligned with the bars (see Appendix \ref{multiimages}). 
It was classified as a \emph{Compton}-thick candidate \citep{guainazzi2005a,bianchi2010,severgnini2012}.

This galaxy was observed four times with \emph{Chandra} between 2006 and 2010, and once with \emph{XMM}--Newton in 2004. \cite{bianchi2010} analysed the \emph{Chandra} data from 2006 and did not report flux variations when they compared their results with the analysis performed by \cite{guainazzi2005a} of the \emph{XMM}--Newton spectrum from 2004. 
\cite{paggi2012} studied the four \emph{Chandra} observations, and fitted the nuclear spectrum with a combination of a two phased photoionized plasma plus a Compton reflection component (PEXRAV), reporting soft X-ray flux variations at 4$\sigma$ of confidence level that they attributed to intrinsic variations of the source. We did not detect variations for this source, the difference most probably because we did not use two of these observations since they are affected by a pileup fraction larger than 10\%.

\cite{ramosalmeida2008} analyzed the \emph{XMM}--Newton light curve and found variations of $\sim$ 300 s. They argued that this is an obscured narrow-line Seyfert 1 galaxy instead of a type 2 Seyfert, based on near-IR data. We analysed two \emph{Chandra} light curves but variations were not found.

\subsection{NGC\,788}

This galaxy was optically classified as a type 2 Seyfert by \cite{huchra1982}. 
A radio counterpart was detected with \emph{VLA} data \citep{nagar1999}.
At X-rays, it was classified as a \emph{Compton}-thin candidate using \emph{ASCA} data \citep{derosa2012}, and shows a point-like source in the 4.5-8 keV energy band (see Appendix \ref{multiimages}). 

It was observed once with \emph{Chandra} in 2009 and once with \emph{XMM}--Newton in 2010.
Long term variability analyses of this source were not found in the literature. We did not find variations between the observations.

Variations of this source in the 14--195 keV energy band were studied by \cite{soldi2013} using data from the \emph{Swift}/BAT 58-month survey. They reported an amplitude of the intrinsic variability of 15[11-19]\%.

\subsection{ESO\,417-G06}

This galaxy was optically classified as a type 2 Seyfert galaxy \citep{maia2003}.
A radio counterpart was observed with \emph{VLA} data \citep{nagar1999}. It was classified as a \emph{Compton}-thin candidate \citep{trippe2011}. 

This galaxy was observed twice with \emph{XMM}--Newton in 2009. Long-term variability studies were not found in the literature. We found spectral variations due to changes in the absorber at hard X-ray energies.

\cite{trippe2011} reported short-term variations of a factor of $\sim$2 in the count rate in the light curves from \emph{Swift}/BAT during the 22 month survey.

\subsection{MARK\,1066}

MARK\,1066 is an early-type spiral galaxy \citep{afanasev1981} showing a double nucleus \citep{gimeno2004}. It was optically classified as a type 2 Seyfert by \cite{goodrichosterbrock1983}, and broad lines were not detected in polarized light \citep{gu2002}. A radio counterpart showing a jet was found by \cite{ulvestadwilson1989}. 
At X-rays, extended soft emission can be observed, aligned with a nuclear spiral structure observed at optical frequencies, also aligned with the IR emission (see Appendix \ref{multiimages}).
\cite{levenson2001} found this to be a heavily obscured AGN, with $N_H > 10^{24} cm^{-2}$ and an equivalent width of the Fe line $\sim$ 3 keV using \emph{ROSAT} and \emph{ASCA} data, i.e., it was classified as a \emph{Compton}-thick candidate.

The galaxy was observed once with \emph{Chandra} in 2003 and once with \emph{XMM}--Newton in 2005.
Variability studies of this object were not found in the literature. We did not find X-ray variations either.

\subsection{3C\,98.0}

Using the optical line measurements in \cite{costero1977}, it can be optically classified as a type 2 Seyfert (see an optical spectrum in Appendix \ref{multiimages}).
A nuclear core plus jets structure was observed at radio frequencies with \emph{VLA} \citep{leahy1997}. 

3C\,98.0 was observed twice with \emph{XMM}--Newton in 2002 and 2003 and once with \emph{Chandra} in 2008.
\cite{isobe2005} studied the two \emph{XMM}--Newton data, and fitted its spectra with a thermal plus a power law model, reporting X-ray luminosities of log(L(2--10 keV)) = 42.90[42.88-42.93] and 42.66[42.60-42.71], respectively, indicating flux variability. These measurements agree well with ours, where variations due to the nuclear continuum were found. 

\cite{awaki2006} studied short term variations of the \emph{XMM}--Newton observation from 2003 and calculated a normalized excess variance of $\sigma_{NXS}^2=36[1-62] \times 10^{-3} $. We did not find short-term variations from one \emph{Chandra} light curve, where upper limits of the $\sigma_{NXS}^2$ were calculated.

We did not find long-term UV variations in the UVW1 filter.

\subsection{MARK\,3}

It was optically classified as a type 2 Seyfert galaxy \citep[][see an optical spectrum in Appendix \ref{multiimages}]{khachikian1974}. Broad lines have been found in polarized light \citep{millergoodrich1990}. 
A high resolution image at 2 cm with \emph{VLA} data shows a double nucleus at radio frequencies \citep{ulvestadwilson1984}.
This galaxy shows extended soft X-ray emission perpendicular to the IR emission and a point-like source at hard X-rays (see Appendix \ref{multiimages}). It is also a \emph{Compton}-thick source \citep{bassani1999,goulding2012}, with a column density of $1.1 \times 10^{24} cm^{-2}$ measured with \emph{BeppoSAX} \citep{cappi1999}.

It was observed 11 times with \emph{XMM}--Newton between 2000 and 2012, and once with \emph{Chandra} in 2012. \cite{bianchi2005} reported variations of the normalization of the absorbed power law when comparing the \emph{XMM}--Newton from 2001 with \emph{Chandra} and \emph{BeppoSAX} data.
\cite{guainazzi2012} studied the X-ray variability of this nucleus along 12 years of observations with \emph{Chandra}, \emph{XMM}--Newton, \emph{Suzaku}, and \emph{Swift} satellites. Their analysis was performed in the 4-10 keV energy band. To estimate the luminosities, they fitted a pure reflection model plus Gaussian lines to the spectra individually, and reported a variability dynamical range larger than 70\%. They also used alternative models to fit the data; variations found independently of the model used. They estimated the shortest variability timescale to be $\sim$ 64 days from the measurement between two statistically inconsistent measures.
From our analysis, variations due to the nuclear continuum were found, with an upper limit of the variability timescale of $\sim$ five months, thus in agreement with the results presented by \cite{guainazzi2012}.

Short-term variations from \emph{XMM}--Newton data were not found neither by \cite{omairavaughan2012} nor by \cite{cappi2006} from light curves from 2000 and 2001, respectively.

\cite{soldi2013} reported an amplitude of the intrinic variability of 35[26-46]\% in the 14--195 keV energy band using data from the \emph{Swift}/BAT 58-month survey.

\subsection{MARK\,1210}

This galaxy, also called the Phoenix galaxy or UGC\,4203, was optically classified as a type 2 Seyfert by \cite{dessauges2000}. Broad lines have been observed in polarized light using spectropolarimetric data \citep{tran1992, tran1995}. The \emph{HST} image shows a nuclear spiral structure (see Appendix~\ref{multiimages}).
A very compact radio counterpart was found with \emph{VLA} at 3.5 cm, with no evidence of jet structure \citep{falcke1998}.
At X-rays, a point like source is observed in the 4.5-8.0 keV energy band (see Appendix~\ref{multiimages}). It was classified as a \emph{Compton}-thick candidate by \cite{bassani1999}. Furthermore, \cite{guainazzi2002} classified this galaxy as a changing look AGN because transitions from \emph{Compton}-thick (\emph{ASCA} data) to \emph{Compton}-thin (\emph{XMM}--Newton data) were found.

MARK\,1210 was observed with \emph{Chandra} six times between 2004 and 2008, and once with \emph{XMM}--Newton in 2001.
\cite{matt2009} used \emph{Suzaku} data from 2007 to study this source (caught in the \emph{Compton}-thin state), and compared with previous observations from \emph{ASCA} and \emph{XMM}--Newton. They fitted the spectra with a power law, a \emph{Compton} reflection, and a thermal (MEKAL) components, and found a change in the absorber, which was about a factor of 2 higher in \emph{Suzaku} data. They obtained intrinsic X-ray luminosities of log(L(2--10 keV)) = 42.87 and 43.04 for \emph{Suzaku} and \emph{XMM}--Newton data.
\cite{risaliti2010} simultaneously fitted the five \emph{Chandra} observations from 2008 using a model consisting on a doubled temperature plus power law to account for the soft energies, an absorbed power law, and a constant cold reflection component (PEXRAV). They concluded that variations are found in both the intrinsic flux and in the absorbing column density. They reported a variability time scale of $\sim$ 15 days, whereby they estimated the physical parameters of the absorbing material, concluding that they are typical of the broad line region (BLR). Their result agrees well with ours. 

\cite{awaki2006} studied short term variations from the \emph{XMM}--Newton data and found $\sigma_{NXS}^2=5.5[0.0-11.0] \times 10^{-3} $. 

\cite{soldi2013} used data from the \emph{Swift}/BAT 58-month survey to account for the variability amplitude ($S_v = 24[15-32] \%$) in the 14--195 keV energy band.

\subsection{NGC\,3079}

This galaxy was optically classified as a type 2 Seyfert \cite[][based on the spectra presented in Appendix \ref{multiimages}]{ho1997}. Broad lines were not detected in polarized light \citep{gu2002}. The \emph{HST} image shows dust lanes (Appendix \ref{multiimages}).
A water maser and parsec-scale jets were observed at radio frequencies with \emph{VLBI} \citep{trotter1998}. The X-ray image in the 0.6-0.9 keV energy band shows strong diffuse emission, while a point-like source is detected in the 4.5-8.0 keV energy band (see Appendix~\ref{multiimages}). It has been classified as a \emph{Compton}-thick object with \emph{BeppoSAX} data \citep[$N_H = 10^{25} cm^{-2}$,][]{comastri2004} and evidences were found also at lower energies \citep{cappi2006,akylas2009,brightman2011}. 

It was observed once with \emph{Chandra} and once with \emph{XMM}--Newton, both in 2001.
We did not find variability studies of this source in the literature. We did not study its variability because the extranuclear emission in \emph{Chandra} data was too high to properly compare \emph{XMM}--Newton and \emph{Chandra} observations.

It is worth noting that NGC\,3079 is classified as a \emph{Compton}-thin candidate in this work but it has been classified as a \emph{Compton}-thick candidate by \cite{cappi2006} using the same \emph{XMM}--Newton observation.
Since these data have the lowest signal-to-noise ratio, this mismatch is most probably due to a problem related with the sensitivity of the data, because we used only data from the pn detector, while they combined pn, MOS1, and MOS2 data in their study, i.e., \cite{cappi2006} data have higher signal-to noise. We notice that cross-calibration uncertainties between pn and MOS cameras may add systematic to statistical uncertainties that can conceil possible intrinsic variability due to large error bars \citep{kirsch2004, ishida2011, tsujimoto2011}, thus preventing us from doing a variability analysis.

\subsection{IC\,2560}

This galaxy was optically classified as a type 2 Seyfert \citep[][see an optical spectrum in Appendix \ref{multiimages}]{fairall1986}. At hard X-rays it shows a point-like source (see Appendix \ref{multiimages}). It was classified as a \emph{Compton}-thick object \citep{balokovic2014}.

It was observed once with \emph{XMM}--Newton in 2003 and once with \emph{Chandra} in 2004.
Variability studies were not found in the literature. We do not report X-ray variations for this source, neither at short nor at long term.

\subsection{NGC\,3393}

NGC\,3393 was optically classified as a type 2 Seyfert \citep[][see an optical spectrum in Appendix \ref{multiimages}]{diaz1988}. A radio counterpart was found using \emph{VLA} data, the galaxy showing a double structure \citep{morganti1999}. The \emph{HST} image shows a nuclear spiral structure aligned with the soft X-ray emission, where the spiral structure can also be appreciated; this emission is perpendicular to the disc emission, observed at optical wavelenghts and aligned with the IR emission (see Appendix \ref{multiimages}).
A point-like source is observed at hard X-rays (see Appendix \ref{multiimages}). It is a \emph{Compton}--thick object observed by \emph{BeppoSAX} \citep[$N_H > 10^{25} cm^{-2}$,][]{comastri2004}. 

This galaxy was observed once with \emph{XMM}--Newton in 2003 and six times with \emph{Chandra} between 2004 and 2012. Variability studies were not found in the literature. We did not find X-ray variations, neither at short nor at long term.

\subsection{NGC\,4507}

The nucleus of this galaxy was optically classified as a type 2 Seyfert \citep[][see an optical spectrum in Appendix \ref{multiimages}]{corbett2002}.
Broad lines have been detected in polarized light \citep{moran2000}. A radio counterpart was observed with \emph{VLA} data \citep{morganti1999}.
At X-rays, it shows a point-like source in the hard energy band (see Appendix \ref{multiimages}), and it is a \emph{Compton}-thin source \citep{bassani1999, braito2013}. 

NGC\,4507 was observed six times with \emph{XMM}--Newton between 2001 and 2010, and once with \emph{Chandra} in 2010.
\cite{matt2004} studied \emph{Chandra} and \emph{XMM}--Newton data from 2001. They fitted the \emph{XMM}--Newton spectrum with a composite of two power laws, a Compton reflection component (PEXRAV), plus ten Gaussian lines, and the \emph{Chandra} spectrum with a power law plus a Gaussian line (only in the 4--8 keV spectral range). They found that the luminosity of the \emph{Chandra} data was about twice that of \emph{XMM}--Newton.
\cite{marinucci2013} studied five observations from \emph{XMM}--Newton in 2010. They fitted the spectra with two photoionised phases using Cloudy, a thermal component, an absorbed power law, and a reflection component. They reported variations of the absorber in timescales between 1.5--4 months.
\cite{braito2013} studied \emph{XMM}--Newton, \emph{Suzaku}, and \emph{BeppoSAX} data spanning $\sim$ 10 years to study the X-ray variability of the nucleus. They fitted the spectra with the model that best represents the \emph{Suzaku} data, composed by two power laws, a PEXRAV component, and eight Gaussian lines, and found variations mainly due to absorption but also due to the intensity of the continuum level. They also fitted the spectra with the {\sc mytorus} model\footnote{www.mytorus.com}, and obtained similar results, although the continuum varied less.
We found variations in the absorber and the normalization of the power law, in agreement with the results by \cite{braito2013}.

We did not find short-term variations from the analysis of one \emph{XMM}--Newton and another \emph{Chandra} light curves.

\cite{soldi2013} reported an amplitude of the intrinsic variability of 20[16-24]\% in the 14--195 keV energy band using data from the \emph{Swift}/BAT 58-month survey.

\subsection{NGC\,4698}

This galaxy was optically classified as a type 2 Seyfert \citep[][see their spectra in Appendix \ref{multiimages}]{ho1997}. \cite{omaira2009a} classified it as a LINER, but \cite{bianchi2012} re-confirmed the type 2 Seyfert classification using optical observations with the \emph{NOT}/ALFOSC/Gr7.
A radio counterpart was found by \cite{houlvestad2001} at 6 cm with \emph{VLA} data. \cite{georgantopoulos2003} stated that this is an atypical Seyfert 2 galaxy because it showed no absoption and lacks the broad line region. The \emph{Chandra} image revealed point-like sources around the nucleus which can be ultraluminous X-ray sources (ULX), the closest located at $\sim$ 30$\arcsec$ from the nucleus.
At X-rays, \cite{omaira2009a} classified it as an AGN candidate, and \cite{bianchi2012}, based on the log($L_X/L_{[OIII]}$) ratio, classified it as a \emph{Compton}-thick candidate.

This galaxy was observed twice with \emph{XMM}--Newton in 2001 and 2010, and once with \emph{Chandra} in 2010. \cite{bianchi2012} compared the \emph{XMM}--Newton spectra and did not find spectral variations, in agreement with the results reported by us.

We did not find UV variations in the UVM2 filter.

\subsection{NGC\,5194}

NGC\,5194, also known as M\,51, is interacting with NGC\,5195. Optical and radio observations show extended emissions to the north and south of the nucleus, resulting from outflows generated by the nuclear activity \citep{ford1985}. The extended emission can be observed at soft X-ray energies (top-left image in Appendix~\ref{multiimages}). Moreover, the \emph{HST} image shows a dusty nuclear spiral structure that can also be observed at IR frequencies (see Appendix~\ref{multiimages}).
This galaxy was optically classified as a type 2 Seyfert \citep[][see their optical spectra in Appendix~\ref{multiimages}]{ho1997}. Broad lines were not detected in polarized light \citep{gu2002}.
A point-like source is detected at hard X-ray energies (see Appendix~\ref{multiimages}). Around the nucleus, it shows at least seven ultraluminous X-ray sources (ULX); the nearest one located at $\sim$ 28$\arcsec$ from the nucleus \citep{dewangan2005}.
It was classified as a \emph{Compton}-thick source using \emph{BeppoSAX} data, with $N_H = 5.6 \times 10^{24} cm^{-2}$ \citep[][see also \citealt{terashimawilson2001,dewangan2005,cappi2006}]{comastri2004}.

This galaxy was observed 10 times with \emph{Chandra} between 2000 and 2012, and six times with \emph{XMM}--Newton between 2003 and 2011.
\cite{lamassa2011} studied three \emph{Chandra} observations between 2000 and 2003. They simultaneously fitted these spectra with the ME2PL model, with spectral values in very good agreement with our SMF0 fitting, and estimated a luminosity of log(L(2-10 keV))=38.95[38.42,39.45]. They did not report variability between the observations. This result is in agreement with ours.

\cite{fukazawa2001} did not find short-term variability from \emph{BeppoSAX} data. We studied six \emph{Chandra} light curves and did not find short term variations either.

UV variations were not detected from the UVW2 and UVM2 filters, but variations were found in the UVW1 filter. However, since this is a \emph{Compton}-thick source, variations are not expected, so it is most probably that the UV emission does not come from the nucleus. Therefore the variations might be related with, e.g., circumnuclear star formation.

\subsection{MARK\,268}

This galaxy was optically classified as a type 2 Seyfert by \cite{komossa1997}. A radio counterpart was detected with \emph{VLA} data at 6 cm, with a weaker component 1.1 kpc away from the nucleus \citep{ulvestadwilson1984}. \emph{XMM}-Newton data show a compact source at hard X-rays (see Appendix \ref{multiimages}) .

It was observed twice with \emph{XMM}--Newton in 2008.
Variability studies were not found in the literature. We did not find variations, but we notice that observations were obtained separated by only two days.

UV variations are not found from the UVM2 and the UVW1 filters.

\subsection{MARK\,273}

Also called UGC\,8696, this galaxy is an ultraluminous infrared galaxy with a double nucleus that was optically classified as a LINER \citep{veilleux1995}, but later re-classified as a type 2 Seyfert from better S/N data \citep{kim1998}. Optical spectra are presented in Appendix~\ref{multiimages}, together with an \emph{HST} image which shows dust lanes. \emph{VLBA} observations showed a radio counterpart \citep[e.g.,][]{carilli2000}. Extended emission to the south is observed at soft X-rays, while it shows a point-like source at hard energies (Appendix \ref{multiimages}). It was classified as a \emph{Compton}-thick candidate \citep{teng2009}.

It was observed once with \emph{Chandra} in 2000, and five times with \emph{XMM}--Newton between 2002 and 2013.
\cite{balestra2005} fitted the \emph{Chandra} and \emph{XMM}--Newton spectra with a composite of three thermal plus an absorbed PL components and found similar spectral parameters, except in the value of the column densities (41[35-47] and 69[50-85] $\times 10^{22} cm^{-2}$, respectively). This result is compatible with ours, with $N_{H2}$ being responsible for the observed variations. In the same sense, \cite{teng2009} studied \emph{Suzaku} data from 2006 and found spectral variations when comparing with \emph{Chandra} and \emph{XMM}--Newton data. They attributed the changes to the covering fraction of the absorber.

We did not find short-term variations from the \emph{Chandra} light curve, neither UV variations from the UVW1 filter.

\subsection{\label{circi}Circinus}

It was optically classified as a type 2 Seyfert galaxy \citep{oliva1994} and it shows broad lines in polarized light \citep{oliva1998}. The \emph{HST} image shows dust lanes (Appendix \ref{multiimages}).
\emph{ATCA} observations show a radio counterpart, a water maser, and large radio lobes \citep{elmouttie1998}.
Circinus is a \emph{Compton}-thick source \citep{bassani1999}, which in fact was observed by \emph{BeppoSAX} \citep[$N_H = 4.3 \times 10^{24} cm^{-2}$,][]{matt1999}.

This galaxy was observed eight times with \emph{Chandra} between 2000 and 2010, and twice with \emph{XMM}--Newton in 2001 and 2014.
The most comprehensive analysis of this source has recently been performed by \cite{arevalo2014b}, who analysed 26 observations from \emph{NuSTAR}, \emph{Chandra}, \emph{XMM}--Newton, \emph{Swift}, \emph{Suzaku}, and \emph{BeppoSAX} satellites spanning 15 years and the energy range 2--79 keV. They used different models to fit the data, based on PEXMON, MyTorus, and Torus models (in XSPEC). Since different appertures were used for the analysis, they decontaminated the extranuclear emission. They concluded that the nucleus did not show variations, in agreement with our result when comparing \emph{Chandra} data.
Moreover, \cite{arevalo2014b} found that extranuclear sources included in the larger apertures showed variations (an ultraluminous X-ray source and a supernova remnant), also in agreement with our results when comparing \emph{XMM}--Newton data, where the extranuclear sources were included, and
we found variations in both the normalizations at soft and hard energies. 

We analysed one \emph{Chandra} light curves, but variations were not detected.

The analysis of light curves from the \emph{Swift}/BAT 58-month survey by \cite{soldi2013} showed a small variability amplitude of 11[10-12]\% in the 14--195 keV energy band.

\subsection{NGC\,5643}

This galaxy was optically classified as a type 2 Seyfert \citep[][see an optical spectrum in Appendix \ref{multiimages}]{philips1983}, and broad lines were not detected in polarized light \citep{gu2002}. The \emph{HST} image shows a nuclear spiral structure (see Appendix \ref{multiimages}).
\emph{VLA} data show a nuclear counterpart alongside fainter features extending to the east and west at radio frequencies \citep{morris1985}.
The \emph{XMM}--Newton image shows a compact source at hard X-ray energies. This is a \emph{Compton}-thick object observed with \emph{BeppoSAX} \citep[$N_H > 10^{25} cm^{-2}$,][]{comastri2004}.

It was observed twice with \emph{XMM}--Newton in 2003 and 2009, and once with \emph{Chandra} in 2004. \cite{matt2013} analyzed the two observations from \emph{XMM}--Newton, who found that the spectra are well reproduced by reflection from warm and cold matter. The spectral parameters were consistent with the same values for the two observations. Thus, variations are not observed. These results agree well with ours, where variations are not found.

\subsection{MARK\,477}

This object was classified as a type 2 Seyfert \citep{veron1997}, and broad lines have been detected in polarized light \citep{tran1992,tran1995}. The \emph{HST} image reveals a structure around the nucleus, that could be a spiral or a circumnuclear ring (see Appendix \ref{multiimages}).
A nuclear counterpart was found at 6 cm using \emph{VLA} data \citep{ulvestadwilson1984}.
It was classified as a \emph{Compton}-thick candidate \citep{bassani1999}.

The source was observed twice with \emph{ASCA} in December 1995; variations were not found when fitting a scattered power law plus a narrow line \citep{levenson2001}.

It was observed twice with \emph{XMM}--Newton in 2010.
We did not find variations between these observations.

\cite{kinney1991} studied UV variability of this source with \emph{HST}, but variations were not found.
We did not find UV flux variations from the UVW1 filter.

\subsection{IC\,4518A}

This galaxy was optically classified as a type 2 Seyfert galaxy \citep{zaw2009}. The \emph{2MASS} image shows two interacting galaxies (see Appendix \ref{multiimages}).
It is a \emph{Compton}-thin source \citep{bassani1999, derosa2008}.

It was observed twice with \emph{XMM}--Newton in 2006. Variability analyses were not found in the literature. However, comparing the luminosities obtained by \cite{derosa2012} and \cite{pereira-santaella2011} of log(L(2-10 keV)) = 42.60 and 42.34 for the different spectra, their results are suggestive of flux variability.
In fact, these luminosities agreed well with our estimates. Our analysis shows that this variability is related with the nuclear continuum.

\subsection{ESO\,138-G01}

\cite{alloin1992} optically classified this galaxy as a type 2 Seyfert. It shows a jet-like morphology at radio frequencies \citep{morganti1999}. The \emph{XMM}--Newton image shows a compact source at hard X-ray energies (see Appendix \ref{multiimages}). It was classified as a \emph{Compton}-thick candidate \citep{collinge2000}.

This galaxy was observed three times with \emph{XMM}--Newton in 2007 and 2013. Variability analyses were not found in the literature.
We did not find X-ray variations.

\subsection{NGC\,6300}

NGC\,6300 is a barred spiral galaxy, whose type 2 Seyfert classification at optical frequencies was derived from the data reported in \cite{philips1983}. The \emph{HST} image shows dust lanes (see Appendix \ref{multiimages}). A nuclear counterpart was found at radio frequencies, without any jet structure \citep{ryder1996}. 
NGC\,6300 was classified as a changing-look AGN, observed in the \emph{Compton}-thick state with \emph{RXTE} in  1997 and in the \emph{Compton}-thin state with \emph{BeppoSAX} in 1999 \citep{guainazzi2002a}. 

The galaxy was observed once with \emph{XMM}--Newton in 2001, and five times with \emph{Chandra} during 2009.
\cite{guainazzi2002a} found variations due to a difference in the normalization of the power law when comparing \emph{BeppoSAX} and \emph{RXTE} data. All the observations analyzed in this work caught the object in the thin state. Variations in the normalizations at soft and hard energies were found when comparing \emph{Chandra} and \emph{XMM}--Newton data.

\cite{matsumoto2004} and \cite{awaki2005,awaki2006} studied the light curve from \emph{XMM}--Newton data and found rapid variations at hard energies.

Variations in the 14--195 keV energy band were analyzed by \cite{soldi2013} using data from the \emph{Swift}/BAT 58-month survey, who estimated an intrinsic variability amplitude of 17[14-20]\%.

\subsection{NGC\,7172}

NGC\,7172 is an early type galaxy located in the HCG\,90 group, that shows dust lanes \citep[][see also Appendix \ref{multiimages}]{sharples1984}. 
Optically classified as a type 2 Seyfert (see an optical spectrum in Appendix \ref{multiimages}), no broad lines have been observed in polarized light \citep{lumsden2001}. A radio core was detected with \emph{VLA} data \citep{unger1987}. At IR frequencies, \cite{sharples1984} found variations in timesclaes of about three months. The nucleus of this galaxy is not detected at UV frequencies with the OM (see Table \ref{obsSey}). Even if \emph{Chandra} data are available for this source, they suffer from strong pileup. The \emph{XMM}--Newton image shows a compact source (see Appendix \ref{multiimages}). 

\cite{guainazzi1998} first reported X-ray flux variations in this source using \emph{ASCA} data. They found short term variations (hours) from the analysis of a light curve from 1996 and long term variations when comparing the flux of these data with previous data from 1995, when it was about three times brighter.
\cite{risaliti2002} studied two \emph{BeppoSAX} observations taken in October 1996 and November 1997 and fitted the data with an absorbed power-law, a thermal component, a cold reflection, a warm reflection and a narrow gaussian line. They reported very similar spectral parameters for the two spectra. 

This galaxy was observed once with \emph{Chandra} in 2000 and three times with \emph{XMM}--Newton between 2002 and 2007.
\cite{lamassa2011} analyzed the \emph{XMM}--Newton spectra by fitting the data with the ME2PL model and needed to fit independently the normalization of the power law. They reported luminosities of log(L(2-10 keV)) = 42.96$^+_-$0.03 (for the spectrum from 2007) and 42.61$^+_-$0.03 (for the other two spectra).
These results agree well with our SMF1.

\cite{awaki2006} analyzed the \emph{XMM}--Newton light curve from 2002. They did not find significant variability when computing the normalized excess variance.

At higher energies, \cite{beckmann2007} reported an intrinsic variability of $S_{Vc}=12^+_-9 \%$ within 20 days using \emph{Swift}/BAT data, and using data from the \emph{Swift}/BAT 58-month survey, \cite{soldi2013} reported a variability amplitude of 28[25-31]\%, both in the 14--195 keV energy band.

\subsection{NGC\,7212}

This galaxy is interacting with a companion (see the \emph{2MASS} image in Appendix \ref{multiimages}). It was optically classified as a type 2 Seyfert galaxy \citep[][see an optical spectrum in Appendix \ref{multiimages}]{veilleuxosterbrock1987}. Broad lines were detected in polarized light \citep{tran1992}. 
At radio wavelenghts, a nuclear counterpart was found together with the interacting galaxy \citep{falcke1998}.
A point-like source is detected at hard X-rays (see Appendix \ref{multiimages}). It was classified as a \emph{Compton}-thick candidate \citep{severgnini2012}.

It was observed once with \emph{Chandra} in 2003 and once with \emph{XMM}--Newton in 2004. \cite{bianchi2006} reported the same fluxes for the two spectra, also in agreement with our results.

\subsection{NGC\,7319}

NGC\,7319 is a spiral galaxy located in the Stephan's Quintet, a group composed by six galaxies including a core of three galaxies \citep{trinchieri2003}. These three galaxies were also observed at radio wavelenghts with \emph{VLA} \citep{aoki1999} and later with \emph{MERLIN} \citep{xanthopoulos2004}, revealing a jet structure in NGC\,7319. 
It has been optically confirmed as a type 2 Seyfert \citep[][see an optical spectrum in Appendix \ref{multiimages}]{rodriguezbaras2014}. The nucleus of this galaxy is not detected at UV frequencies with the OM (see Table \ref{obsSey}). At X-rays, a point-like source is observed in the 4.5--8.0 keV energy band, and it shows extended emission at soft X-ray energies (Appendix \ref{multiimages}).

It was observed twice with \emph{Chandra} in 2000 and 2007, and once with \emph{XMM}--Newton in 2001.
We did not find variability studies in the literature. We found variations in the nuclear power of the nucleus, accompained by absorber variations at soft energies.

One \emph{Chandra} and the \emph{XMM}--Newton light curves were analysed, but short-term variations were not detected.

\section{Images}

The images in the next sections will be published in the journal.

\subsection{\label{multiimages} Optical spectra, and X-ray, 2MASS and optical \emph{HST} images}

In this appendix we present images at different wavelenghts for each energy, and the optical spectrum when available from NED.
At X-rays we extracted \emph{Chandra} data in four energy bands: 0.6-0.9 keV (top-left), 1.6-2.0 keV (top-middle), 4.5-8.0 keV (top-right), and 0.5-10.0 keV (bottom-left). The {\sc csmooth} task included in CIAO was used to adaptatively smooth the three images in the top panels (i.e., the images in the 0.5-10.0 keV energy band are not smoothed), using a fast Fourier transform algorithm and a minimum and maximum significance level of the signal-to-noise of 3 and 4, respectively. 
When data from \emph{Chandra} was not available, \emph{XMM}--Newton images were extracted in the same energy bands, and the {\sc asmooth} task was used for adaptatively smooth the images.
At infrared frequencies, we retrieved an image from 2MASS in the $K_s$ filter\footnote{http://irsa.ipac.caltech.edu/applications/2MASS/IM/interactive.html}.
At optical frequencies we used images from the \emph{Hubble} Space telescope (\emph{HST})\footnote{http://hla.stsci.edu/}, preferably in the F814W filter but when it was not available we retrieved an image in the F606W filter.
\emph{HST} data have been processed following the sharp dividing method to show the internal structure of the galaxies \citep{marquez1996}.
The red squares in the bottom images represent the area covered by the \emph{HST} image (presented in the bottom-right panel when available).
In all images the gray levels extend from twice the value of the background dispersion to the maximum value at the center of each galaxy. We used IRAF \footnote{http://iraf.noao.edu/} to estimate these values.

\subsection{\label{Ximages} \emph{Chandra} and  \emph{XMM}--Newton images }

In this appendix we present the images from \emph{Chandra} (left) and \emph{XMM}-Newton (right) that were used to compare the spectra from these two instruments in the 0.5-10 keV band. In all cases, the gray scales extend from twice the value of the background dispersion to the maximum value at the center of each galaxy.

\onecolumn

\section{\label{lightcurves} Light curves}

In this appendix the plots corresponding to the light curves are provided. Three plots per observation are presented, corresponding to soft (left), hard (middle), and total (right) energy bands. Each light curve has a minimum of 30 ksec (i.e., 8 hours) exposure time, while long light curves are divided into segments of 40 ksec (i.e., 11 hours). Each segment is enumerated in the title of the light curve. Count rates versus time continua are represented. The solid line represents the mean value, dashed lines the $^+_-$1$\sigma$ from the average.

\end{document}